\documentclass[10pt]{article}
\pdfoutput=1
\usepackage{jheparxiv}
\usepackage[latin1]{inputenc}
\usepackage{latexsym,diagbox,stackrel}
\usepackage{relsize}
\usepackage{shuffle}
\usepackage{amsmath}
\usepackage{mathrsfs}
\usepackage{arydshln,leftidx,mathtools}
\usepackage{bbold}
\usepackage{graphicx}
\usepackage{subcaption}
\usepackage{pdflscape}
\usepackage{amssymb}
\usepackage{multirow}
\usepackage{enumerate}
\usepackage{blkarray}
\usepackage[usenames,dvipsnames,svgnames,table]{xcolor}
\usepackage{cleveref}
\usepackage{tikz}
\usepackage{graphicx}
\usepackage{slashed}
\usetikzlibrary{shapes.geometric, arrows}
\definecolor{light-gray}{gray}{0.95}
\definecolor{light-grayII}{gray}{0.85}
\usepackage{array}
\usepackage{pdfpages}

\def\d{\text{d}}

\newcommand{\C}{\mathbb{C}}

\newcommand{\dbar}{\bar\partial}

\newcommand{\tr}{\mathrm{tr}}

\newcommand{\pf}{\text{Pf} \,}

\newcommand{\px}{\raisebox{.15\baselineskip}{\Large\ensuremath{\wp}}}

\subheader{\hfill \begin{tabular}{r} \texttt{QMUL-PH-18-08} \end{tabular}}
 
\title{\Large Two-Loop Scattering Amplitudes from Ambitwistor Strings: from Genus Two to the Nodal Riemann Sphere}
\author[a,c]{Yvonne Geyer}
\author[b,c]{\& Ricardo Monteiro}
        
\affiliation[a]{School of Natural Sciences, Institute for Advanced Study \\
        Einstein Drive, Princeton, NJ 08540, USA}

\affiliation[b]{Centre for Research in String Theory, School of Physics and Astronomy \\
        Queen Mary University of London, E1 4NS, United Kingdom}
        
\affiliation[c]{Kavli Institute for Theoretical Physics \\
        University of California Santa Barbara, CA 93106-4030, USA}
        
\emailAdd{yvonnegeyer@ias.edu}
\emailAdd{ricardo.monteiro@qmul.ac.uk}

\abstract{We derive from ambitwistor strings new formulae for two-loop scattering amplitudes in supergravity and super-Yang-Mills theory, with any number of particles. We start by constructing a formula for the type II ambitwistor string amplitudes on a genus-two Riemann surface, and then study the localisation of the moduli space integration on a degenerate limit, where the genus-two surface turns into a Riemann sphere with two nodes. This leads to scattering amplitudes in supergravity, expressed in the formalism of the two-loop scattering equations. For super-Yang-Mills theory, we import `half' of the supergravity result, and determine the colour dependence by considering a current algebra on the nodal Riemann sphere, thereby completely specifying the two-loop analogue of the Parke-Taylor factor, including non-planar contributions. We also present in appendices explicit expressions for the Szeg\H{o} kernels and the partition functions for even spin structures, up to the relevant orders in the degeneration parameters, which may be useful for related investigations in conventional superstring theory.}

\begin{document}

\maketitle

\section{Introduction}\label{sec:intro}

\subsection{Motivation}

String theory has provided a variety of crucial insights into quantum field theory, opening the way to remarkable dualities and motivating new principles to constrain effective field theories. Nonetheless, one of the oldest insights, and perhaps the most basic one, remains as powerful as ever: the striking rearrangement of perturbative field theory, seen as the low-energy limit of perturbative string theory. In fact, certain scattering amplitudes in field theory, particularly in the presence of supersymmetry, were first computed with the aid of string theory, as in Ref.~\cite{Green:1982sw} and many others. The appeal of this programme is that the string theory worldsheet allows for the use of powerful techniques of two-dimensional conformal field theory, leading to a formalism that is strikingly different from the traditional Feynman diagram expansion of perturbative field theory. The worldsheet provides, for instance, a picture for the scattering of closed strings as the `double copy' of the scattering of open strings \cite{Kawai:1985xq}. This leads directly to formulae that relate scattering amplitudes in gravity and in gauge theory, which have been explored to great effect, especially since a diagrammatic version of this double copy was proposed \cite{Bern:2008qj,Bern:2010ue}.

While the lessons from perturbative string theory are very encouraging, the calculations are challenging beyond the first few orders; at two loops, see e.g.~\cite{DHoker:1988ta,DHoker:2002hof,DHoker:2001kkt,DHoker:2001qqx,DHoker:2001foj,DHoker:2001jaf,DHoker:2005dys,DHoker:2005vch} for the impressive RNS superstring results and \cite{Magnea:2013lna,Magnea:2015fsa} for recent examples of bosonic string calculations. The computation of loop corrections requires higher-genus string worldsheets, whose mathematical description is highly elaborate and not fully developed. An understanding of this description seems necessary or at least very helpful even if we are only interested in the low-energy field theory limit.

A new application of string methods to field theory has recently come to fruition, following the understanding that at least certain massless perturbative field theories can be described directly by a string-type theory. The spectrum of such strings coincides with that of the corresponding field theory, and no low-energy limit is required. These are the {\it ambitwistor strings} proposed by Mason and Skinner \cite{Mason:2013sva}. They are inspired by Witten's seminal model of a twistor string associated to four-dimensional gauge theory \cite{Witten:2003nn}. The twistor string leads to beautiful expressions for tree-level scattering amplitudes in gauge theory as residue integrals in the moduli space of a Riemann sphere \cite{Roiban:2004yf}. These expressions were more recently extended into an elegant formalism to describe tree-level massless scattering in any spacetime dimension, for a variety of theories, by Cachazo, He and Yuan (CHY) \cite{Cachazo:2013gna,Cachazo:2013hca,Cachazo:2013iea,Cachazo:2014xea}. The construction of ambitwistor strings was guided by the requirement of reproducing the CHY formulae \cite{Mason:2013sva,Ohmori:2015sha,Casali:2015vta}. 

Given that ambitwistor strings are supposed to directly describe perturbative field theories, an obvious question is what happens at loop level. In conventional string theory, the field theory limit ($\alpha'\to0$) is associated to a degeneration limit of the moduli space; for instance, at one loop the limit is such that $\alpha'\,\text{Im}(\tau)$ stays finite, where $\tau$ is the torus modulus \cite{Green:1982sw}. Higher-genus mathematical objects, like theta functions, give way to much simpler expressions in that limit. How is this to happen for ambitwistor strings, which are already field theories, and possess no $\alpha'$ parameter? The answer was given in Refs.~\cite{Geyer:2015bja,Geyer:2015jch,Geyer:2016wjx}, following genus one \cite{Adamo:2013tsa,Casali:2014hfa} and genus two \cite{Adamo:2015hoa} studies: the residue integral in moduli space localises on a degenerate limit simply via the use of the residue theorem. The resulting worldsheet is a Riemann sphere with nodes (pairs of identified points), through which flow the loop momenta. This provides a new formalism that extends the CHY representation from tree-level amplitudes to loop-level integrands. The type of formula for the loop integrands is naturally interpreted as a forward limit of tree-level amplitudes \cite{He:2015yua,Baadsgaard:2015twa,Geyer:2015jch,Cachazo:2015aol,Feng:2016nrf,He:2016mzd,He:2017spx,Roehrig:2017gbt,Geyer:2017ela}, in the spirit of the Feynman tree theorem.

In this paper, we will construct the two-loop formulae obtained from ambitwistor strings for loop integrands in type II supergravity and in super-Yang-Mills theory. The detailed derivation from the genus-two ambitwistor string will put into firm footing some heuristic aspects of our earlier analysis \cite{Geyer:2016wjx}, and will extend the  four-point formulae presented there to any number of particles. The elaborate technical content of our analysis indicates that the precise approach that we employ here may be too challenging at higher loops. We hope, however, that our results will be sufficient to identify an easier generalisation route. The long term goal is to develop a formalism based directly on the nodal Riemann sphere, without any reference to higher-genus surfaces. The first steps of such a formalism were accomplished at one loop in \cite{Roehrig:2017gbt}, where formulae previously obtained via the degeneration of the torus were reproduced on the sphere using a `gluing operator'. Moreover, we will see in this work another important advantage of the nodal sphere approach: we propose formulae for two-loop super-Yang-Mills theory amplitudes based on the nodal sphere, without starting from a genus-two expression.

Before proceeding with a summary of our main results, we provide here a brief survey of work on ambitwistor strings, for the benefit of the reader unfamiliar with this topic. As we mentioned, they were proposed in  \cite{Mason:2013sva} as worldsheet chiral conformal field theories reproducing the CHY formulae for tree-level scattering amplitudes \cite{Cachazo:2013hca}. The first examples of ambitwistor strings described the tree-level amplitudes of type II supergravity, super-Yang-Mills theory (in a heterotic-type model) and the bi-adjoint $\phi^3$ scalar theory. Later on, in \cite{Ohmori:2015sha,Casali:2015vta}, a variety of other models -- distinguished by the worldsheet matter content and symmetries -- were engineered in order to reproduce CHY formulae for several interesting theories of massless particles \cite{Cachazo:2014nsa,Cachazo:2014xea}, including Einstein-Yang-Mills, Dirac-Born-Infeld and the non-linear sigma model. Other variations on the models of Mason and Skinner include: a pure spinor version of the supergravity and super-Yang-Mills models \cite{Berkovits:2013xba,Gomez:2013wza}; a version based on twistor variables for theories in four spacetime dimensions \cite{Geyer:2014fka}, with preliminary work at one loop \cite{Farrow:2017eol}; a derivation of the anomalies of the type II theory on a curved background \cite{Adamo:2014wea}, leading to the supergravity equations of motion as the consistency condition for the background; studies of the soft behaviour of amplitudes based on the relation of ambitwistor space to null infinity \cite{Geyer:2014lca,Adamo:2014yya,Adamo:2015fwa}; an ambitwistor string field theory construction \cite{Reid-Edwards:2017goq}; a class of models adapted to the projective null cone, describing certain conformal field theories \cite{Adamo:2017zkm}; models describing certain higher-derivative theories \cite{Azevedo:2017lkz};
and a calculation of the three-point amplitude for scattering on plane wave backgrounds \cite{Adamo:2017sze}. Along with these studies, there is important work on the precise connection of ambitwistor strings -- chiral theories with a massless spectrum -- to conventional string theory, in particular to the null string \cite{Casali:2016atr, Casali:2017zkz,Casali:2017mss, Siegel:2015axg, Yu:2017bpw, Azevedo:2017yjy, Azevedo:2018dgo}.

\subsection{Summary of results} \label{subsec:summary}

We present here a summary of our final formulae for type II supergravity and super-Yang-Mills theory at two loops. 
In both cases, the amplitude is expressed as\footnote{In this summary of results, we have chosen to extract the form degree of $\mathscr{I}^{(2)}_n$, $\mathcal{E}_A$ and other objects into the overall $\d^{n+4}\sigma_{A}$ in \cref{equ:initial_formula}. This is the most common notation in the scattering equations literature. In the body of this paper, however, we keep the form degrees of each object, so the reader should bear this in mind when comparing the expressions here with those in other sections.}
\begin{equation} \label{equ:initial_formula}
 \mathcal{M}_n=\int \!\frac{\d^{d}\ell_1\,\d^{d}\ell_2}{\ell_1^2\,\ell_2^2}
 \int  \hspace{-15pt}{\phantom{\Bigg(}}_{\mathfrak{M}_{0,n+4}}\hspace{-10pt}
\frac{\d^{n+4}\sigma_{A}}{\text{vol SL}(2,\mathbb{C})^2}\,
\,\,\prod_{A}\,\bar\delta\big( \mathcal{E}_A\big)\,\,\mathscr{I}^{(2)}_n\,,
\end{equation}
where $\sigma_A\in\{\sigma_{1^\pm},\sigma_{2^\pm},\sigma_i\}$ are punctures on the sphere associated to loop momenta insertions ($\pm\ell_I$ for $\sigma_{I^\pm}$) and the external particles ($i=1,\cdots,n$). The loop integrand is therefore written as a CHY-type integral, with the integration completely localised on the solutions to the two-loop scattering equations:
\begin{subequations} \label{equ:2loopSE}
\begin{align}
 \mathcal{E}_i&=k_i\cdot\ell_1\left(  \frac{1}{\sigma_i-\sigma_{1^+}} - \frac{1}{\sigma_i-\sigma_{1^-}} \right)
 +k_i\cdot\ell_2\left(  \frac{1}{\sigma_i-\sigma_{2^+}} - \frac{1}{\sigma_i-\sigma_{2^-}} \right)+\sum_{j\neq i} \frac{k_i\cdot k_j}{\sigma_i-\sigma_j} \,,\\
\pm \mathcal{E}_{1^\pm}&=\frac{1}{2}\left(\ell_1+\ell_2\right)^2 \left(  \frac{1}{\sigma_{1^\pm}-\sigma_{2^+}} - \frac{1}{\sigma_{1^\pm}-\sigma_{2^-}} \right)+\sum_{j} \frac{\ell_1\cdot k_j}{\sigma_{1^\pm}-\sigma_j}\,,\\
\pm \mathcal{E}_{2^\pm}&=\frac{1}{2}\left(\ell_1+\ell_2\right)^2 \left(  \frac{1}{\sigma_{2^\pm}-\sigma_{1^+}} - \frac{1}{\sigma_{2^\pm}-\sigma_{1^-}} \right)+\sum_{j} \frac{\ell_2\cdot k_j}{\sigma_{2^\pm}-\sigma_j}\,.
\end{align}
\end{subequations}

\begin{figure}[t]
	\centering 
	  \includegraphics[width=7cm]{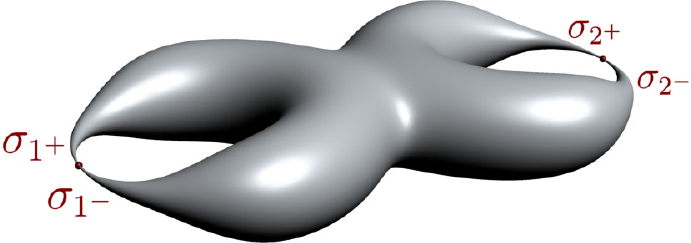}
	\caption{The bi-nodal Riemann sphere, with nodes parametrised by $\sigma_{1^\pm}$ and $\sigma_{2^\pm}$ representing the two loops of field theory. }
	\label{fig:binodal}
\end{figure}

In the supergravity case, we derive our formula from the genus-two ambitwistor string amplitude, which is based on the {\it genus-two scattering equations} involving the period matrix. The crucial ingredient in the derivation is the residue theorem on moduli space. We use it to turn the genus-two formula into a formula on the bi-nodal Riemann sphere; see \cref{fig:binodal}. The latter formula is based on what are more appropriately called the {\it two-loop scattering equations} \eqref{equ:2loopSE}.
The result for supergravity follows from the asymptotics of the (maximal non-separating) degeneration limit leading to the bi-nodal Riemann sphere. While the genus-two origin of the supergravity formula requires $d=10$, the formula \eqref{equ:initial_formula} on the bi-nodal sphere can be dimensionally reduced as usual to, for example, maximal $\mathcal N=8$ supergravity in $d=4$. We focused on the even spin structures contribution (the full result for $d<10$), and on NS-NS external states with polarisation tensors $\epsilon_i^\mu\tilde\epsilon^\nu_i$, which form a basis for general NS-NS states. The supergravity result is
\begin{equation}\label{equ:int_sugra_final}
\mathscr{I}^{(2),\, \text{sugra}}_n = \mathcal{I}^{(2)}_n(\epsilon)\,\,{\mathcal{I}}^{(2)}_n(\tilde\epsilon)\,\, \frac{(1^+2^-)(1^-2^+)}{(1^+1^-)(2^+2^-)} \,,
\end{equation}
where $\mathcal{I}^{(2)}_n$ is the analogue of the chiral integrand in conventional superstring theory, receiving contributions from all spin structures,
\begin{equation}\label{equ:int_chiral_final}
 \mathcal{I}_n^{(2)}=\mathcal{I}_n^{\text{NS}}+\mathcal{I}_n^{\text{R}2}+\mathcal{I}_n^{\text{R}1}+\mathcal{I}_n^{\text{RR}}\,.
\end{equation}
The ten even spin structures are naturally grouped into contributions corresponding to states running in the loops:
\begin{subequations} \label{equ:integrand_NS-R_intro}
\begin{align}
  \mathcal{I}_n^{\text{NS}}&=4J\,\sum_{n_1,n_2\in\{0,1\}}\,\mathcal{Z}_{\text{NS}}^{(-n_1,-n_2)}\, \pf\big(M_{\text{NS}}\big)\big|_{q_1^{n_1}q_2^{n_2}}\,,\\
  \mathcal{I}_n^{\text{R}2}&=2J\,\Bigg(\mathcal{Z}_{\text{R}2}^{(0,0)}\, \pf\big(M_{\text{R}2}\big)\big|_{q_1^0q_2^0}+\mathcal{Z}_{\text{R}2}^{(-1,0)}\, \pf\big(M_{\text{R}2}\big)\big|_{q_1^1q_2^0}\Bigg)\,,\\
  \mathcal{I}_n^{\text{R}1}&=2J\,\Bigg(\mathcal{Z}_{\text{R}1}^{(0,0)}\, \pf\big(M_{\text{R}1}\big)\big|_{q_1^0q_2^0}+\mathcal{Z}_{\text{R}1}^{(0,-1)}\, \pf\big(M_{\text{R}1}\big)\big|_{q_1^0q_2^1}\Bigg)\,,\\
  \mathcal{I}_n^{\text{RR}}&=J\,\mathcal{Z}_{\text{RR}_9}^{(0,0)}\, \pf\big(M_{\text{RR}_9}\big)\big|_{q_1^0q_2^0}+J\,\mathcal{Z}_{\text{RR}_0}^{(0,0)}\, \pf\big(M_{\text{RR}_0}\big)\big|_{q_1^0q_2^0}\,,
\end{align}
\end{subequations}
where we have, respectively, NS states running in both loops, NS state in loop 1 and Ramond state in loop 2, NS state in loop 2 and Ramond state in loop 1, and finally Ramond states in both loops. Here, $J^{-1} = (1^+2^+)(1^+2^-)(1^-2^+)(1^-2^-)$, and the partition function factors $\mathcal{Z}$ are described in \cref{sec:partition}. Moreover, $\pf(M)|_{q_1^{a}q_2^{b}}$ denotes the ${q_1^{a}q_2^{b}}$ coefficient in the Taylor expansion of the pfaffian of $M$ around $q_1=q_2=0$; the vanishing of the modular parameters $q_1$ and $q_2$ is the degeneration limit corresponding to the bi-nodal Riemann sphere. The definition of the matrices $M$, dependent on the states running in the loop and on the external polarisations, is given in \cref{sec:int_nodalRS}. Finally, the cross ratio in \cref{equ:int_sugra_final} ensures the absence of certain unphysical poles that are allowed by the two-loop scattering equations. While ref.~\cite{Geyer:2016wjx} first pointed out the need for this cross ratio, here we provide a derivation from first-principles based on the degeneration from the genus-two surface. In particular, the introduction of the cross ratio enables the extension of the domain of integration of the remaining genus-two modular parameter $q_3$, allowing for the final formula to be expressed as a moduli integral on the bi-nodal Riemann sphere, constrained only by the two-loop scattering equations.

Apart from the cross ratio, which is a new feature at two loops, we want to emphasise the similarity of our type II supergravity formula to the tree-level formula of CHY \cite{Cachazo:2013hca}, based on the pfaffian of a matrix analogous to our matrices, and to the one-loop formula of ref.~\cite{Geyer:2015bja}, which also includes contributions from different spin structures.

The super-Yang-Mills result is closely related to the final supergravity formula. In this case, however, we propose an expression directly on the bi-nodal Riemann sphere, instead of performing a delicate degeneration limit from a genus-two super-Yang-Mills formula (which may not even exist). The colour dependence is determined from a current algebra correlator on the sphere, in the spirit of the heterotic string, and the two nodes are represented by a sum over the colour indices of the corresponding current algebra insertions. The formula for the scattering of gluons with polarisation vectors $\epsilon_i^\mu$ is
\begin{equation}
\mathscr{I}^{(2),\, \text{sYM}}_n = \mathcal{I}^{(2)}_n(\epsilon)\,\,{\mathcal I}^{\text{PT}(2)}_n \,,
\end{equation}
where $\mathcal{I}^{(2)}_n$ was introduced in \cref{equ:int_chiral_final}, and the colour dependence is carried by the two-loop `Parke-Taylor factor',
\begin{equation}
{\mathcal I}^{\text{PT}(2)}_n = \sum_{\gamma \in S'_{n+2}} 
\frac{\text{tr}([[\cdots[[[T^{a_{1^+}},T^{a_{\gamma(1)}}],T^{a_{\gamma(2)}}],T^{a_{\gamma(3)}}],\cdots],T^{a_{\gamma(n+2)}}]T^{a_{1^-}}) \, \delta^{a_{1^+},a_{1^-}} \, \delta^{a_{2^+},a_{2^-}}}
{\big(1^+\,\gamma(1)\,\gamma(2)\,\gamma(3)\cdots \gamma(n+2)\, 1^-\big)}\,,
\end{equation}
where $(ijk\cdots l)\equiv(\sigma_i-\sigma_j)(\sigma_j-\sigma_k)\cdots (\sigma_l-\sigma_i)$. The sum is over permutations of the $n+2$ punctures $\{\sigma_{2^\pm},\sigma_i\}$, i.e., the punctures $\sigma_1^{\pm}$ are fixed.  We denote the set of permutations by $S'_{n+2}$ (and not $S_{n+2}$) because we restrict the permutations to satisfy the following ordering of the nodal punctures: $(1^+\cdots 2^+\cdots2^-\cdots1^-)$; there are therefore $(n+2)!/2$ valid permutations. This restriction plays a role analogous to that of the cross ratio in \cref{equ:int_sugra_final}: it ensures the absence of unphysical poles. As in the supergravity case, our super-Yang-Mills formula is strongly reminiscent of the tree-level formula of CHY \cite{Cachazo:2013hca} and the one-loop formula of ref.~\cite{Geyer:2015bja}.

The detailed definition of the ingredients in \cref{equ:int_chiral_final} leading to $\mathcal{I}^{(2)}_n$ as described in this paper makes use of two extra marked points, $x_1$ and $x_2$, which are not part of the CHY-type integration in \eqref{equ:initial_formula}. These are associated with a gauge choice, the location of the supersymmetry picture-changing operators at genus two, analogous to conventional superstring theory. We prove that our formulae do not depend on this gauge choice, but leave for future work the possibility of simplifying the formulae with a smart choice of these physically irrelevant marked points. In this paper, we merely check how this simplification occurs in practice for the four-point formula. 

The expressions given here describe type II supergravity and super-Yang-Mills theory in $d=10$ (except for odd spin structures, which we did not consider). Formulae for theories in fewer spacetime dimensions are obtained via dimensional reduction as usual. In the case of reduction on a 6-torus, the corresponding four-dimensional theories are ${\mathcal N}=8$ supergravity and ${\mathcal N}=4$ super-Yang-Mills theory. While all four-dimensional supergravities are expected to be ultraviolet divergent in perturbation theory, and therefore do not possess an S-matrix, one can still define a loop integrand at any loop order. Indeed, this has been the subject of intense work that aims to study in detail the ultraviolet properties; see \cite{Bern:2018jmv} for recent results in ${\mathcal N}=8$ supergravity. Our ten-dimensional `amplitudes' are understood in this context -- the result is the loop integrand itself.

The amplitude formulae for both supergravity and the super-Yang-Mills reproduce known expressions for two-loop four-particle scattering amplitudes \cite{Geyer:2016wjx}. Moreover, we verify that only physical factorisation channels contribute to the amplitude, and the amplitude is independent of the gauge choice associated to the two extra marked points, $x_1$ and $x_2$, as indicated above. However,  a direct comparison of our results for $n>4$ to known formulae using factorisation  is beyond the scope of this paper and left for future work. Instead, our focus throughout the paper lies on deriving \eqref{equ:initial_formula} from the ambitwistor string correlator at genus two.

\subsection{Outline of paper}

The paper is organised as follows. In \cref{sec:review}, we review the type II ambitwistor string, its relation to the tree-level CHY formulae for scattering amplitudes, and the one-loop extension of this story. Section~\ref{sec:rev_basics} is a brief introduction to Riemann surfaces, with particular emphasis on genus two. We construct the type II ambitwistor string amplitude on a genus-two surface in \cref{sec:g=2}. In  \cref{sec:nodalRS}, we discuss in detail the localisation of the genus-two amplitude on a degenerate limit of the moduli space, via the residue theorem. This procedure leads to an expression for the type II amplitude on a bi-nodal Riemann sphere, which we develop in full detail in \cref{sec:contour_argument}. Section~\ref{sec:sym} presents the analogous formula on the bi-nodal Riemann sphere for super-Yang-Mills amplitudes. We conclude in \cref{sec:discussion} with a discussion of future directions.

%%%%%%%%%%%%%%%%%%%%%%%%%%%%%%
%%%%%%%%%%%%%%%%%%%%%%%%%%%%%%
\section{Review of the ambitwistor string}\label{sec:review}

Ambitwistor strings are two-dimensional chiral conformal field theories, which are conjectured to describe the perturbative interactions of quantum field theories of massless particles. Their construction in \cite{Mason:2013sva} was guided by the CHY formulae for scattering amplitudes \cite{Cachazo:2013hca}. For most of this work, except for a later section where we consider colour degrees of freedom, we will focus on the RNS ambitwistor string, which is a string-like formulation of type II supergravity.

\subsection{Type II ambitwistor string}

The action of the type II ambitwistor string can be written as
\begin{equation}
\label{equ:typeII}
 S=\frac{1}{2\pi}\int_\Sigma P\cdot \dbar X +\frac{1}{2}\psi\cdot\dbar\psi+\frac{1}{2}\tilde\psi\cdot\dbar\tilde\psi
  - e \bigg(P\cdot \partial X+\frac{1}{2}\psi\cdot\partial\psi+\frac{1}{2}\tilde\psi\cdot\partial\tilde\psi\bigg)
   -\frac{\tilde{e}}{2}\,P^2 -\chi P\cdot\psi-\tilde\chi P\cdot\tilde\psi \,.
\end{equation}
The fields take values in the following line bundles:
\begin{subequations}\label{equ:fields_def}
\begin{align}
 X&: \Sigma\rightarrow M\,, && e\,,\,\tilde{e} \in \Omega^{0,1}\big(\Sigma, T_\Sigma\big)\,,\\
 P&\in \Omega^{1,0}\big(\Sigma, T^*M\big)\,, && \chi\,,\,\tilde\chi\in\Pi\Omega^{0,1}\big(\Sigma,T_\Sigma^{1/2}\big)\,,\\
  \psi, &\,\tilde\psi\in \Pi\Omega^{0}\big(\Sigma, K_\Sigma^{1/2}\otimes TM\big)\,.
\end{align}
\end{subequations}
In CFT language, this means that these worldsheet fields have a single component (hence sections of line bundles) with the following conformal weight: $(0,0)$ for $X$, $(1,0)$ for $P$, $(1/2,0)$ for $\psi,\tilde\psi$, $(-1,1)$ for $e,\tilde{e}$, and $(-1/2,1)$ for $\chi,\tilde{\chi}$. Moreover, $\Pi\Omega$ denotes fermionic form-fields. Notice that, in our notation, $\dbar=\d\bar z \partial_{\bar z}$, so that each term in the action is a top form on the Riemann surface $\Sigma$.

The bosonic fields $e,\tilde{e}$ (known as Beltrami differentials) and the fermionic fields $\chi,\tilde{\chi}$ are Lagrange multipliers enforcing the constraints $P^2=0$ and $P\cdot\psi=P\cdot\tilde\psi=0$ that are associated to symmetries of the action. The constraint enforced by $e$ is the vanishing of the chiral stress-energy tensor, generating holomorphic diffeomorphisms,
\begin{subequations}
\begin{align}
 &  \delta_v X^\mu =v\,\partial X^\mu\,,&& \delta_v P_\mu =\partial(v\,P_\mu)\,, && \delta_v e = \dbar v +v\partial e - e\partial v\,, && \delta_v \tilde{e} = v\partial\tilde{e} - \tilde{e}\partial v\,,  \nonumber \\
 &  \delta_v \psi^\mu =v\,\partial \psi^\mu + \frac{1}{2} \psi^\mu\partial v\,, &&  \delta_v \tilde\psi^\mu =v\,\partial\tilde \psi^\mu + \frac{1}{2}\tilde\psi^\mu\partial v\,, &&  \delta_v \chi =v\,\partial \chi - \frac{1}{2} \chi\partial v\,, && \delta_v \tilde\chi =v\,\partial\tilde \chi - \frac{1}{2}\tilde \chi\partial v \nonumber \,,
\end{align}
\end{subequations}
On the other hand, $\tilde e$ is associated to the `ambitwistor gauge transformation', affecting only the bosonic fields,
$$
\delta_{\alpha}X^\mu = \alpha \,\eta^{\mu\nu}P_\nu\,, \qquad \delta_\alpha P_\mu = 0\,, \qquad \delta_{\alpha}e = 0 \,, \qquad \delta_{\alpha}\tilde{e} = \dbar \alpha -\alpha \partial e + e\partial\alpha \,.
$$
The fermionic symmetries are a supersymmetric extension of this ambitwistor gauge transformation. In particular, the constraint $P \cdot\psi$ associated to $\chi$ generates
\begin{subequations}
\begin{align}
 &\delta_{\epsilon}X^\mu = \epsilon\psi^\mu &&  \delta_\epsilon P_\mu = 0\,, && \delta_\epsilon \psi^\mu = \epsilon \eta^{\mu\nu}P_\nu \,, &&\delta_\epsilon \tilde\psi_\mu = 0\,, \nonumber\\
 & \delta_\epsilon e = 0\,, &&\delta_\epsilon\tilde{e} = 2\epsilon \chi\,, &&\delta_\epsilon\chi = \dbar\epsilon +e\partial \epsilon - \frac{1}{2}\epsilon\partial e\,, && \delta_\epsilon\tilde{\chi}= 0 \,, \nonumber
\end{align}
\end{subequations}
and analogously for $\tilde\chi$.

Ambitwistor space is the space of null geodesics of complexified spacetime, which in this paper is simply complexified Minkowski spacetime. The features of the action \eqref{equ:typeII} that effectively lead to a supersymmetrised version of ambitwistor space as the target space are (i) the constraint $P^2=0$, together with the associated `ambitwistor gauge transformation', which identifies points in the cotangent bundle that lie along the same geodesic, and (ii) the $\mathcal N =2$ supersymmetric extension of the constraint $P^2=0$ and the associated transformations. Notice that there is a crucial difference with respect to the conventional type II string. The `square' of each supersymmetry transformation is not the transformation generated by the stress-tensor, but the one generated by $P^2$,
\begin{equation}
\{ \delta_{\epsilon_1},\delta_{\epsilon_2} \} X^\mu = - \{ \epsilon_1,\epsilon_2 \} \eta^{\mu\nu}P_\nu\,, \qquad 
\{ \delta_{\epsilon_1},\delta_{\epsilon_2} \} \psi^\mu =0 \,,
\end{equation}
and analogously for $\tilde\epsilon$. Equivalently, this can be expressed in terms of the constraint algebra $\{P^2, \, P\cdot\psi,\,P\cdot\tilde\psi\}$ as \footnote{This algebra also plays an important role in the formuation of the ambitwistor string on curved backgrounds. Requiring the algebra to remain consistent at the quantum level  directly gives rise to the $d=10$ supergravity equations of motion, as explained beautifully in \cite{Adamo:2014wea}.}
\begin{equation}
 \big(P\cdot\psi\big)(z)\,\big(P\cdot\psi\big)(w)\sim\frac{P^2}{z-w}\,,\qquad  \big(P\cdot\tilde\psi\big)(z)\,\big(P\cdot\tilde\psi\big)(w)\sim\frac{P^2}{z-w}\,,\quad  \big(P\cdot\psi\big)(z)\,\big(P\cdot\tilde\psi\big)(w)\sim 0\,.
\end{equation}
While this algebra of constraints strongly resembles the RNS superstring agebra, the ambitwistor constraint $P^2/2$ bears no relation to the worldsheet stress-energy tensor $T$.\footnote{In contrast to the RNS superstring, where the role of $P$ is played by $\partial X$.} The fermionic constraints $P\cdot \psi$ and $P\cdot\tilde\psi$  therefore do \emph{not} generate worldsheet superdiffeomorphisms, but rather the supersymmetric extension of the worldsheet gauge theory constraint $P^2/2$. The symmetry group of the ambitwistor string thus consists of (non-supersymmetric) worldsheet diffeomorphisms and the  worldsheet gauge supergroup  PSL$(1,1|\C)$. In contrast to the superstring, all supersymmetries of the ambitwistor string thus reside in the gauge supergroup, and consequently the theory is  formulated over a Riemann surface, \emph{not} a super-Riemann surface.

A more obvious distinction between the action \eqref{equ:typeII} and its  string theory counterpart is that it has no dimensionful parameter -- no $\alpha'$. We can therefore anticipate that the spectrum is massless.

\subsection{BRST quantization}\label{sec:BRST}

We now proceed to quantise the ambitwistor string, according to the BRST procedure. We follow closely the presentation in \cite{Adamo:2013tsa,Roehrig:2017gbt}. We start by introducing two $bc$ and two $\beta\gamma$ ghost systems for the gauge symmetries,
\begin{subequations}\label{equ:ghosts_def}
 \begin{align}
  &b, \tilde{b}\in \Pi\Omega^0\big(\Sigma, K_\Sigma^2\big) \,, && \beta,\tilde \beta \in\Omega^0\big(\Sigma, K_\Sigma^{3/2}\big) \,,\\
  &c, \tilde{c}\in \Pi\Omega^0\big(\Sigma, T_\Sigma\big) \,, && \gamma,\tilde \gamma \in\Omega^0\big(\Sigma, T_\Sigma^{1/2}\big) \,.
 \end{align}
\end{subequations}
In CFT language, the conformal weights for the fermionic ghosts are $(2,0)$ for $b, \tilde{b}$, and $(-1,0)$ for $c, \tilde{c}$, while for the bosonic ghosts we have $(3/2,0)$ for $\beta,\tilde\beta$, and $(-1/2,0)$ for $\gamma,\tilde\gamma$.

For worldsheet gravity, we proceed in a similar manner as in string theory: we simply set $e=0$ and integrate over the moduli space of the Riemann surface. Moreover, the ghosts $bc$ play the usual role in vertex operators. While $\tilde e$ is not the complex conjugate of $e$, the gauge fixing of both $e$ and $\tilde{e}$ still leads to a measure on moduli space, albeit one that completely localises the integration, as we shall see in a moment.

After setting $e=0$, the symmetry transformations associated to $\tilde{e}$ and $\chi$, $\tilde\chi$ vary these fields only within a fixed Dolbeault ($\dbar$) cohomology class. Since these cohomology classes are finite dimensional, the functional integrations over these fields are effectively reduced to finite-dimensional integrals after gauge fixing (apart from the ghosts). In anticipation of the inclusion of vertex operators, we consider the cohomology classes of a Riemann surface with $n$ maked points $\{z_i\}$ at which the gauge transformations are required to vanish. Now, since $\tilde{e} \in \Omega^{0,1}\big(\Sigma, T_\Sigma\big)$, it is $\dbar$-closed. Its cohomology class is denoted as $H^{0,1}\big(\Sigma, T_\Sigma(-z_1...-z_n)\big)$ and has dimension $n+3g-3$, so we can span it with a basis of Beltrami differentials $\{\mu_r\}$, with $r=1,\ldots,n+3g-3$. Similarly, the cohomology class for $\chi$ or $\tilde\chi$ is $H^{0,1}\big(\Sigma, T_\Sigma^{1/2}(-z_1...-z_n)\big)$ and has dimension $n+2g-2$, so we can span it with a basis $\{\chi_\alpha\}$, with $\alpha=1,\ldots,n+2g-2$. Gauge fixing the $\dbar$-exact part of the fields to zero corresponds to adding a gauge-fixing term to the action of the form
\begin{equation}\label{equ:S_GF}
 S_{\text{GF}}=\frac{1}{2\pi}\int_\Sigma \big\{Q\,,\,\tilde b\, (\tilde{e}-\tilde{e}_0) +\beta\, (\chi-\chi_0)+\tilde\beta\, (\tilde\chi-\tilde\chi_0)\big\} \,,
\end{equation}
where
\begin{equation}
\tilde e_0 = \sum_{r=1}^{n+3g-3} s_r\mu_r\,, \qquad 
\chi_0=\sum_{\alpha=1}^{n+2g-2} \zeta_\alpha\chi_\alpha\,, \qquad 
\tilde\chi_0=\sum_{\alpha=1}^{n+2g-2} \tilde\zeta_\alpha \chi_\alpha\,,
\end{equation}
where $s_r$ are bosonic parameters and $\zeta_\alpha,\tilde\zeta_\alpha$ are fermionic parameters. The gauge-fixing procedure introduces finite-dimensional integrations over the $s_r$ and the $\zeta_\alpha,\tilde\zeta_\alpha$, as well as over the fermionic parameters $q_r=Q\circ s_r$ and the bosonic parameters $\varrho=Q\circ \zeta_\alpha$, $\tilde\varrho=Q\circ \tilde \zeta_\alpha$. Moreover, it introduces functional integrations over the Nakanishi-Lautrup fields $H=Q\circ \tilde b$ and $G=Q\circ \beta$, $\tilde G=Q\circ \tilde \beta$. All these parameters and fields arise from the gauge-fixing term \eqref{equ:S_GF}.\footnote{Alternatively, following \cite{Witten:2012bh}, we can define  the action of $Q$ on the moduli space directly as an exterior derivative: $Q\circ \{ s_r,\zeta_\alpha,\tilde\zeta_\alpha\}=\{\d  s_r,\d \zeta_\alpha,\d \tilde\zeta_\alpha\}$. The later differentials on moduli space already provide the appropriate measure, since only the contributions that build up the complete moduli space measure give a non-vanishing contribution to the path integral in view of the ghost integrations.}

Let us consider the parts of the path integral associated with $\tilde e$. The important terms in the complete action are
\begin{equation}
\frac{1}{2\pi} \int_\Sigma - \frac{1}{2} \tilde e P^2 + H \tilde e - \sum_{r=1}^{n+3g-3} \left( s_r \mu_r H  + q_r \mu_r \tilde b
\right)\,.
\end{equation}
Integrating out $\tilde e$ field fixes $H=P^2/2$. The integrations over $s_r$ and $q_r$ then lead to the insertions of picture changing operators (PCOs)
\begin{equation}\label{equ:PCOint}
\prod_{r=1}^{n+3g-3} \bar{\delta} \left(\int_\Sigma \mu_r P^2\right) \,\, \left(\int_\Sigma \mu_r \tilde b\right)\,.
\end{equation}
The role of the insertions of $\int_\Sigma \mu_r \tilde b$ is similar to that in conventional string theory. In particular, they (i) absorb the $\tilde c$ ghosts in vertex operators, for $\mu_r$ chosen to extract the residue at a marked point $y_r$, giving $\oint_{y_r} \tilde b$, and (ii) at higher genus, saturate the zero-modes integration, for $\mu_r$ chosen to extract the value of the field at a point, giving $\tilde b(y_r)$. The delta functions, for which the definition is $2\pi i\, \bar \delta (z) = \bar\partial(1/z)$, are the novel feature of ambitwistor strings. They impose the constraint $P^2=0$, which, as we shall see later, fully localises the measure on the Riemann surface moduli space and leads to the scattering equations.    

A comment is in order regarding the asymmetry of gauge fixing between $e$ and $\tilde e$. The role played by the Beltrami differentials that span the deformation of the complex structure $e$, which we will call $\hat\mu_r$, is to provide the conventional measure on (the chiral bosonic) moduli space of the Riemann surface. Therefore, the insertions of $\int_\Sigma \hat\mu_r b$ differ from those of $\int_\Sigma \mu_r \tilde b$ beyond the change of chirality, and this affects the measure of the path integral, as we shall discuss in section~\ref{sec:correlator}.

The parts of the path integral associated with $\chi$ and $\tilde\chi$ are treated in a similar manner to that in  type II string theory. In particular, they lead to the insertions of PCOs,
\begin{equation}\label{equ:PCOsusy}
\prod_{\alpha=1}^{n+2g-2} \big(\delta(\beta) \delta(\tilde\beta) \, P\cdot\psi \, P\cdot\tilde\psi\big)(x_\alpha) \,,
\end{equation}
at locations $\{x_\alpha\}$ picked up by the choice of basis $\{\chi_\alpha\}$.

Finally, we can write down the gauge-fixed action, which is linear in all fields and  includes the kinetic terms for the ghosts,
\begin{equation}
\label{equ:typeII_gf}
 S_\text{g.f.}=\frac{1}{2\pi}\int_\Sigma P\cdot \dbar X +\frac{1}{2}\psi\cdot\dbar\psi+\frac{1}{2}\tilde\psi\cdot\dbar\tilde\psi + b \dbar c + \tilde b \dbar \tilde c + \beta \dbar \gamma + \tilde\beta\dbar  \tilde\gamma \,.
\end{equation}
We are left with the following OPEs:
\begin{equation}
\label{equ:opes}
P_\mu (z) X^\nu(0) \sim - \delta^\nu_\mu \, \frac{\d z}{z} \,, \quad \psi^\mu (z) \psi^\nu(0) \sim \eta^{\mu\nu}\,\frac{\d z}{z} \,,\quad b(z)c(0) \sim \frac{\d z}{z} \,,\quad \beta(z)\gamma(0) \sim -\frac{\d z}{z} \,,
\end{equation}
and similarly for `tilded' fields. In the remainder of this paper, we will be typically drop the differential symbols, where it should be obvious how these should be reinstated to provide for expressions of the appropriate weight.

The central charge is computed in a similar manner as in conventional type II string theory, giving $3(d-10)$. This is twice the result in the conventional string, because it is effectively the sum of its chiral and anti-chiral central charges. The critical dimension is the same, $d=10$. Only in $d=10$ is the BRST operator nilpotent:
\begin{equation}
\label{equ:brstQ}
Q=\frac{1}{2\pi}\oint c\, \left(T^{\text{m}}+\frac{1}{2}T^{bc}\right) + \tilde c \,\frac{P^2}{2} + \gamma \, P\cdot \psi + \tilde\gamma \, P\cdot \tilde\psi - \tilde b\, (\gamma^2+\tilde\gamma^2)\,,
\end{equation}
where
\begin{align}
T^{\text{m}}&=P\cdot \partial X + \frac1{2} \psi\cdot\partial\psi + \frac1{2} \tilde\psi\cdot\partial\tilde\psi 
- (\partial \tilde b) \tilde c + 2 \partial(\tilde b \tilde c )
- (\partial  \beta)  \gamma + \frac{3}{2} \partial( \beta  \gamma )
 - (\partial \tilde \beta) \tilde \gamma + \frac{3}{2} \partial(\tilde \beta \tilde \gamma )
 \,, \nonumber \\
T^{bc}&=- (\partial  b)  c + 2 \partial( b  c ) \,,
\end{align}

\subsection{Vertex operators}

Vertex operators are elements of the BRST cohomology, and in an ambitwistor string these always correspond  to massless states. Notice that there is no mass scale, whereas in standard string theory this is provided by the inverse string length.
In the case of the type II ambitwistor string, the states (and their interactions) are those of type II supergravity. Before we proceed, let us point out that, since all the fields are left-moving, there are more options in the GSO projection \cite{Azevedo:2017lkz}. We are only considering here the GSO projection analogous to that in type II string theory, where the projection is applied independently to left-moving and right-moving states; the analogue states in our case are left-moving `untilded' and `tilded' states.

A basis for fixed vertex operators can be built from elements
\begin{equation}
\label{equ:vop}
{\mathcal O}(z)=c(z)\tilde c(z) \,U(z) \tilde U (z)\,e^{ik\cdot X(z)} \,, \quad k^2=0\,,
\end{equation}
where $U$ and $\tilde U$ take the forms familiar from conventional string theory for the Neveu-Schwarz (NS) sector and the Ramond (R) sector. In contrast to standard string theory however,  all the operators are left-moving in the ambitwistor case. We will only consider here the scattering of NS-NS external states, so that
 \begin{equation}
\label{equ:NSvop}
U_\text{NS} = \delta(\gamma) \,\epsilon\cdot \psi \,, \qquad \tilde U_\text{NS} = \delta(\tilde\gamma)\,\tilde \epsilon\cdot \tilde\psi \,.
\end{equation}
General NS-NS polarisation tensors can be obtained from linear combinations of these $\epsilon_\mu\tilde\epsilon_\nu$ states. Along with the massless condition, $k^2=0$, BRST closure requires that $\epsilon\cdot k=\tilde\epsilon\cdot k=0$. These constraints follow respectively from the contributions of $\tilde c\,P^2/2$ and $\gamma\, P\cdot\psi+\tilde\gamma\, P\cdot\tilde\psi$ in the BRST operator $Q$. Gauging the worldsheet supergroup thus projects out negative-norm states from the ambitwistor string spectrum.
For the Ramond sector vertex operators, see e.g.~\cite{Adamo:2013tsa,Roehrig:2017gbt}. Of course, even though we will only consider NS-NS external states, all states run in the loops. Indeed, the Ramond vertex operators would be crucial if we tried to reproduce the results of the present paper using a gluing operator, as was accomplished in \cite{Roehrig:2017gbt} at one loop.

The effect of the supersymmetry-related PCOs \eqref{equ:PCOsusy} is familiar from superstring theory. For marked points ${x_\alpha}$ coinciding with fixed vertex operator locations ${z_i}$, we get
$$
\lim_{x_\alpha\rightarrow z_i}\big(\delta(\beta)\,\delta(\tilde\beta)\,P\cdot\psi\,P\cdot\tilde\psi\big)(x_\alpha) \;{\mathcal O}_\text{NS-NS}(z_i)=c\tilde{c}\,(\epsilon\cdot P+k\cdot \psi\,\epsilon\cdot\psi)
(\tilde\epsilon\cdot P+k\cdot \tilde\psi\,\tilde\epsilon\cdot\tilde\psi)e^{ik\cdot X}(z_i)
=c\tilde c\, V(z_i) .
$$
On the other hand, for the PCOs \eqref{equ:PCOint}, there is a crucial difference with respect to string theory. Let us use the notation
$$
\langle \lambda_1 \lambda_2 \rangle = \int_\Sigma \lambda_1 \lambda_2
$$
for the standard Serre duality pairing. If a Beltrami differential $\mu_r$ is such that it extracts a residue at a marked point ${y_r}$, and if we take this point to coincide with the vertex operator locations ${z_i}$, we get, taking into account also the gauge fixing related to $e$,
$$
\lim_{y_r\rightarrow \mu_i} \Big(
\big\langle\mu_r b\big\rangle\big\langle\mu_r \tilde{b}\big\rangle\,\,\bar\delta\big(\langle\mu_r P^2\rangle\big)\Big)\,\big(c \tilde{c} \, V\big)(z_i) =
\bar\delta\big(\langle\mu_i P^2\rangle\big)\,V(z_i)\,, \quad \text{with} \quad \langle\mu_i P^2\rangle =\text{Res}_{z_i}P^2 \,,
$$
and the integrated vertex operator is
 \begin{equation}
 \label{typeIIintvo}
\mathcal{V} = \int_\Sigma \bar\delta\big(\text{Res}_{z_i}P^2\big)\,V(z_i) \,.
\end{equation}
{}

\subsection{Scattering equations and CHY formulae}

We discussed above how the constraint $P^2=0$ is imposed in terms of PCOs after gauge fixing. There are $n+3g-3$ of such PCOs, and this is precisely the dimension of the moduli space $\mathfrak{M}_{g,n}$ of the genus-$g$ Riemann surface with $n$ marked points. Therefore, the integration over $\mathfrak{M}_{g,n}$ is fully localised at a set of critical points in moduli space. The problem of finding the complete critical set has only been addressed at genus zero, and this is an important motivation for turning the problem for $\mathfrak{M}_{g,n}$ into one for $\mathfrak{M}_{0,n+2g}$, as we will achieve in this paper for $g=2$.

In this section, let us consider the case of the Riemann sphere  for illustration. The $n-3$ moduli are associated to the locations $\{z_i\}$ of the $n-3$ integrated vertex operators, and the basis $\{\mu_r\}$ is naturally chosen so as to extract the residues of $P^2$ at those points, as in \eqref{typeIIintvo}. The important observation is that $P_\mu$ is determined, up to zero modes, by integrating out $X^\mu$ in the path integral. Suppose we have $n$ vertex operators, each depending on $X^\mu$ only through the plane wave factor $e^{i k_i X(z_i)}$. Then we can integrate out the $PX$ system exactly. For the zero mode of $X^\mu$, we get a delta function imposing momentum conservation, $\sum_i k_i=0$, as in  string theory. For the non-zero modes of $X^\mu$, we get a delta functional imposing the constraint
\begin{equation} \label{eq:Pconstraint}
 \dbar P_\mu = 2\pi i\sum_ik_{i\,\mu}\,\bar\delta(z-z_i)\,\d z \,.
\end{equation}
It says that $P_\mu$ is a meromorphic differential with simple poles at $z=z_i$ with residues $k_i$.
This constraint holds at any genus, but only at genus zero does it fully determine $P_\mu$ due to the absence of zero modes (solutions to the homogeneous equation). We get\footnote{We will use $\sigma$ as a coordinate on the sphere, and $z$ as a coordinate on a genus-$g$ surface. This will be useful later for clarity, when we relate a degenerate genus-$g$ surface to a sphere with nodes.}
\begin{equation}
P_\mu = \d \sigma \sum_{i=1}^n \frac{k_i}{\sigma-\sigma_i}\,.
\end{equation}
Since $k_i^2=0$, it is clear that $P^2$ is a meromorphic quadratic differential with only simple poles at $\sigma=\sigma_i$. Then the statement that $P^2=0$ on the sphere is equivalent to the statement that $\text{Res}_{\sigma_i}P^2=0,\;\forall \sigma_i$. These residues give the {\it scattering equations},
\begin{equation}
\text{Res}_{\sigma_i}P^2=\mathcal{E}_i= 2\, \d \sigma \sum_{j \neq i} \frac{k_i\cdot k_j}{\sigma_i-\sigma_j} =0 \,, \;\forall \sigma_i \,.
\end{equation}
There are only $n-3$ linearly independent equations, due to the 3 identities $\sum_i\mathcal{E}_i\sigma_i^q=0$ for $q=0,1,2$. This is consistent with the fact that $\{\sigma_i\}$ is only meaningful up to SL$(2,\mathbb C)$ coordinate transformations on the sphere. Up to these SL$(2,\mathbb C)$ transformations, the scattering equations determine $(n-3)!$ solutions. These are the critical points at which the integration over $\mathfrak{M}_{0,n}$ in the ambitwistor string amplitude is fully localised. The direct way of evaluating the amplitude is to sum over the contributions from each solution $\{\sigma_i^\text{sol}\}$ to the scattering equations,
\begin{equation} \label{eq:Msphere}
{\mathcal M}_n^\text{sphere} = \int_{\mathfrak{M}_{0,n}} \Big(\prod_i {}'\,  \bar\delta(\mathcal{E}_i)\Big) \; {\mathscr I} \, = \sum_{\{\sigma_i\}=\{\sigma_i^\text{sol}\}} \frac{\mathscr I}{J} \,,
\end{equation}
where, after the first equality, the delta functions factor is independent of the chosen $n-3$ linearly independent equations, and, after the second equality, $1/J$ represents a factor coming from the measure. This is precisely the CHY representation of the amplitude, which motivated the construction of the ambitwistor string. For the type II ambitwistor string, the result is the CHY formula for a gravity amplitude, presented in \cite{Cachazo:2013hca}, where all details can be found. 

At tree level, i.e., on the Riemann sphere, the amplitude for NS-NS external states is the same in type II supergravity and in the bosonic Einstein--dilaton--B-field gravity (or NS-NS gravity). In fact, if for the external states we take linear combinations of basis states $\epsilon_\mu\tilde\epsilon_\nu$ corresponding to gravitons, then the amplitude is the same as in pure Einstein gravity. For the factorisable external states $\epsilon_\mu\tilde\epsilon_\nu$, the CHY integrand factorises, ${\mathscr I}_\text{NS-NS}={\mathcal I}(\epsilon_i)\,{\mathcal I}(\tilde\epsilon_i)$. The object $\mathcal I$, dependent on the momenta $k_i$ and polarisations $\epsilon_i$ of the external states, as well as on the marked points $\sigma_i$, has a beautiful expression in terms of the Pfaffian of a matrix, and we will construct its two-loop analogue later on.

The formula for ${\mathscr I}_\text{NS-NS}$ exhibits a {\it double copy} relation between gravity and gauge theory, since  an amplitude in Yang-Mills theory has the same building block ${\mathcal I}(\epsilon_i)$ in its CHY integrand: ${\mathscr I}_\text{YM}={\mathcal I}(\epsilon_i)\,{\mathcal I}^\text{colour}(a_i)$, where the $a_i$ are the Lie algebra indices of the external gluons. 

Finally, notice that the CHY formula \eqref{eq:Msphere} turns out to be valid in any number of dimensions, even though  the type II ambitwistor string is only critical in $d=10$ dimensions,  since this is the only dimension where  the BRST operator is nilpotent.

\subsection{One loop: from the torus to the nodal Riemann sphere}\label{sec:1-loop}

The scattering equations on a genus-one Riemann surface (torus) were first discussed in \cite{Adamo:2013tsa}. The main difference with respect to the genus-zero case discussed above is that the PCOs imposing $P^2=0$ cannot all be chosen to extract the residue at a marked point. While there are $n$ PCOs for $n$ vertex operator marked points, only $n-1$ of the latter are associated to the moduli space $\mathfrak{M}_{1,n}$, due to translation invariance (analogous to SL$(2,\mathbb C)$ on the sphere). Therefore, only $n-1$ of the PCOs can be of the type \eqref{typeIIintvo}. The remaining PCO may be chosen to set $P^2(z_0)=0$ for a point $z_0$ not coincident with the other marked points. We get an amplitude of the form
\begin{equation}  \label{eq:Mtorus}
{\mathcal M}_n^\text{torus} = \int \d^{10}\ell \int_{\mathfrak{M}_{1,n}}\hspace{-8pt}\d\tau \,\,\,\bar\delta\big(P(z_0)^2\big) \prod_{i=2}^n  \bar\delta\big(\text{Res}_{z_i}P^2\big) \; {\mathscr I}^{(1)} \, \,,
\end{equation}
where we chose to deal with translation invariance by fixing $z_1$ (due to linear dependence, the residue of $P^2$ at $z_1$ vanishes if the residues at $z_{i>1}$ vanish). The relation of this formula to \eqref{eq:Msphere} is clear, but there are new features. One is that the integration over $\mathfrak{M}_{1,n}$ includes an integration over the modular parameter $\tau$ of the torus, and the new scattering equation, $P(z_0)^2=0$, may be thought of as being associated to this modulus, in the same way as the others are associated to the vertex operator locations. The other new feature is the integration over the zero mode of $P_\mu$, which is required in the path integral by the fact that, on the torus, $P_\mu$ is determined by the equation \eqref{eq:Pconstraint} only up to a zero mode $\ell_\mu \d z$, with $\ell_\mu$ constant.

The new type of constraint, $P(z_0)^2=0$, can be expressed in an alternative way, as was also pointed out in \cite{Adamo:2013tsa}. This will be more useful for us at genus two. After imposing the scattering equations of type $\text{Res}_{z_i}P^2=0$, we are left with a holomorphic $P^2$, i.e., it has no poles. The only possibility is that $P^2=u\, \d z^2$, and one can show that $u\to\ell^2$ as $\tau\to i\infty$. Defining $u$ in this manner, we can substitute the insertion $\bar\delta\big(P(z_0)^2\big)$ in \eqref{eq:Mtorus} by the insertion $\bar\delta(u\, \d z^2)$. At higher genus, the same argument can be used to write $P^2=u^{IJ}\omega_I\omega_J$ in terms of holomorphic differentials $\omega_I$, after imposing the residue-type scattering equations.

Ref.~\cite{Adamo:2013tsa} determined the type II supergravity integrand ${\mathscr I}^{(1)}_\text{typeII}={\mathcal I}^{(1)}(\epsilon_i)\,{\mathcal I}^{(1)}(\tilde\epsilon_i)$. This is reminiscent of the tree-level result, but now ${\mathcal I}^{(1)}(\epsilon_i)$ is not related to a single Pfaffian, but to a linear combination of these, since there are contributions from the four spin structures of the torus. We will briefly discuss the spin structures of  Riemann surfaces below. Moreover, ref.~\cite{Adamo:2013tsa} also checked the modular invariance of the amplitude, i.e., the invariance under the identifications $\tau\sim\tau+1\sim-1/\tau$, where the inversion identification requires $\ell_\mu \to \tau\ell_\mu$.

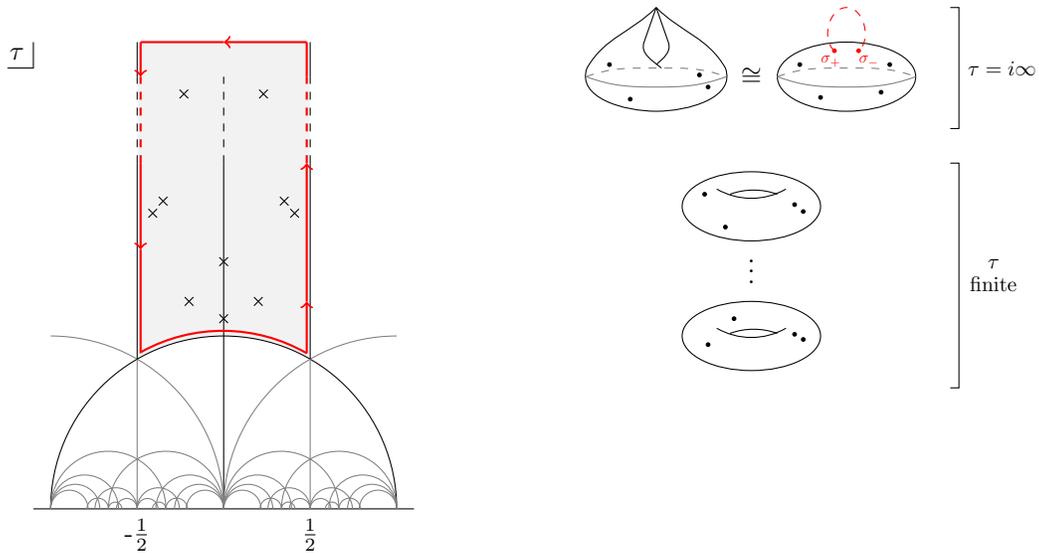
\begin{figure}[ht]
\centering
 \begin{tikzpicture} [scale=2.3]
%background
  \filldraw [fill=light-gray, draw=white] (0.5,0.866) arc [radius=1, start angle=60, end angle= 120] -- (-0.5,2.7) -- (0.5,2.7) -- (0.5,0.866);
%axes
  \draw [black] (1.1,0) -- (-1.1,0);
  \draw [black] (0,0) -- (0,2);
  \draw [black,dashed] (0,2) -- (0,2.5);
  \node at (0.5,-0.15) {$\frac{1}{2}$};
  \node at (-0.5,-0.15) {-$\frac{1}{2}$};
%label
  \node at (-1.195,2.63) {$\tau$};
  \draw (-1.1,2.7) -- (-1.1,2.55) -- (-1.25,2.55);
%basic outline of fundamental domain
  \draw [gray] (0,0) arc [radius=1, start angle=0, end angle= 90];
  \draw [black] (1,0) arc [radius=1, start angle=0, end angle= 180];
  \draw [gray] (1,1) arc [radius=1, start angle=90, end angle= 180];
  \draw [gray] (0.5,0) -- (0.5,0.866);
  \draw [gray] (-0.5,0) -- (-0.5,0.866);
  \draw (0.5,0.866) -- (0.5,2);
  \draw (-0.5,0.866) -- (-0.5,2);
  \draw [dashed] (0.5,2) -- (0.5,2.5);
  \draw [dashed] (-0.5,2) -- (-0.5,2.5);
  \draw (0.5,2.5) -- (0.5,2.7);
  \draw (-0.5,2.5) -- (-0.5,2.7);
%contour
  \draw [red,thick,->] (0.48,2.7) -- (0,2.7);
  \draw [red,thick] (0,2.7) -- (-0.48,2.7);
  \draw [red,thick] (0.48,2.5) -- (0.48,2.7);
  \draw [red,thick,<-] (-0.48,2.5) -- (-0.48,2.7);
  \draw [red,thick,dashed] (0.48,2) -- (0.48,2.5);
  \draw [red,thick,dashed] (-0.48,2) -- (-0.48,2.5);
  \draw [red,thick,->] (0.48,1.2) -- (0.48,2);
  \draw [red,thick,->] (0.48,0.9) -- (0.48,1.2);
  \draw [red,thick,<-] (-0.48,1.5) -- (-0.48,2);
  \draw [red,thick] (-0.48,0.9) -- (-0.48,1.5);
  \draw [red,thick] (0.48,0.9) arc [radius=0.97, start angle=60, end angle= 119];
%solutions
  \draw (-0.025,1.125) -- (0.025,1.075);
  \draw (-0.025,1.075) -- (0.025,1.125);
  %\draw [fill] (0,1.1) circle [radius=.3pt];
  \draw (0.225,1.225) -- (0.175,1.175);
  \draw (0.225,1.175) -- (0.175,1.225);
  %\draw [fill] (0.2,1.2) circle [radius=.3pt];
  \draw (-0.225,1.225) -- (-0.175,1.175);
  \draw (-0.225,1.175) -- (-0.175,1.225);
  %\draw [fill] (-0.2,1.2) circle [radius=.3pt];
  \draw (-0.025,1.455) -- (0.025,1.405);
  \draw (-0.025,1.405) -- (0.025,1.455);
  %\draw [fill] (0,1.43) circle [radius=.3pt];
  \draw (-0.385,1.735) -- (-0.435,1.685);
  \draw (-0.385,1.685) -- (-0.435,1.735);
  %\draw [fill] (-0.41,1.71) circle [radius=.3pt];
  \draw (0.385,1.735) -- (0.435,1.685);
  \draw (0.385,1.685) -- (0.435,1.735);
  %\draw [fill] (0.41,1.71) circle [radius=.3pt];
  \draw (0.325,1.805) -- (0.375,1.755);
  \draw (0.325,1.755) -- (0.375,1.805);
  %\draw [fill] (0.35,1.78) circle [radius=.3pt];
  \draw (-0.325,1.805) -- (-0.375,1.755);
  \draw (-0.325,1.755) -- (-0.375,1.805);
  %\draw [fill] (-0.35,1.78) circle [radius=.3pt];
  \draw (0.205,2.425) -- (0.255,2.375);
  \draw (0.255,2.425) -- (0.205,2.375);
  %\draw [fill] (0.23,2.4) circle [radius=.3pt];
  \draw (-0.205,2.425) -- (-0.255,2.375);
  \draw (-0.205,2.375) -- (-0.255,2.425);
  %\draw [fill] (-0.23,2.4) circle [radius=.3pt];
% rest of fundamental domain
  \draw [gray] (0,0) arc [radius=0.333, start angle=0, end angle= 180];
  \draw [gray] (-0.334,0) arc [radius=0.333, start angle=0, end angle= 180];
  \draw [gray] (0.666,0) arc [radius=0.333, start angle=0, end angle= 180];
  \draw [gray] (1,0) arc [radius=0.333, start angle=0, end angle= 180];
  \draw [gray] (0,0) arc [radius=0.196, start angle=0, end angle= 180];
  \draw [gray] (-0.608,0) arc [radius=0.196, start angle=0, end angle= 180];
  \draw [gray] (0.392,0) arc [radius=0.196, start angle=0, end angle= 180];
  \draw [gray] (1,0) arc [radius=0.196, start angle=0, end angle= 180];
  \draw [gray] (0,0) arc [radius=0.142, start angle=0, end angle= 180];
  \draw [gray] (-0.716,0) arc [radius=0.142, start angle=0, end angle= 180];
  \draw [gray] (0.284,0) arc [radius=0.142, start angle=0, end angle= 180];
  \draw [gray] (1,0) arc [radius=0.142, start angle=0, end angle= 180];
  \draw [gray] (0,0) arc [radius=0.108, start angle=0, end angle= 180];
  \draw [gray] (-0.784,0) arc [radius=0.108, start angle=0, end angle= 180];
  \draw [gray] (0.216,0) arc [radius=0.108, start angle=0, end angle= 180];
  \draw [gray] (1,0) arc [radius=0.108, start angle=0, end angle= 180];
  \draw [gray] (0.5,0) arc [radius=0.122, start angle=0, end angle= 180];
  \draw [gray] (-0.5,0) arc [radius=0.122, start angle=0, end angle= 180];
  \draw [gray] (0.744,0) arc [radius=0.122, start angle=0, end angle= 180];
  \draw [gray] (-0.256,0) arc [radius=0.122, start angle=0, end angle= 180];
  \draw [gray] (0.5,0) arc [radius=0.060, start angle=0, end angle= 180];
  \draw [gray] (-0.5,0) arc [radius=0.060, start angle=0, end angle= 180];
  \draw [gray] (0.620,0) arc [radius=0.060, start angle=0, end angle= 180];
  \draw [gray] (-0.380,0) arc [radius=0.060, start angle=0, end angle= 180];
  \draw [gray] (0.666,0) arc [radius=0.039, start angle=0, end angle= 180];
  \draw [gray] (0.412,0) arc [radius=0.039, start angle=0, end angle= 180];
  \draw [gray] (-0.588,0) arc [radius=0.039, start angle=0, end angle= 180];
  \draw [gray] (-0.334,0) arc [radius=0.039, start angle=0, end angle= 180];
  \draw [gray] (0.334,0) arc [radius=0.062, start angle=0, end angle= 180];
  \draw [gray] (0.790,0) arc [radius=0.062, start angle=0, end angle= 180];
  \draw [gray] (-0.666,0) arc [radius=0.062, start angle=0, end angle= 180];
  \draw [gray] (-0.210,0) arc [radius=0.062, start angle=0, end angle= 180];
%fat torus
  \draw (3.05,1) circle [x radius=0.4, y radius=0.2];
  \draw (2.85,1.045) arc [radius=0.4, start angle=240, end angle=300];
  \draw (3.2,1.015) arc [radius=0.4, start angle=70, end angle=110];
  \draw [fill] (2.95,1.1) circle [radius=.3pt];
  \draw [fill] (2.8,0.95) circle [radius=.3pt];
  \draw [fill] (3.35,0.98) circle [radius=.3pt];
  \draw [fill] (3.3,1.01) circle [radius=.3pt];
%thinning torus
  \draw (3.05,1.75) circle [x radius=0.4, y radius=0.2];
  %\draw (3.05,1.92) to [out=180, in=60] (2.9,1.85) to [out=250, in=180] (3.05,1.72) to [out=0, in=290] (3.2,1.85) to [out=120, in=0] (3.05,1.92);
  \draw (2.85,1.85) arc [radius=0.4, start angle=240, end angle=300];
  \draw (3.2,1.82) arc [radius=0.4, start angle=70, end angle=110];
  \draw [fill] (2.78,1.82) circle [radius=.3pt];
  \draw [fill] (2.9,1.63) circle [radius=.3pt];
  \draw [fill] (3.35,1.72) circle [radius=.3pt];
  \draw [fill] (3.3,1.76) circle [radius=.3pt];
% dots
 %\node at (3.05,2) {$\vdots$};
 \node at (3.05,1.42) {$\vdots$};
%pinched torus
  \draw (2.5,2.9) to [out=225, in=60] (2.1,2.55) to [out=250, in=180] (2.5,2.3) to [out=0, in=290] (2.9,2.55) to [out=120, in=315] (2.5,2.9);
  \draw[gray] (2.095,2.5) to [out=340, in=180] (2.5,2.44) to [out=0, in=200] (2.905,2.5);
  \draw[gray, dashed] (2.095,2.5) to [out=20, in=180] (2.5,2.55) to [out=0, in=160] (2.905,2.5);
  \draw (2.5,2.9) to [out=250, in=70] (2.43,2.7) to [out=250, in=140] (2.53,2.55);
  \draw (2.5,2.9) to [out=290, in=110] (2.57,2.7) to [out=290, in=40] (2.5,2.57);
  %\draw (3,2.8) to [out=160, in=90] (2.4,2.5) to [out=280, in=180] (3,2.2) to [out=00, in=270] (3.6,2.5) to [out=90, in=20] (3,2.8);
  \draw [fill] (2.23,2.57) circle [radius=.3pt];
  \draw [fill] (2.35,2.37) circle [radius=.3pt];
  \draw [fill] (2.8,2.44) circle [radius=.3pt];
  \draw [fill] (2.75,2.51) circle [radius=.3pt];
%equivalent to  
  \draw (3.6,2.5) circle [x radius=0.4, y radius=0.2];
  \draw[gray] (3.2,2.5) to [out=340, in=180] (3.6,2.44) to [out=0, in=200] (4,2.5);
  \draw[gray, dashed] (3.2,2.5) to [out=20, in=180] (3.6,2.55) to [out=0, in=160] (4,2.5);
  \draw [dashed,red] (3.53,2.65) to [out=140, in=180] (3.6,2.9) to [out=0, in=40] (3.67,2.65);
  \draw [fill] (3.33,2.57) circle [radius=.3pt];
  \draw [fill] (3.45,2.38) circle [radius=.3pt];
  \draw [fill] (3.8,2.41) circle [radius=.3pt];
  \draw [fill] (3.85,2.57) circle [radius=.3pt];
  \draw [fill,red] (3.53,2.65) circle [radius=.3pt];
  \draw [fill,red] (3.67,2.65) circle [radius=.3pt];
  \node at (3.51,2.59) {\scalebox{0.6}{\textcolor{red}{$\sigma_+$}}};
  \node at (3.73,2.59) {\scalebox{0.6}{\textcolor{red}{$\sigma_-$}}};
  \node at (3.05,2.5) {$\cong$};
  %description
  \draw (4.2,2.9) -- (4.25,2.9) -- (4.25,2.2) -- (4.2,2.2);
  \node at (4.5,2.55) {\scalebox{0.8}{$\tau=i\infty$}};
  \draw (4.2,2) -- (4.25,2) -- (4.25,0.7) -- (4.2,0.7);
  \node at (4.45,1.42) {\scalebox{0.8}{$\tau$}};
  \node at (4.45,1.3) {\scalebox{0.8}{finite}};
 \end{tikzpicture}
 
\caption{The residue theorem on the fundamental domain.}
\label{fig:fund-dom}
\end{figure}  

\begin{figure}[ht]
	\centering 
	\includegraphics[width=5cm]{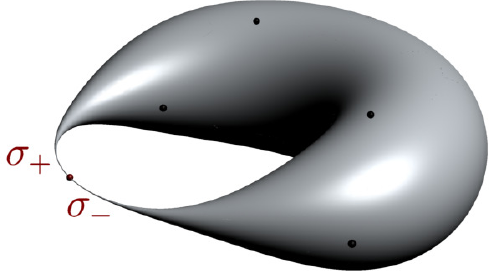}
	\caption{The nodal Riemann sphere, including the labels of the node.}
	\label{fig:nodalRS}
\end{figure}

While the formula \eqref{eq:Mtorus} satisfies all tests, its evaluation is very hard due to the appearance of theta functions in the genus-one scattering equations. Ref.~\cite{Geyer:2015bja} provided a major simplification, by noticing that the integration over $\tau$, which is part of the integration over ${\mathfrak{M}_{1,n}}$, can be localised at $\tau=i\infty$, or equivalently at $q=0$ for $q=e^{2\pi i \tau}$. This is accomplished via an integration by parts, moving the derivative $\dbar\big(1/P(z_0)^2\big)$ away from this constraint. Equivalently, it can be seen as an application of the residue theorem to the fundamental domain of $\tau$, and it relies on the modular invariance of the original genus-one amplitude. The localisation gives a degenerate torus, equivalent to a Riemann sphere with a pair of identified points called a node; see figure \ref{fig:fund-dom} and \ref{fig:nodalRS} for illustration. Changing to coordinates $\sigma$ more appropriate to describe a sphere, the final result is
\begin{equation}  \label{eq:M1loop}
{\mathcal M}_n^\text{torus} = {\mathcal M}_n^\text{nodal sphere} = \int \d^{d}\ell\,\frac1{\ell^2} \int_{\mathfrak{M}_{0,n+2}} \Big(\prod_A {}'\,  \bar\delta(\mathcal{E}_A)\Big) \; {\mathscr I}^{(1)}_0  \,, \qquad  
{\mathscr I}^{(1)}_0={\mathscr I}^{(1)}_{q\to0}\,,
\end{equation}
where $\{\sigma_A\}=\{\sigma_i,\sigma_+,\sigma_-\}$, with the two extra marked points representing loop momentum insertions. In the limit $q\to0$, we have
\begin{equation}
P_{\mu} = \ell_\mu\, \omega + \d \sigma \sum_{i=1}^n \frac{k_i}{\sigma-\sigma_i}\,,
\qquad \omega = \d \sigma \left(\frac{1}{\sigma-\sigma_+}- \frac{1}{\sigma-\sigma_-} \right) \,.
\end{equation}
After applying the residue theorem, the original constraint $P(z_0)^2=0$ is no longer enforced, and therefore $P^2$ does not vanish on the nodal sphere. Indeed, $P^2$ has double poles at $\sigma_+$ and $\sigma_-$, and therefore the quadratic differential of interest with only simple poles is $\mathfrak{P}_1=P^2-\ell^2\omega^2$. Using this differential, the one-loop scattering equations can be written compactly as $\text{Res}_{\sigma_A}\mathfrak{P}_1=0,\,\forall \sigma_A$. The end result is a CHY-type formula of the loop integrand. In this formula, we can actually take the loop momentum to lie in $d$ dimensions, whereas on the torus $d=10$ was essential for modular invariance.

The type II supergravity formula on the nodal sphere takes the form ${\mathscr I}^{(1)}_{0\;\text{typeII}}={\mathcal I}^{(1)}_0(\epsilon_i)\,{\mathcal I}^{(1)}_0(\tilde\epsilon_i)$. While no ambitwistor string model for super-Yang-Mills theory has been studied on the torus, ref.~\cite{Geyer:2015bja} took the one-loop formula \eqref{eq:M1loop} in the same spirit as the CHY approach, and proposed a formula for super-Yang-Mills theory based on the principle of the double copy. With a suitable one-loop generalisation of the colour part, it takes the form ${\mathscr I}_{0\;\text{SYM}}={\mathcal I}^{(1)}_0(\epsilon_i)\,{\mathcal I}^\text{PT(1)}(a_i)$. Ref.~\cite{Geyer:2015jch} extended these formulae to the cases of non-supersymmetric Yang-Mills theory and gravity.

In this paper, we shall follow the same steps, now from genus two to the bi-nodal Riemann sphere. We leave the non-supersymmetric extension for a future publication.

To conclude, let us also mention work on an alternative approach to the loop-level scattering equations, based on the (hyper)elliptic parametrisation of the Riemann surfaces \cite{Gomez:2016bmv,Cardona:2016bpi,Cardona:2016wcr,Gomez:2016cqb,Ahmadiniaz:2018nvr}.

%%%%%%%%%%%%%%%%%%%%%%%%%%%%%%
%%%%%%%%%%%%%%%%%%%%%%%%%%%%%%
\section{The toolkit at genus two}\label{sec:rev_basics}

In this section, we review  the main tools used to study conformal field theories on higher-genus Riemann surfaces: the Green's functions and partition functions for chiral $bc$ and $\beta\gamma$ systems of any conformal weight.  To this end, we discuss basic objects of the theory of (compact) Riemann surfaces, with particular attention to the genus-two case. This lays the basis for the review of Szeg\H{o} kernels and the Verlinde formulas for the partition function.  We refer the reader to Fay's classic reference \cite{fay1973theta} and to the string theory references \cite{DHoker:1988ta,Verlinde:1986kw,DHoker:2002hof} for detailed expositions.

\subsection{The basics} 

\begin{figure}[ht]
	\centering 
	\includegraphics[width=6cm]{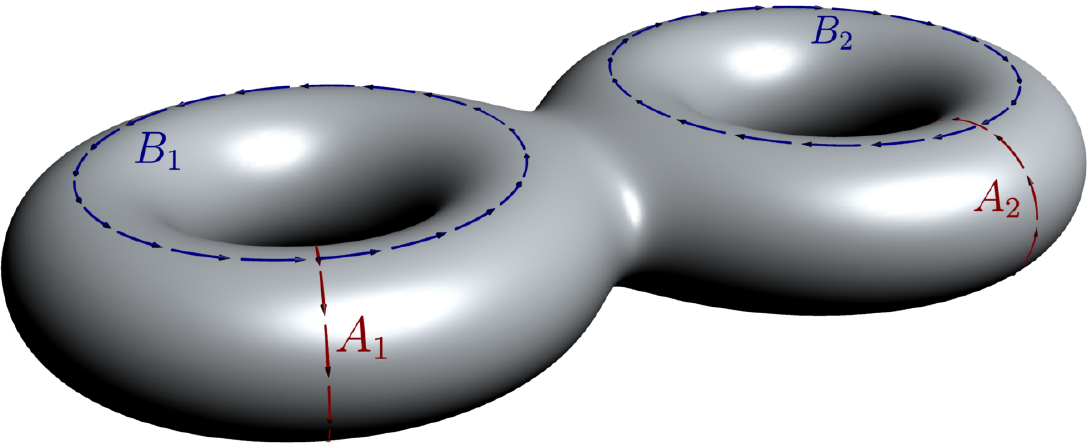}
	\caption{Homology basis of cycles at genus two. The orientation of the cycles ensures that the intersection form is canonical.}
	\label{fig:homologybasis}
\end{figure}

For a genus-$g$ Riemann surface, we choose a {\it homology basis} of cycles $A_I$ and $B_I$, $I=1,\ldots,g$, such that the intersection form is canonical, $\#(A_I,B_J)=\delta_{IJ}=-\#(B_J,A_I)$; see figure \ref{fig:homologybasis} for $g=2$.
The modular group $\text{Sp}(2g,\mathbb{Z})$,
\begin{equation}
\label{eq:defmodular}
 \begin{pmatrix}a&b\\c&d\end{pmatrix}\begin{pmatrix}0&\mathbb{1}\\-\mathbb{1}&0\end{pmatrix}\begin{pmatrix}a&b\\c&d\end{pmatrix}^T=\begin{pmatrix}0&\mathbb{1}\\-\mathbb{1}&0\end{pmatrix}\,,\qquad\quad M=\begin{pmatrix}a&b\\c&d\end{pmatrix} \;\in\; \text{Sp}(2g,\mathbb{Z})\,,
\end{equation}
is a discrete group that acts on the homology basis as $M\binom BA$, leaving the intersection form invariant. 

There are $g$ linearly independent holomorphic 1-forms $\omega_I$ on a genus-$g$ Riemann surface.
These are known as {\it holomorphic Abelian differentials} or as {\it Abelian differentials of the first kind}. They can be chosen to have normalised $A$-periods,\footnote{For any Riemann surface given by a hyperelliptic curve, $y^2=\prod_{a=1}^{2 g-2}(x-x_a)$, a (non-normalised) basis of holomorphic Abelian differentials is given by $x^{I-1} \d x/y$, with $I=1,\ldots,g$. All genus-two Riemann surfaces are hyperelliptic, but this is not true at higher genus.}
\begin{equation}
\label{eq:periods}
 \oint_{A_I}\omega_J=\delta_{IJ}\,,\qquad\qquad \oint_{B_I}\omega_J=\Omega_{IJ}\,.
\end{equation}
The matrix $\Omega_{IJ}$ defined in this manner can be proven to be symmetric, and it is known as the {\it period matrix}. Under a modular transformation \eqref{eq:defmodular}, the period matrix transforms as
\begin{equation}\label{equ:mod_Omega}
\Omega\;\; \to \;\; \widetilde\Omega= \big(a\,\Omega +b\big)\,\big(c\,\Omega+d\big)^{-1}\,.
\end{equation}

At genus two, $\Omega$ has 3 independent components, and we will find it convenient to define the variables\footnote{We follow a standard convention, used for example in \cite{DHoker:2001jaf}, where the first two $q$'s are defined without a factor of 2 in the exponential. For this choice, important expansions used later depend only on integer powers of the $q$'s, rather than on square roots.}
\begin{equation}\label{equ:q_II_def}
q_{11}=e^{i\pi \,\Omega_{11}} \,,\quad  q_{22}=e^{i\pi \,\Omega_{22}} \,, \quad q_{12}=e^{2 i\pi\, \Omega_{12}} \,.
\end{equation}
A standard choice of fundamental domain representing the genus-two moduli space is defined by the following conditions:
\begin{align}\label{equ:fund-domain}
(i)& \quad -\frac{1}{2}\leq \text{Re}(\Omega_{11}),\text{Re}(\Omega_{12}),\text{Re}(\Omega_{22}) \leq \frac{1}{2}\,, \nonumber \\
(ii)& \quad 0<2 \text{Im}(\Omega_{12})\leq\text{Im}(\Omega_{11})\leq\text{Im}(\Omega_{22}) \,, \\
(iii)& \quad \left|\text{det}(c\,\Omega+d)\right| >1 \quad \forall
\begin{pmatrix}a&b\\c&d\end{pmatrix} \in \text{Sp}(4,\mathbb{Z}) \,.
\nonumber
\end{align}

Later on, we will study in detail a singular limit of the moduli space, where both $q_{11}$ and $q_{22}$ vanish. This corresponds to a non-separating degeneration of the surface, with the pinching of both $A_I$ cycles, leading to a genus-zero degenerate surface. This surface is a Riemann sphere with two nodes (pairs of identified points), one per collapsed $A_I$ cycle. 

The $g$ holomorphic Abelian differentials $\omega_I$ also define the Abel map, given a base point $z_0$ on the Riemann surface. For a divisor\footnote{A divisor is mainly used to represent zeros or singularities of meromorphic functions or differentials. In particular, $d_1 z_1 + d_2 z_2 + \ldots + d_m z_m$ denotes behaviour of order $(z-z_r)^{d_r}$ at the points $z_r$ of the surface.} $d_1 z_1 + d_2 z_2 + \ldots + d_m z_m$ of degree $d_1+d_2+\ldots+d_m$, the Abel map takes the form 
\begin{equation}
d_1 z_1 + d_2 z_2 + \ldots + d_m z_m \;\; \mapsto \;\; \sum_{r=1}^m d_r \int_{z_0}^{z_r} \omega_I  
\;\; \in \; {\mathbb C}^g \,.
\end{equation}
In particular, $z_1-z_2\,\mapsto\,\int_{z_2}^{z_1} \omega_I$\,. The integration is over any curve connecting the initial and final points, and so the map is naturally thought of modulo the integration over cycles $A_I$ and $B_I$, otherwise it is multiple valued. Given the periods of $\omega_I$ in \eqref{eq:periods}, the Abel map can be seen as a single-valued map from a point or a divisor on the Riemann surface $\Sigma$ into the Jacobian variety, defined as $J(\Sigma)\equiv{\mathbb C}^g/\{{\mathbb Z}^g+\Omega{\mathbb Z}^g\}$.

\subsection{Theta functions and spin structures}

The theta functions are defined on $\zeta\in{\mathbb C}^g$ as 
\begin{equation}
\vartheta[\kappa](\zeta) \equiv \sum_{n\in {\mathbb Z}^g} \text{exp} \left( 
i \pi (n+\kappa')^T\Omega(n+\kappa') + 2i \pi (n+\kappa')^T(\zeta+\kappa'')
\right)\,,
\end{equation}
where $\Omega$ is the period matrix and $\kappa = (\kappa'|\kappa'')$ denotes the theta characteristic, with $\kappa',\kappa''\in{\mathbb C}^g$. We are interested in characteristics corresponding to {\it spin structures}, i.e., such that $\kappa',\kappa''\in({\mathbb Z}/2{\mathbb Z})^g$; we will be more explicit below. The parity property of theta functions,
\begin{equation}
\vartheta[\kappa](-\zeta) = (-1)^{4 \kappa'\cdot\kappa''} \vartheta[\kappa](\zeta) \,,
\end{equation}
agrees with the designation of spin structures as even/odd according to whether $4 \kappa'\cdot\kappa''$ is even/odd. 

The argument $\zeta\in{\mathbb C}^g$ of interest for the theta functions is typically related to a point or a divisor of the Riemann surface $\Sigma$ via the Abel map, defined above. Throughout the paper, we will often denote the argument directly as a divisor, with the Abel map implicit. The theta functions are quasi-periodic on the lattice $\{{\mathbb Z}^g+\Omega{\mathbb Z}^g\}$. For $M,N\in{\mathbb Z}^g$, we have
\begin{equation}
\vartheta[\kappa](\zeta+M+\Omega N) = \text{exp}
\left( -i \pi N^T\Omega N - 2i \pi N^T(\zeta+\kappa') + 2i \pi M^T \kappa'' \right)\, \vartheta[\kappa](\zeta) \,.
\end{equation}
Since the exponential factor is nowhere vanishing, the divisor of a theta function is well defined on the Jacobian $J(\Sigma)$.

An important result in the theory of theta functions is the {\it Riemann vanishing theorem}. Let us denote $\vartheta(\zeta)=\vartheta[0](\zeta)$. The theorem states that
\begin{equation}
\vartheta(\zeta)=0 \quad \Leftrightarrow \quad  \zeta = \Delta - z_1- z_2 \cdots - z_{g-1} \,,
\end{equation}
for some $g-1$ points $z_r$ on $\Sigma$. The divisor $- z_1- z_2 \cdots - z_{g-1}$ maps to ${\mathbb C}^g$ via the Abel map, while $\Delta\in{\mathbb C}^g$ is the {\it Riemann vector of constants}, which is defined as
 \begin{equation}
 \label{eq:Riemannvector}
\Delta_I = \frac{1-\Omega_{II}}{2} + \sum_{J\neq I} \oint_{A_J} \omega_J(z) \int_{z_0}^z \omega_I \,, \qquad I=1,\ldots,g \,.
\end{equation}

Before proceeding, let us return to the spin structures. There exist $4^{g}$ spin structures at genus $g$, of which $2^{g-1}(2^g+1)$ are even and $2^{g-1}(2^g-1)$ are odd. They label the choice of periodic/anti-periodic boundary conditions of a 1/2-form (world-sheet spinor) on the $A_I$ and $B_I$ cycles. At genus two, there are 16 spin structures. We write them here explicitly for illustration, in the form $\kappa = (\kappa'|\kappa'')$, using the conventions of \cite{DHoker:2001jaf}: the 10 even spin structures, for which we reserve the label $\delta$,
\begin{align}
2\delta_1=\left( \begin{array}{rrr} 0 & \vrule & 0 \\ 0 & \vrule & 0 \end{array} \right) \qquad
2\delta_2=\left( \begin{array}{rrr} 0 & \vrule & 0 \\ 0 & \vrule & 1 \end{array} \right) \qquad
2\delta_3=\left( \begin{array}{rrr} 0 & \vrule & 1 \\ 0 & \vrule & 0 \end{array} \right) \qquad
2\delta_4=\left( \begin{array}{rrr} 0 & \vrule & 1 \\ 0 & \vrule & 1 \end{array} \right) \qquad
\nonumber \\
2\delta_5=\left( \begin{array}{rrr} 0 & \vrule & 0 \\ 1 & \vrule & 0 \end{array} \right) \qquad
2\delta_6=\left( \begin{array}{rrr} 0 & \vrule & 1 \\ 1 & \vrule & 0 \end{array} \right) \qquad
2\delta_7=\left( \begin{array}{rrr} 1 & \vrule & 0 \\ 0 & \vrule & 0 \end{array} \right) \qquad
2\delta_8=\left( \begin{array}{rrr} 1 & \vrule & 0 \\ 0 & \vrule & 1 \end{array} \right) \qquad
\nonumber \\
2\delta_9=\left( \begin{array}{rrr} 1 & \vrule & 0 \\ 1 & \vrule & 0 \end{array} \right)  \qquad
2\delta_0=\left( \begin{array}{rrr} 1 & \vrule & 1 \\ 1 & \vrule & 1 \end{array} \right) \,,\quad \,\,
\end{align}
and the 6 odd spin structures, for which we reserve the label $\nu$,
\begin{align}
2\nu_1=\left( \begin{array}{rrr} 0 & \vrule & 0 \\ 1 & \vrule & 1 \end{array} \right) \qquad
2\nu_3=\left( \begin{array}{rrr} 0 & \vrule & 1 \\ 1 & \vrule & 1 \end{array} \right) \qquad
2\nu_5=\left( \begin{array}{rrr} 1 & \vrule & 0 \\ 1 & \vrule & 1 \end{array} \right) \qquad
\nonumber \\
2\nu_2=\left( \begin{array}{rrr} 1 & \vrule & 1 \\ 0 & \vrule & 0 \end{array} \right) \qquad
2\nu_4=\left( \begin{array}{rrr} 1 & \vrule & 1 \\ 0 & \vrule & 1 \end{array} \right) \qquad
2\nu_6=\left( \begin{array}{rrr} 1 & \vrule & 1 \\ 1 & \vrule & 0 \end{array} \right)  \,.\quad \,\,
\end{align}
For the various relations between even and odd spin structures, see \cite{DHoker:2001jaf}. The ambitwistor string path integral contains, just as its standard superstring counterpart, a sum over spin structures of world-sheet spinors. Certain combinations of the spin structures correspond to states propagating along each $B_I$ cycle: Neveu-Schwarz (NS) states for $\kappa'_I=0$ and Rammond (R) states for $\kappa'_I=1$.
The following table shows which spin structures contribute to the four types of states (NS/R along cycles $B_1$/$B_2$):
 \begin{equation}\label{eq:R-NS}
 \begin{array}{|c|c|c|c|c|}
\hline
& \vrule & \text{NS}_1 & \vrule & \text{R}_1 \\
\hline
\quad \text{NS}_2 \;\, \phantom{.} & \vrule & \delta_1, \delta_2, \delta_3, \delta_4 & \vrule &  \delta_7, \delta_8, \nu_2,\nu_4  \\
\hline
\quad \text{R}_2 \;\, \phantom{.} & \vrule &  \delta_5, \delta_6, \nu_1,\nu_3 & \vrule &  \delta_9, \delta_0, \nu_5,\nu_6 \\
\hline
\end{array}
 \end{equation} \\
This splitting of the sum over spin structures will allow us to select the propagating states and potentially to consider theories with or without sypersymmetry.
 
In the limit that will be important to us later, where both $q_{11}$ and $q_{22}$ vanish, it is straightforward to extract the first few orders in $q_{11}$ and $q_{22}$ of the genus-two theta functions, which will be relevant for our calculations,
\begin{equation}\label{equ:theta_expansion}
\vartheta[\kappa](\zeta) = \sum_{n_1,n_2\in {\mathbb Z}} 
q_{11}^{(n_1+\kappa'_1)^2} q_{22}^{(n_2+\kappa'_2)^2} q_{12}^{(n_1+\kappa'_1)(n_2+\kappa'_2)}
\;\text{exp} \;2 \pi i \big(  (n_1+\kappa'_1)(\zeta_1+\kappa''_1)+(n_2+\kappa'_2)(\zeta_2+\kappa''_2) \big)\,.
\end{equation}

\subsection{Prime form, \texorpdfstring{Szeg\H{o}}{Szego} kernels and meromorphic differentials} \label{sec:primeSzegodiffs}

We are now in a position to define several types of differentials on a Riemann surface that will be useful. Let us first define the {\it prime form}. Consider an odd spin structure $\nu$. The 1-form\, $\sum_{I=1}^g \partial_I \vartheta[\nu](0) \;\omega_I$\, is holomorphic, its $2g-2$ zeros are quadratic, and its square root defines (up to an overall sign) a holomorphic 1/2-form $h_\nu$. The prime form is defined as
\begin{equation}
E(z,w)\equiv \frac{\vartheta[\nu](z-w)}{h_\nu(z)h_\nu(w)} \,.
\end{equation}
It is a holomorphic $(-1/2)$-form in both $z$ and $w$, with a unique simple zero at $z=w$,
\begin{equation}
E(z,w) \approx \frac{z-w}{\sqrt{\d z}\sqrt{\d w}} \qquad \text{for} \; z\approx w\,.
\end{equation}
The prime form is independent of the choice of odd spin structure $\nu$ used for its definition. We recall that the Abel map is implicit in the argument of the theta function, $z-w\,\mapsto\,\int_{w}^{z} \omega_I$.

For each even spin structure $\delta$, the {\it Szeg\H{o} kernel} is defined as
\begin{equation}
\label{equ:szegodef}
S_\delta(z,w)\equiv \frac{\vartheta[\delta](z-w)}{\vartheta[\delta](0) E(z,w)}  \,.
\end{equation}
It is a (1/2)-form in both $z$ and $w$, with a simple pole at $z=w$,
\begin{equation}
S_\delta(z,w) \approx \frac{\sqrt{\d z}\sqrt{\d w}}{z-w} \qquad \text{for} \; z\approx w\,,
\end{equation}
and it is holomorphic elsewhere. The Szeg\H{o} kernel plays the role of fermionic Green's function for a $bc$-system with weight 1/2 and even spin structure $\delta$, and thus arises in the correlation functions of the world-sheet spinor fields $\psi$ and $\tilde \psi$ of the ambitwistor string. In particular,
\begin{equation}
\langle \psi^\mu(z)\psi^\nu(w)  \rangle_{\delta} = \eta^{\mu\nu} S_\delta(z,w) \,.
\end{equation}

The prime form is single valued when $z$ goes around an $A_I$-cycle, but it has non-trivial monodromy around a $B_I$-cycle,
\begin{equation}
E(z,w) \,\mapsto\, -\exp \left( -i\pi\Omega_{II}+2\pi i \int_w^z \omega_I \right) E(z,w) \,.
\end{equation}
It can, however, be used to define single-valued differentials of interest.\footnote{Abelian differentials are the holomorphic or meromorphic 1-forms on a Riemann surface. The Abelian differentials of the {\it first kind} are the holomorphic differentials, and are said to be normalised if their $A$-periods obey the first condition in \eqref{eq:periods}. The Abelian differentials of the {\it second kind} are the meromorphic differentials with only poles without residues, i.e.~no simple poles. The Abelian differentials of the {\it third kind} are the meromorphic differentials with only simple poles. The Abelian differentials of second and third kinds are said to be normalised if their $A$-periods vanish.}
There is a normalised {\it Abelian differential of the second kind} defined by 
\begin{equation}
\omega(z,w) \equiv  \d z \d w \, \partial_z\partial_w \log{E(z,w)} = \omega(w,z) \,.
\label{eq:nAd2nd}
\end{equation}
In this case, it is actually a 1-form in both $z$ and $w$, with a double pole at $z=w$,
\begin{equation}
\omega(z,w) \approx  \frac{\d z \d w}{(z-w)^2} \qquad \text{for} \; z\approx w\,,
\end{equation}
and it is holomorphic elsewhere. Its $A$-periods vanish and its $B$-periods are $2\pi i \,\omega_I(w)$ for $z$ around $B_I$, where $\omega_I$ are the holomorphic (i.e., first-kind) Abelian differentials from \eqref{eq:periods}.
The prime form also defines a class of normalised {\it Abelian differentials of the third kind} as 
\begin{equation}
\omega_{w_1,w_2}(z) \equiv  \d z\,\partial_z \log\frac{E(z,w_1)}{E(z,w_2)} =- \omega_{w_2,w_1}(z) \,.
\end{equation}
This is a 1-form with a pair of simple poles with $\pm 1$ residues,
\begin{equation}
\omega_{w_1,w_2}(z) \approx  (-1)^a \frac{\d z}{z-w_a} \qquad \text{for} \; z\approx w_a\,,\;\; a=1,2\,,
\end{equation}
 and it is holomorphic elsewhere. Again, its $A$-periods vanish. 
 
 Suppose that we want to solve the following equation for a differential $p=p(z)dz$,
 \begin{equation}
 \label{eq:qtoy}
 \dbar p = 2\pi i \sum_i q_i \,\bar\delta(z-z_i)\,\d z \,,
\end{equation}
for some constants $q_i$ satisfying $\sum_i q_i=0$\,. We recall that $2\pi i \bar\delta(z)=\bar\partial(1/z)$\,. We can re-express the equation as
 \begin{equation}
 \dbar p = 2\pi i \sum_i q_{i}\,[\bar\delta(z-z_i)-\bar\delta(z-z_\ast)]\,\d z \,,
\end{equation}
where $z_*$ is an arbitrary point. So $p$ is a meromorphic differential with simple poles at $z_i$ of residue $q_i$, and with no pole at $z_*$ due to $\sum_i q_i=0$\,. On a genus-$g$ Riemann surface, the general solution can be written as
\begin{equation}
\label{eq:soldbarp}
p = c^I \omega_I + \sum_i q_i \,\omega_{i,*}\,,
\end{equation}
where the $c^I$ are the $g$ constants of integration of the homogeneous equation, each associated to an Abelian differential of the first kind, and the $\omega_{i,*}$ are Abelian differentials of the third kind, with residue 1 at $z_i$ and residue $-1$ at $z_*$. We will later on make use of this result, with the ambitwistor worldsheet field $P_\mu$ playing the role of $p$, and with the external momenta ${k_i}_\mu$ playing the role of the charges $q_i$.

\subsection{Zero modes and partition functions of chiral \texorpdfstring{$bc$}{bc} and \texorpdfstring{$\beta\gamma$}{beta gamma} systems}\label{sec:Verlinde}

Using the holomorphic Abelian differentials, the theta functions and the prime form, it is possible to construct a prominent class of objects relevant to the study of conformal field theories on Riemann surfaces, namely the partition functions of chiral fermionic $bc$ or bosonic $\beta\gamma$ systems.

The number of zero modes of $\dbar_\lambda$, i.e., the operator $\dbar$ acting on a worldsheet field of integer or half-integer weight $\lambda$, is given by, for $g\geq2$,
\begin{equation}
\label{eq:zmcount}
\Upsilon(\lambda) = 
\left\{
\begin{array}{cl} 0 &\text{for $\lambda<0$ or $\lambda=1/2$ with even spin structure},\\ 1&\text{for $\lambda=0$ or $\lambda=1/2$ with odd spin structure},\\ g&\text{for $\lambda=1$}, \\ (2\lambda-1)(g-1) &\text{for $\lambda\geq3/2$}.
    \end{array} \right.
\end{equation}
The cases with $\lambda\in\{0,1/2,1\}$ actually apply at any genus. For $\lambda=0$, the zero mode is the constant function. For $\lambda=1$, the $g$ zero modes correspond to the Abelian holomorphic differentials. For even (odd) spin structures, there is no (one) zero mode of $\dbar_{1/2}$, the worldsheet Dirac operator.

Consider a system with weights $\lambda$ for $b$ (or $\beta$) and $1-\lambda$ for $c$ (or $\gamma$), and denote $Q=2\lambda-1$. The partition function is defined such that the zero modes are saturated. For instance, for a $bc$ system with $\lambda>1$,  the partition function is the determinant of $\dbar$ acting on $c$,\footnote{Notice that, in our notation, the partition function is a differential form, whereas in some works, such as \cite{Verlinde:1986kw}, only the determinant is defined as the differential form.}
\begin{equation}
{\det}' \,\dbar _{1-\lambda}= \int \mathcal{D}b\,\mathcal{D}c \;\; e^ {-S_{b,c}}\hspace{-.1cm}\prod_{i=1}^{Q(g-1)} \hspace{-.1cm} b(z_i) \,,
\end{equation}
where the prime in $\det '$ denotes the saturation of zero modes, without which the path integral would vanish; the saturation is produced by the $Q(g-1)$ insertions of $b$, which absorb the zero modes. The partition function therefore depends on the $z_i$. The manner in which the various ingredients of the amplitude will appear, among them the partition functions of several chiral systems, leads to the cancellation of all dependences of this type. In the case $\lambda=1/2$, which is relevant for the $\Psi^\mu$ and $\tilde\Psi^\mu$ systems of the ambitwistor string, the partition function is $(\det \dbar _{1/2;[\kappa]})^{1/2}$, where $\kappa$ denotes the spin structure.  For a $\beta\gamma$ system with $\lambda\geq 3/2$, the partition function is $(\det \dbar _{1-\lambda})^{-1}$. Each of these determinants is `primed' whenever zero modes require care, for either of the conjugate fields.

With the help of bosonisation, Ref.~\cite{Verlinde:1986kw} computed the determinant of $\det \dbar _{1-\lambda}$ in all these cases.\footnote{See also Ref.~\cite{Knizhnik:1986kf} for an alternative approach.} Here, we just quote the results.  For $\lambda\neq 1$, and specifying the spin structure (only relevant for half-integer $\lambda$),
\begin{equation}
\det \dbar _{1-\lambda;[\kappa]}=Z^{-1/2} \widehat{Z}_{\lambda}[\kappa] \,,
\end{equation}
with 
\begin{subequations}\label{equ:Verlinde}
\begin{align}
 \widehat Z_\lambda[\kappa]&=\vartheta[\kappa]\Bigg(\sum_{i=1}^{Q(g-1)} z_i-Q\Delta \Bigg)\prod_{i<j}E(z_i,z_j)\prod_i \sigma(z_i)^Q  \,,\\
 Z^{3/2}&=\vartheta\Bigg(\sum_{I=1}^g z_I-w-\Delta \Bigg)\frac{\prod_{I<J}E(z_I,z_J)\prod_I \sigma(z_I)}{\det(\omega_I(z_J))\prod_{I}E(z_I,w) \,\sigma(w)} \,,
\end{align}
\end{subequations}
where $\sigma$ is a $g/2$-form defined by the ratio
\begin{equation}
\frac{\sigma(z)}{\sigma(w)}= \frac{\vartheta(\sum_{I=1}^g r_I -z-\Delta)}{\vartheta(\sum_{I=1}^g r_I -w-\Delta)} \, \prod_{I=1}^g \frac{E(r_I,w)}{E(r_I,z)}\,,
\end{equation}
which is independent of the points $r_I$.
It follows from the Riemann vanishing theorem that $\sigma$ has neither zeroes nor poles. In the special case $\lambda=1$, relevant for the $PX$ system of the ambitwistor string, the partition function is given by $(\det ' \dbar_0)^{-1}=Z^{-1}$.

\subsection{Deligne-Mumford compactification and non-separating degenerations}\label{sec:degen-review}
Later on, we will see that the full two-loop amplitude localises on a singular boundary of the moduli space, where both $q_{11}$ and $q_{22}$ vanish. This boundary divisor describes a non-separating degeneration of the surface where both $A_I$-cycles collapse to a point. The resulting surface is a Riemann sphere  with a node  (pair of identified points) corresponding to the each  pinched $A_I$-cycle. 

To this end, we review briefly the Deligne-Mumford compactification of the moduli space of Riemann surfaces, with special focus on the non-separating degenerations. More details can be found in the original papers \cite{Deligne:1969, fay1973theta}.\\

\begin{figure}[ht]
	\centering 
	\begin{subfigure}[t]{0.49\textwidth} 
        \centering
        \includegraphics[width=6cm]{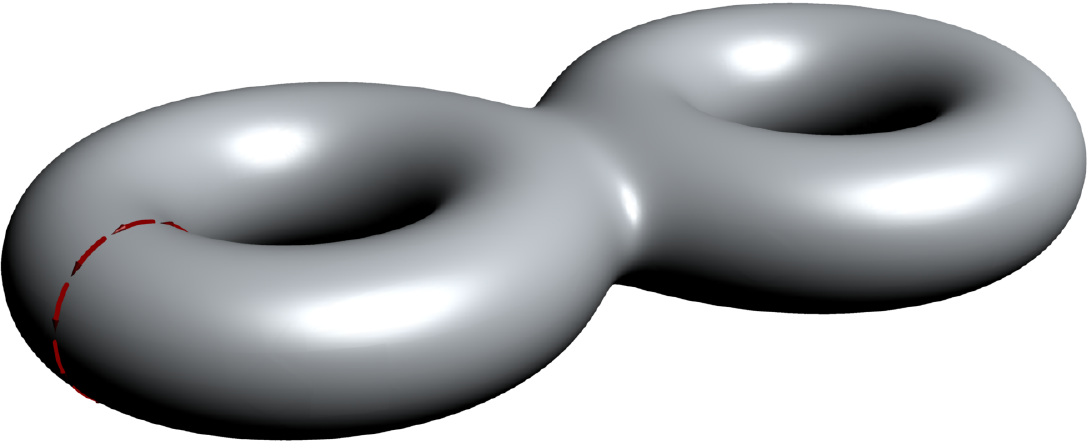}\\\vspace{10pt}
        \includegraphics[width=6cm]{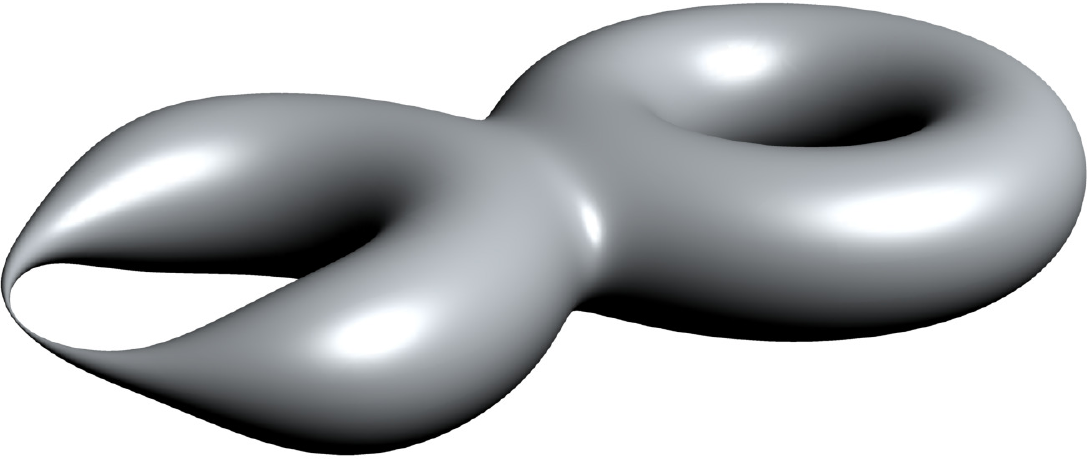}\vspace{5pt}
        \caption{Original genus two surface and its non-separating degeneration $\mathfrak{D}^{\text{non-sep}}_2$, corresponding to a nodal torus with two additional punctures from the pinched $A$-cycle.}
        \label{fig:Deligne-Mumford_non-sep}
    \end{subfigure}\hfill
    \begin{subfigure}[t]{0.48\textwidth} 
        \centering
        \includegraphics[width=6cm]{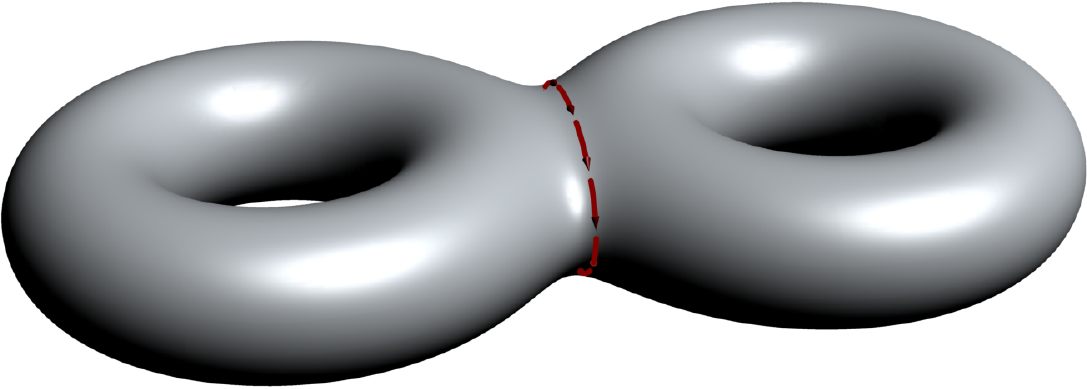}\\\vspace{18pt}
        \includegraphics[width=6cm]{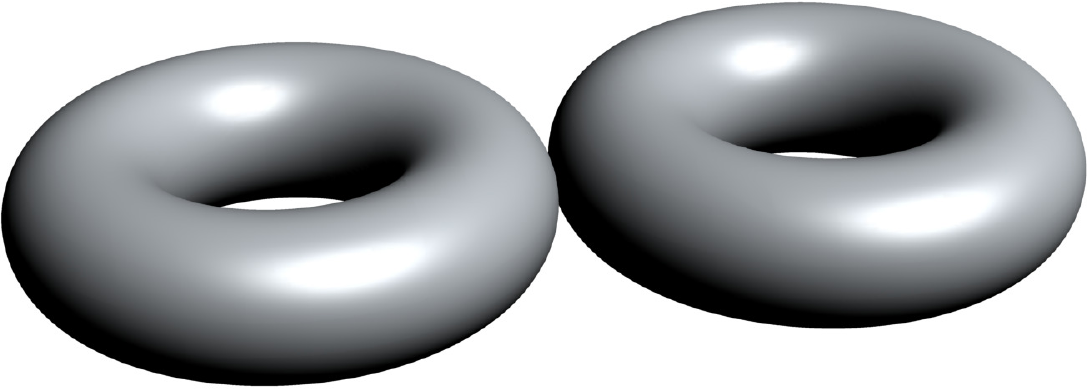}\vspace{5pt}
        \caption{Original genus two surface and its separating degeneration $\mathfrak{D}^{\text{sep}}_2$, corresponding to two tori, each with an additional puncture corresponding to the connecting node.}
        \label{fig:Deligne-Mumford_sep}
    \end{subfigure}%
	\caption{Separating and non-separating boundary divisors. The pinched cycles are indicated in red.}
	\label{fig:Deligne-Mumford}
\end{figure}

\paragraph{A lightning review of the Deligne-Mumford compactification.} %Implicitly, the residue theorems make extensive use of the Deligne-Mumford compactification of the moduli space. Before proceeding, we will give here a very brief review of the main idea.
The moduli space $\mathfrak{M}_{g,n}$ of Riemann surfaces with punctures is not compact because nodal surfaces arising from the contraction of a homology cycle are not included. The Deligne-Mumford compactification $\widehat{\mathfrak{M}}_{g,n}$  of the moduli space  \cite{Deligne:1969} is obtained by adding these nodal curves as ``divisors at infinity'' \cite{Witten:2012bh}. These divisors correspond to the possible degenerations of the Riemann surface $\Sigma$, and are characterised by whether the contracted homology cycle is trivial over $\mathfrak{M}_{g,0}$ or not. In the former case, they are known as separating degenerations $\mathfrak{D}^{\text{sep}}_{g,n}$, and they split $\Sigma$ into two components while partitioning the punctures accordingly. The nodal  singularity  adds an additional puncture on each surface, so that
\begin{equation}\label{equ:sep-degen}
 \mathfrak{D}^{\text{sep}}_{g,n} \cong\widehat{\mathfrak{M}}_{g_1,n_1+1}\times \widehat{\mathfrak{M}}_{g_2,n_2+1}\,,
\end{equation}
where $g=g_1+g_2$ and $n=n_1+n_2$; see figure \ref{fig:Deligne-Mumford_sep} for illustration. Non-separating degenerations, on the other hand, give rise to a surface of lower genus $g-1$, while adding two (identified) punctures corresponding to the node,
\begin{equation}
  \mathfrak{D}^{\text{non-sep}}_{g,n} \cong\widehat{\mathfrak{M}}_{g-1,n+2}\,,
\end{equation}
as illustrated in  figure \ref{fig:Deligne-Mumford_non-sep} .This behaviour of the moduli space plays a crucial role in worldsheet theories, where it corresponds to a factorisation behaviour similar to the cut of a Feynman diagram; see e.g. \cite{Witten:2012bh} for a recent review in the context of superstring theory.

\begin{figure}[ht]
	\centering 
	\includegraphics[width=6cm]{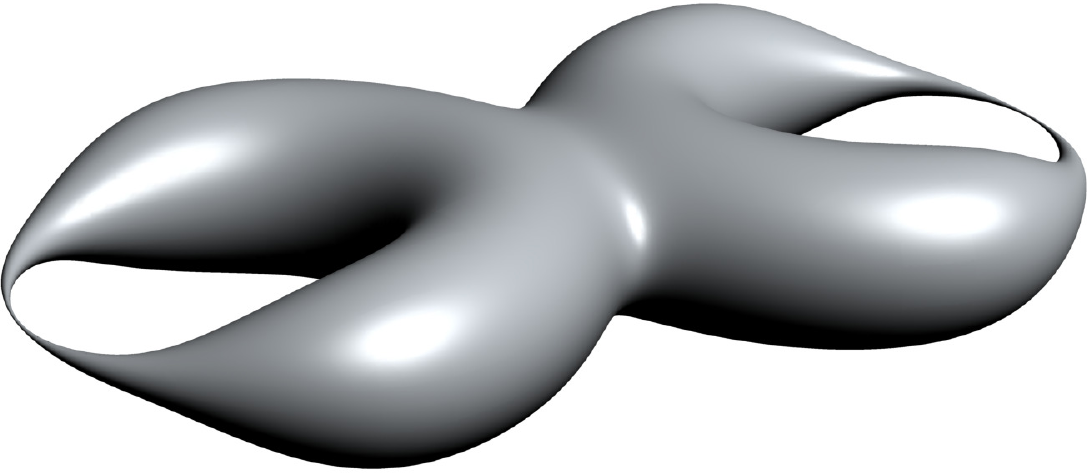}
	\caption{The maximal non-separating degeneration $\mathfrak{D}^{\text{max}}_{2,n}$ at genus two, corresponding to a bi-nodal Riemann sphere.}
	\label{fig:max_non-sep}
\end{figure}

For the ambitwistor string, we are most interested in the maximal non-separating divisor $\mathfrak{D}^{\text{max}}_{g,n}$, defined as the divisor degenerating $g$ non-trivial homology cycles,
\begin{equation}
  \mathfrak{D}^{\text{max}}_{g,n} \cong\widehat{\mathfrak{M}}_{0,n+2g}\,,
\end{equation}
as in \cref{fig:max_non-sep}.
In the second part of this article, we prove that the two-loop supergravity amplitude localises on this boundary divisor, and can thus be formulated over a bi-nodal Riemann sphere.\\

\paragraph{The non-separating degeneration. } We now discuss the non-separating degeneration of a Riemann surface in terms of its holomorphic differentials and its period matrix.

Consider the non-separating degeneration of the $A_g$-cycle, and denote the corresponding modular parameter by $ q_{gg} = \text{exp}(i\pi \Omega_{gg})$.  At the boundary divisor, the cycle $A_g$ shrinks to a single point and forms a node, whose locations 
we denote by $z_{g^+}$ and $z_{g^-}$. Due to the relation \eqref{eq:periods} between the normalised holomorphic differentials and the period matrix, the parameter $q_{gg}$ must tend to zero, and we can thus give the asymptotics of both the holomorphic differentials $\omega_I$ and the period matrix $\Omega_{IJ}$ as a series expansion in $q_{gg}$.
In particular, the  holomorphic differentials  $\omega_{I<g}$ approach the basis of holomorphic differentials $\omega_I^{(g-1)}$ on the lower-genus Riemann surface, while $\omega_g$ turns into the normalised (on the lower-genus surface) Abelian differential of the third kind with simple poles at the node, $\omega_{g^+,g^-}$. The precise asymptotics are given by Fay's degeneration formula \cite{fay1973theta},
\begin{subequations}\label{equ:Fay-hol-g}
\begin{align}
 \omega_I (z)& = \omega_I^{(g-1)}(z) + q_{gg}^2\,\left(\frac{\omega_I^{(g-1)}(z_{g^+})}{\d z_{g^+}}-\frac{\omega_I^{(g-1)}(z_{g^-})}{\d z_{g^-}}\right)\big(v_{g^+}(z) - v_{g^-}(z)\big)+\mathcal{O}(q_{gg}^4)\,,\\
 \omega_g (z)& = \frac1{2\pi i}\,\omega_{g^+,g^-}(z) +q_{gg}^2\, \widehat{\omega}_g(z) +\mathcal{O}(q_{gg}^4)\,.
\end{align}
\end{subequations}
Here, $\widehat{\omega}_g$ is a meromorphic differential with poles of order three at $z_{g^+}$ and $z_{g^+}$, and $v_{g^+},\,v_{g^-}$ are differentials of the second kind with a double pole at the nodal points $z_{g^+}$ and $z_{g^-}$, respectively; see \cite{fay1973theta,DHoker:1988ta}  for details.\footnote{The differentials $v_{g^\pm}$ are such that $v_{g^\pm}(z)=\frac1{4}\,\omega(z,z_{g^\pm})/\d z_{g^\pm}$, with $\omega(z,w)$ given in \eqref{eq:nAd2nd}. Fortunately, we will not need the precise form of the subleading terms in \eqref{equ:Fay-hol-g} in this paper.} For the period matrix, the asymptotics read
\begin{equation}\label{equ:Fay_period_matrix_g}
 \Omega = \begin{pmatrix}\Omega_{IJ}^{(g-1)} & \int_{z_{g^-}}^{z_{g^+}}\omega_I \\\int_{z_{g^-}}^{z_{g^+}}\omega_J & \;\;\frac{1}{i\pi}\ln q_{gg}+\text{const} \end{pmatrix}+\mathcal{O}(q_{gg}^2)\,.
\end{equation}
Just as the holomorphic differentials,  the period matrix thus descends to the lower-genus Riemann surface, while the entries $\Omega_{Ig}$ and $\Omega_{gI}$ encode the Abel map image of the node divisor.

When studying non-separating degenerations, it is often convenient to choose a parametrisation of the period matrix adapted to the problem. This will be especially important in the ambitwistor string, where non-separating boundary divisors associated to the pinching of different $A$-cycles contribute. We will thus frequently make use of the following parametrisation:
\begin{equation}\label{equ:paramet_Omega}
 \Omega = \begin{pmatrix} \tau_1+\tau_3 & \tau_3 \\ \tau_3 & \tau_2+\tau_3 \end{pmatrix}\,.
\end{equation}
This parametrisation has the advantage of isolating the contribution from each non-separating  boundary divisor. In particular, the limits $\tau_1\rightarrow i\infty$ or $\tau_2\rightarrow i\infty$ directly correspond, respectively, to the pinching of the $A_{1}$ or $A_2$-cycle in figure \ref{fig:homologybasis}. Moreover, pinching the cycle $A_1+A_2$ implies that $\tau_3\rightarrow i\infty$, as can be seen from a modular transformation exchanging the roles of $\tau_3$ and $\tau_2$. In the original representation \eqref{equ:q_II_def} of the period matrix, the latter degeneration requires an additional blow-up procedure to resolve $\tau_{1,2}$ remaining finite.
In analogy with \cref{equ:q_II_def}, we may further define 
\begin{equation}
 q_1 = e^{i\pi\tau_1}\,,\qquad q_2 = e^{i\pi\tau_2}\,,\qquad q_3 = e^{2i\pi\tau_3}\,.
\end{equation}
The benefit of this parametrisation of the moduli is that it neatly identifies the non-separating boundary divisors as $q_r=0$.

%%%%%%%%%%%%%%%%%%%%%%%%%%%%%%
%%%%%%%%%%%%%%%%%%%%%%%%%%%%%%
\section{The genus-two type II amplitude}\label{sec:g=2}

With the tools introduced in the last section at hand, we can now return to the type II ambitwistor string. Picking up where we left off in \cref{sec:review}, we calculate the $n$-point  correlator on a genus two Riemann surface. The calculation closely mirrors  the analogous procedure in the RNS superstring \cite{DHoker:2001kkt, DHoker:2001qqx, DHoker:2001foj, DHoker:2001jaf, DHoker:2005dys, DHoker:2005vch},  and leads to modular invariant expressions for the amplitude; which we prove in \cref{sec:mod-inv}. This close similarity may come as a surprise, given the conceptual differences: the ambitwistor string is inherently chiral\footnote{So no analogue of the chiral splitting procedure \cite{DHoker:1989cxq, DHoker:2001qqx}  of the RNS superstring is necessary.} and formulated over a bosonic -- not supersymmetric -- Riemann surface. We will see the details of how this plays out throughout this section, both in general and for the simplest non-trivial example, the four-point amplitude.

In addition to modular invariance, another feature familiar from the one-loop amplitude persists at genus two: the localisation of the moduli integral\footnote{Excluding the $P_\mu$ zero modes, whose integration corresponds to the loop integration.} on the scattering equations. As we will see throughout the next few sections, these two properties -- modular invariance and localisation on the scattering equations -- jointly localise the amplitude on the non-separating boundary divisor through the use of a residue theorem. The resulting formulation on a (bi-)nodal Riemann sphere will be the focus of \cref{sec:nodalRS} and \cref{sec:contour_argument}.

For simplicity, we will restrict ourselves to type II amplitudes with NS-NS external states, i.e., external states corresponding to linear superpositions of graviton, dilaton and B-field. Moreover, we will consider only the contribution from the 10 even spin structures. The contribution from the 6 odd spin structures -- that we do not consider here -- will obviously also satisfy modular invariance and localisation on the scattering equations. The odd spin structures do not contribute to the four-point amplitude, which we will analyse in detail. {\it After} the degeneration to the bi-nodal Riemann sphere, to be performed in later sections, we can easily consider theories in $d<10$ obtained by dimensional reduction. In this case, the odd spin structures do not contribute for any number of external particles. This applies to four-dimensional ${\mathcal N}=8$ supergravity and other supergravities obtained by dimensional reduction; likewise for the study of four-dimensional ${\mathcal N}=4$ super-Yang-Mills, from the super-Yang-Mills expressions to be constructed later. The absence of contributions from the odd spin structures can be understood as follows. An odd spin structure contains one zero mode of $\psi^\mu$ and one zero mode of $\tilde\psi^\mu$ (for each $\mu$), as discussed in \eqref{eq:zmcount}. The fermionic integration over these two sets of 10 zero modes leads to two 10-dimensional Levi-Civita symbols whose indices must be contracted into external polarisations, external momenta or loop momenta, due to the structure of the correlator (see \cite{Adamo:2013tsa} at genus one). If the latter quantities only span 9 or fewer dimensions, then the contribution from the odd structures to the amplitude vanishes.\footnote{Similarly, the odd spin structures do not contribute to the four-point amplitude because the two loop momenta, the four polarisations and the four external momenta only span 9 dimensions, due to momentum conservation.} The reason why the dimensional reduction of our amplitude formulae should be performed after the degeneration to the bi-nodal sphere is that, at genus two, $d=10$ is required by modular invariance. On the bi-nodal sphere, however, there is no notion of modular invariance, and therefore the formulae can be dimensionally reduced.

\subsection{The correlator}\label{sec:correlator}
The main object of this section is the $n$-point genus-two correlator of the ambitwistor string. Formally, this correlator is
\begin{equation}
 \mathcal{M}_n=\int \mathcal{D}\big[X,P,\psi,\tilde\psi,e,\tilde{e},\chi,\tilde{\chi}\big]\,e^{-S}\,\prod_{i=1}^n{\mathcal O}_i \hspace{5pt}\Bigg|_{g=2}\,,
 \end{equation}
 where the ${\mathcal O}_i$ are vertex operators representing external particles. The proper BRST definition of the correlator was discussed in section \ref{sec:review}, and the result is
 \begin{align}
 \mathcal{M}_n=\int \d^3\Omega\,\sum_{\kappa,\tilde\kappa}\eta_\kappa\,\eta_{\tilde\kappa}\,  &
 \left\langle \;\prod_{s=1}^3 \big\langle\hat\mu_s b\big\rangle\,\prod_{r=1}^3 \big\langle\mu_r \tilde{b}\big\rangle\,\,\bar\delta\Big(\big\langle\mu_r P^2\big\rangle\Big) \times \right.\nonumber \\ & \quad
 \left.\times
 \prod_{\alpha=1}^2 \delta\Big(\big\langle\chi_\alpha\beta \big\rangle\Big)\,\big\langle\chi_\alpha P\cdot\psi \big\rangle \,\,
 \delta\Big(\big\langle\tilde\chi_\alpha\tilde\beta \big\rangle\Big)\,\big\langle\tilde\chi_\alpha P\cdot\tilde\psi \big\rangle
 \,\prod_{i=1}^n\mathcal{V}_i \;\right\rangle_{\kappa,\tilde\kappa}\,,
 \label{equ:Z_unsimpl}
\end{align}
where the integrated vertex operators are given by
 \begin{equation}
 \label{typeIIintvo2}
\mathcal{V}_i = \int_\Sigma \bar\delta\big(\text{Res}_{z_i}P^2\big)\,(\epsilon_i\cdot P+k_i\cdot\psi\, \epsilon_i\cdot \psi)
(\tilde\epsilon_i\cdot P+k_i\cdot\tilde\psi\, \tilde\epsilon_i\cdot \tilde\psi)e^{ik_i\cdot X} \,.
\end{equation}
In the following, we will focus first on three salient features of this expression: (i) the double sum over spin structures, which incorporates the GSO projection, (ii) the distinct choice of Beltrami differentials $\hat\mu_s$ and $\mu_r$, which arose from the gauge fixing of $e$ and $\tilde e$, respectively, and (iii) the scattering equations, both those included in the integrated vertex operators $\mathcal{V}_i$, given by \eqref{typeIIintvo}, and the remaining three equations, which together impose $P^2=0$. 

\paragraph{GSO projection.}  As in the conventional RNS superstring, we project onto the correct degrees of freedom using the GSO projection, which amounts to summing over spin structures in the path integral formalism. We implement the GSO projection independently for $\psi$ and $\tilde\psi$, and denote the corresponding spin structures by $\kappa $ and $\tilde\kappa$, respectively.  This fixes the amplitude up to relative phases $\eta_\kappa$ between spin structures,  which are determined by modular invariance and unitarity \cite{Seiberg:1986by}.\footnote{In summary, modular invariance determines the relative phases among the even and among the odd spin structures, whereas unitarity fixes the relative phase between the two sectors (NS and R) to be $\pm1$.} We will see later in detail for even spin structures how modular invariance fixes the relative phases.

For simplicity, we will only discuss  the contribution from the even spin structures $\delta$ to the amplitude. This restriction is possible because  modular invariance preserves the distinction between even and odd spin structures, and thus each individual  sector of spin structures (even or odd) is modular invariant.\footnote{Though of course the resulting amplitudes from just the even or the odd sector are not unitary in $d=10$ \cite{Seiberg:1986by}.} Moreover, as discussed above, the odd spin structures do not contribute to amplitudes in dimensions $d\leq 9$, or to amplitudes with $n\leq 4$  external particles.\footnote{In the chiral ambitwistor string, there is no subtlety in reducing dimensions due to the absence of winding modes; see \cite{Geyer:2015jch}, Appendix D.}

\paragraph{Choice of beltrami differentials.} The amplitude \eqref{equ:Z_unsimpl} can be simplified further by a judicious choice of basis for the genus-two Beltrami differentials and their fermionic counterparts.  We recall from our discussion of PCOs in section \ref{sec:review} that $3g-3$ Beltrami differentials, which at genus two we denote as $\{\mu_r\}_{r=1}^3$, can be conveniently chosen to evaluate the field they are paired with at points $y_r$ not coincident with the particle punctures. The same applies to the fermionic counterparts $\{\chi_\alpha\}_{\alpha=1}^2$ and $\{\tilde\chi_\alpha\}_{\alpha=1}^2$ at points $x_\alpha$. That is,\footnote{This is a slight abuse of notation, the rhs is understood to carry no form degree in this particular instance.}
\begin{equation}\label{equ:beltrami_yx}
 \big\langle \mu_r\, \phi\big\rangle = \phi(y_r)\,,\qquad\qquad \big\langle \chi_\alpha\, \beta\big\rangle = \beta(x_\alpha)\,,\qquad\qquad \big\langle \tilde\chi_\alpha\, \tilde\beta\big\rangle =\tilde \beta(x_\alpha)\,.
\end{equation}
The full amplitude must of course be independent of the choice of $x_\alpha$ and $y_r$, and this will serve as an important check for our final expressions.

Recall also that the choice of Beltrami differentials $\hat\mu_s$ associated to the gauge fixing of $e$ was distinct from that for the gauge fixing of $\tilde e$, $\mu_r$. In particular, the choice of the three extra Beltrami differentials $\{\hat\mu_s\}_{r=1}^3$ relates them to the genus-two period matrix, while for  $\{\mu_r\}_{r=1}^3$ we took \eqref{equ:beltrami_yx}. Since we have reviewed the basic facts on holomorphic differentials in the last section, we can explicitly relate the factors  $\big\langle\hat\mu_s b\big\rangle$ and $\big\langle\mu_r b\big\rangle$ in \eqref{equ:Z_unsimpl}. This will make the correlator symmetric between the $bc$ and the $\tilde b \tilde c$ systems. In particular,
\begin{equation}
 \prod_{s=1}^3 \big\langle\hat\mu_s b\big\rangle = \frac{\det \big\langle\hat\mu_s \phi_t\big\rangle}{\det \big\langle\mu_r \phi_t\big\rangle}\prod_{r=1}^3 \big\langle\mu_r b\big\rangle= \frac{\det \big\langle\hat\mu_s \phi_t\big\rangle}{\det \phi_t(y_r)}\prod_{r=1}^3 b(y_r) = \frac{1}{\det \omega_I\omega_J(y_r)}\prod_{r=1}^3 b(y_r)\,.
\end{equation}
Here, $\phi_t$ denotes a basis of holomorphic quadratic differentials, and in the last step we have chosen $\phi_t=\omega_I\omega_J$ in order to simplify the expression. We conclude that\footnote{Notice that the factors on either side of \cref{equ:simppcos} have different form degrees. While on the left hand side, every factor has form degree zero (e.g. $\langle\chi_\alpha P\cdot\psi \rangle$), on the right hand side all factors carry form degree ($\chi_\alpha P\cdot\psi (x_\alpha)$ has form degree 3/2). The full expressions  are of course equal, the form degree cancels appropriately on the right hand side.}
\begin{align}\label{equ:simppcos}
&\prod_{s=1}^3 \big\langle\hat\mu_s b\big\rangle\,\prod_{r=1}^3 \big\langle\mu_r \tilde{b}\big\rangle\,\,\bar\delta\Big(\big\langle\mu_r P^2\big\rangle\Big)  \prod_{\alpha=1}^2 \delta\Big(\big\langle\chi_\alpha\beta \big\rangle\Big)\,\big\langle\chi_\alpha P\cdot\psi \big\rangle
  \,\, \delta\Big(\big\langle\tilde\chi_\alpha\tilde\beta \big\rangle\Big)\,\big\langle\tilde\chi_\alpha P\cdot\tilde\psi \big\rangle = \nonumber \\
 & =  \frac{1}{\det \omega_I\omega_J(y_r)}\prod_{r=1}^3 b(y_r)\,\tilde b(y_r) \,\,\bar\delta\big(P^2(y_r)\big)  \prod_{\alpha=1}^2 \delta\Big(\beta(x_\alpha) \Big)\, \delta\Big(\tilde\beta(x_\alpha) \Big)\, P\cdot\psi(x_\alpha)\,P\cdot\tilde\psi(x_\alpha) \,.
\end{align} 

\vspace{0cm}

\subsection{The scattering equations}\label{sec:SE}

Let us now focus on the scattering equations and the $PX$-system. The only dependence of the correlator on $X$ is in the kinetic term $\int P\cdot \dbar X$  and in the plane wave factors $e^{ik_i\cdot X}$ of the vertex operators. Similarly to conventional string theory, the integration over the (constant) zero mode of $X$ leads to a delta function, which imposes the constraint of momentum conservation, $\sum_ik_i=0$\,. The integration over the non-zero modes of $X$ leads to another delta function that localises $P$ to its classical value through
\begin{equation}
 \dbar P_\mu = 2\pi i\sum_ik_{i\,\mu}\,\bar\delta(z-z_i)\,\d z \,.
\end{equation}
As discussed at the end of section \ref{sec:primeSzegodiffs}, on a genus-two Riemann surface this is solved by
\begin{equation}\label{equ:ell}
 P_\mu = \ell^I_\mu \omega_I +\sum_{i=1}^n k_{i\,\mu}\omega_{i,*} \,,
\end{equation}
where $\omega_{i,*}$ is a meromorphic differential of the third kind with residues $\pm 1$ at the point $z_i$ and at an arbitrary reference point $z_*$. The residue at $z_*$ vanishes from \cref{equ:ell} due to momentum conservation.
We suggestively denote the zero mode parameters of $P_\mu$ by ${\ell_1}_\mu$ and ${\ell_2}_\mu$. If the meromorphic differential $\omega_{i,*}$ is normalised (i.e., has vanishing A-periods), then ${\ell_I}_\mu=\oint_{A_I}P_\mu$. Naturally, the path integral will involve an integration over ${\ell_1}_\mu$ and ${\ell_2}_\mu$. Moreover, the localisation of $P$ introduces a Jacobian factor of $(\det' \dbar_0)^{-10}=Z^{-10}$, as discussed in \cref{sec:Verlinde}.

The constraint $P^2=0$ is imposed in the gauge fixing procedure via $n+3g-3$ delta functions, which are of two types, according to the choice of Beltrami differentials $\mu_r$ in $\bar\delta(\langle\mu_r\,P^2\rangle)$. The first is the type included in the $n$ integrated vertex operators \eqref{typeIIintvo}. In this case, $\mu_i$ extracts the residue at the puncture $z_i$, for $i=1,\cdots,n$,
\begin{equation}\label{equ:particleSE_g=2}
\bar\delta\Big( \big\langle \mu_i\,P^2\big\rangle \Big) = \bar\delta\big( \text{Res}_{z_i}P^2 \big)\,.
\end{equation}
On the support of these $n$ scattering equations, $P^2$ is holomorphic, and therefore it can be expressed in terms of the holomorphic differentials $\omega_I$ as $P^2=u^{IJ}\omega_I\omega_J$, for some $u^{IJ}$. In our choice leading to the simplification \eqref{equ:simppcos}, the remaining $3g-3=3$ scattering equations are associated to Beltrami differentials that extract the value of the field at a point $y_r$,   
\begin{equation}
\bar\delta\Big( \big\langle \mu_r\,P^2\big\rangle \Big) = \bar\delta\big(P^2(y_r)\big) = \frac{1}{\det \omega_I\omega_J(y_r)}\prod_{I\leq J} \bar\delta\big(u^{IJ}\big)\,.
\end{equation}
Notice that, in the absence of vertex operators, this implies that $u^{IJ}=\ell_I\cdot\ell_J=0$ for all $I$, $J$.

Putting all this together, the amplitude is given by
\begin{equation}\label{equ:Z_presplit}
\begin{split}
 \mathcal{M}_n=&\;\delta^{10}\left(\sum_{i=1}^n k_i\right)\,\int
 \frac{ \d^{20}\ell\,\d^3\Omega}{(\det \omega_I\omega_J(y_r))^2\,Z^{10}} \prod_{I\leq J}\bar\delta\big(u^{IJ}\big)\times \\
 &\qquad\times\sum_{\delta,\tilde\delta}\eta_\delta\eta_{\tilde\delta} \,\left\langle\prod_{r=1}^3 b(y_r)\,\tilde{b}(y_r)\,\prod_{\alpha=1}^2 \delta\Big(\beta(x_\alpha) \Big)\, \delta\Big(\tilde\beta(x_\alpha) \Big)\, P\cdot\psi(x_\alpha)\,P\cdot\tilde\psi(x_\alpha) \,\prod_{i=1}^n\mathcal{V}_i\right\rangle_{\delta,\tilde\delta}
 \end{split}
\end{equation}
where we have pulled the $\bar\delta$-functions out of the path integral with the understanding that the $PX$ integral  has been performed, and has localised $P$ to its classical value \eqref{equ:ell}. In the following sections, we will evaluate the remaining correlator.

In principle, the integration over the moduli space of the Riemann surface $\mathfrak{M}_{2,n}$ in \eqref{equ:Z_presplit} is completely localised on the solutions to the genus-two scattering equations,
\begin{equation}
\text{Res}_{z_i}P^2 = 0 \,,\quad i=1\cdots,n\,, \qquad u^{IJ}=0 \,,\quad I,J=1,2\,.
\end{equation}
In practice, it seems hopeless to solve these equations explicitly. The formula for the genus-two $n$-point amplitude studied in this section seems, therefore, impractical. We will see in later sections, however, how it can be turned into a much more manageable formula on the Riemann sphere. In the meantime, we will describe several simplifications of \eqref{equ:Z_presplit} at genus two.

\subsection{The moduli space of the ambitwistor string at genus two}   \label{sec:moduli_space_A}

As seen in the correlator \eqref{equ:Z_presplit}, the complete moduli space of the ambitwistor string not only includes the moduli space of marked Riemann surfaces $\mathfrak{M}_{g,n}$, but also the moduli corresponding to the zero modes of the field $P_\mu\in\Omega^0(\Sigma,K_\Sigma)$. At genus two, the latter consist of $\ell_\mu^1$ and $\ell_\mu^2$, as in \eqref{equ:ell}, which are both integrated over the full ten-dimensional momentum space. Clearly, this should be interpreted as the loop momenta integration, but there are two important subtleties, which we discuss now.

The first subtlety is related to the contour of integration. Since the ambitwistor string target space is the space of complexified null geodesics, the zero mode coefficients $\ell_\mu^1$ and $\ell_\mu^2$ are integrated over the \emph{complexified} ten-dimensional momentum space. To make contact with a  field-theory-like loop integration, we must thus choose a reality prescription corresponding to a  contour selecting a middle-dimensional slice of ${\mathbb C}^{20}$. Clearly, the most natural choice for this contour would be the real slice $\mathbb{R}^{20}\subset\mathbb{C}^{20}$, with an appropriate $i\epsilon$-prescription. However, the zero mode coefficients $\ell_\mu^I$ are not unconstrained: recall that under a modular transform, the period matrix transforms as $\Omega\rightarrow\widetilde\Omega= \big(a\Omega +b\big)\,\big(c\Omega+d\big)^{-1}$. This implies that the normalised holomorphic differentials $\omega_I$ transform as
\begin{equation}\label{equ:mod_transform_holdiff}
  \omega\rightarrow\widetilde\omega=\omega\big(c\Omega+d\big)^{-1}\,.
\end{equation}
The meromorphic differentials, however, are invariant under modular transformations, as we will discuss in detail in \cref{sec:mod_inv_sub}. In order for $P_\mu$ to transform homogeneously the modular group, the loop momenta $\ell_\mu^I$ must compensate\footnote{The non-trivial transformation \eqref{eq:mod_loop} of the loop momenta plays a crucial role for modular invariance. In particular, it ensures that the scattering equations transform with homogeneous modular weight, see \cref{sec:mod_inv_sub} for details.} for the transformation of the holomorphic differentials, 
\begin{equation}\label{eq:mod_loop}
 \ell\!\to\!(c\Omega+d)\ell\,.
\end{equation}
A real loop integration contour for one fundamental domain therefore corresponds to a  different, generally complex contour for other fundamental domains. We should thus only require the loop integration contour to be real for \emph{one} fundamental domain, with its behaviour for other parametrisations of the moduli space determined by the modular transformation \eqref{eq:mod_loop}.\footnote{In other words, there is an equivalence class of integration cycles related by modular transformations, and the prescription is that a correct integration cycle is in the equivalence class of the real cycle (with $i\epsilon$-prescription). The differential form that is integrated is modular invariant, as we check in section~\ref{sec:mod-inv}.} Since we will consider a singular limit of the period matrix in the degeneration to the bi-nodal sphere, it is simpler to just take the contour as the real section for the fundamental domain we have chosen to work with, given by \eqref{equ:fund-domain}. We see no obstruction to the validity of this prescription, but it would be important to investigate it further. 

Recall in this context that the ten-dimensional supergravity amplitude is of course not defined, even if the loop integrand can be constructed --  so when we talk of  an ``amplitude" here, this is an abuse of language. However, as outlined in the introduction, the final formula \eqref{equ:initial_formula} for the amplitude on the bi-nodal sphere (after applying two residue theorems on the moduli space) is valid in any dimension $d\leq 10$. At this point, the structure of a field-theory-like integrand becomes clear \cite{Geyer:2015bja}, and it is possible to use dimensional regularisation and to define an appropriate $i\epsilon$-prescription \cite{Baadsgaard:2015twa} for the loop integration.  \\

The second subtlety is that the way in which $\ell_\mu^1$ and $\ell_\mu^2$, the moduli of $P_\mu$, appear in the correlator is asymmetric. Recall that we defined the fundamental domain of the period matrix according to a set of inequalities in \eqref{equ:fund-domain}, including the condition (ii):
\begin{equation}\label{equ:fund-dom-ii}
 0<2 \text{Im}(\Omega_{12})\leq\text{Im}(\Omega_{11})\leq\text{Im}(\Omega_{22})\,.
 \end{equation}
This leads to an asymmetry in $\ell_\mu^1$, $\ell_\mu^2$ and $\ell_\mu^1+\ell_\mu^2$, which is unnatural from the point of view of the field-theory loop interpretation.\footnote{
 To see this asymmetry in action, consider the subset of scattering equations $u^{IJ}=0$ and the modular parameters \eqref{equ:q_II_def}. As we shall see in \cref{sec:nodalRS}, it is possible to show that, for the maximal non-separating boundary divisor $q_{II}=e^{i\pi\Omega_{II}}\to 0$, we have $u^{II}=\ell_I^2+q_{II}\mathcal{F}_{I}+\mathcal{O}(q_{II}^2)$. Now, the inequality \eqref{equ:fund-dom-ii} implies that $|q_{22}|\leq|q_{11}|$. Therefore, in a double-degenerate limit $|q_{22}|\ll |q_{11}|\ll 1$, the existence of solutions to the scattering equations implies that $|\ell^2_2|\ll |\ell^2_1|$. In a further degeneration $q_{12}\to 0$, we also get $|\ell^2_2|\ll |\ell^2_1|\ll |(\ell_1+\ell_2)^2|$.
}
To address this, we can symmetrise over the different parametrisations of the zero modes of $ P_\mu$, namely
\begin{align}\label{equ:all_P}
 P_\mu^{(1)} &= \ell_{1\,\mu} \omega_1 +\ell_{2\,\mu} \omega_2+\sum_{i=1}^n k_{i\,\mu}\omega_{i,*} \,, && P_\mu^{(4)}  = \ell_{2\,\mu} \omega_1 + \big(\ell_{1}  +\ell_{2} \big)_\mu \omega_2 +\sum_{i=1}^n k_{i\,\mu}\omega_{i,*} \,,\nonumber\\
 P_\mu^{(2)} &= \ell_{2\,\mu} \omega_1 +\ell_{1\,\mu} \omega_2+\sum_{i=1}^n k_{i\,\mu}\omega_{i,*} \,,&&P_\mu^{(5)} = \ell_{1\,\mu} \omega_1 + \big(\ell_{1}  +\ell_{2} \big)_\mu\omega_2 +\sum_{i=1}^n k_{i\,\mu}\omega_{i,*} \,,\\
 P_\mu^{(3)} &= \big(\ell_{1}  +\ell_{2} \big)_\mu\omega_1+\ell_{2} \omega_2+\sum_{i=1}^n k_{i\,\mu}\omega_{i,*} \,,&& P_\mu^{(6)} = \big(\ell_{1}  +\ell_{2} \big)_\mu\omega_1 +\ell_{1\,\mu} \omega_2 +\sum_{i=1}^n k_{i\,\mu}\omega_{i,*} \,.\nonumber
\end{align}
Effectively, we are symmetrising over the orderings of the inequalities \eqref{equ:fund-dom-ii}.
The full amplitude is then 
\begin{equation}\label{equ:patches_ampl}
 \mathcal{M}_n=\sum_{\alpha=1}^6\mathcal{M}_n^{(\alpha)}\,,
\end{equation}
where each term $\mathcal{M}_n^{(\alpha)}$ is evaluated at $P_\mu=P^{(\alpha)}_\mu$. To define this expression more rigorously, we can solve $P_\mu$ by $P_\mu=c^I_\mu \omega_I +\sum_{i=1}^n k_{i\,\mu}\omega_{i,*}$, and insert an identity of the form 
\begin{equation}
 1=\int \d^{20}\ell \,\,\delta\Big(\ell_\mu^I-\oint_{\widehat{A}_I}P_\mu\Big)
\end{equation}
in the amplitude. The different charts are then given by different choices of cycles $\widehat{A}_I$, with e.g. $(\widehat{A}_1,\widehat{A}_2) = (A_1,A_2)$ for the parametrisation $\alpha=1$, and $(\widehat{A}_1,\widehat{A}_2) = (A_2,A_1)$ for the parametrisation $\alpha=2$. We recover the same expressions given above after integrating out the charges $c^I_\mu$.

While this is a cumbersome prescription for the amplitude, it is actually equivalent to a much simpler representation. To see this recall that \cref{equ:all_P} forces the zero modes $\ell_\mu^I$ to transform non-trivially under modular transformations to ensure that $P_\mu$ is of homogeneous (vanishing) modular weight. Using this, we can apply a modular transformation to relate all six terms in the amplitude to $\mathcal{M}_n^{(1)}$ with $P^{(\alpha=1)}_\mu$, but now with different inequalities for the imaginary parts of period matrix. This is most easily established for the modular transformation relating $\mathcal{M}_n^{(2)}$ to $\mathcal{M}_n^{(1)}$, which just exchanges the cycles $(A_1,B_1)\leftrightarrow(A_2,B_2)$. This implies $\omega_1\leftrightarrow \omega_2$, and so we conclude that this modular transformation also  exchanges $\ell_1\leftrightarrow\ell_2$ and $\Omega_{11}\leftrightarrow\Omega_{22}$. Applying these transformations to $\mathcal{M}_n^{(2)}$, we recover $\mathcal{M}_n^{(1)}$ integrated over the copy of the fundamental domain defined by $2\text{Im}(\Omega_{12})\leq\text{Im}(\Omega_{22})\leq \text{Im}(\Omega_{11})$, instead of \eqref{equ:fund-dom-ii}. There are suitable modular transformations relating each term in the sum \eqref{equ:patches_ampl} to $\mathcal{M}_n^{(1)}$, which are described in \cref{app:proof_modspace}. The reader may find it easier to understand this discussion after taking a look at \cref{sec:mod-inv}, where modular transformations are studied in detail. To summarise, the terms $\mathcal{M}_n^{(\alpha)}$ in \eqref{equ:patches_ampl} all localise $P_\mu$ on $ P_\mu^{(1)} = \ell^I_\mu \omega_I +\sum_{i=1}^n k_{i\,\mu}\omega_{i,*} $, but are formulated over six different copies of the fundamental domain $\mathfrak{M}_g$,
\begin{align}
\mathcal{M}_n^{(1)} &\equiv  \mathcal{M}_n^{(1)}\bigg|_{0<2\text{Im}(\Omega_{12})\leq\text{Im}(\Omega_{11})\leq \text{Im}(\Omega_{22})} && \mathcal{M}_n^{(4)} =  \mathcal{M}_n^{(1)}\bigg|_{0<\text{Im}(\Omega_{11})\leq\text{Im}(\Omega_{22})\leq 2\text{Im}(\Omega_{12})}\nonumber\\
  \mathcal{M}_n^{(2)} &=  \mathcal{M}_n^{(1)}\bigg|_{0<2\text{Im}(\Omega_{12})\leq\text{Im}(\Omega_{22})\leq \text{Im}(\Omega_{11})} && \mathcal{M}_n^{(5)} =  \mathcal{M}_n^{(1)}\bigg|_{0<\text{Im}(\Omega_{22})\leq\text{Im}(\Omega_{11})\leq 2\text{Im}(\Omega_{12})}\\
   \mathcal{M}_n^{(3)} &=  \mathcal{M}_n^{(1)}\bigg|_{0<\text{Im}(\Omega_{11})\leq2\text{Im}(\Omega_{12})\leq \text{Im}(\Omega_{22})} && \mathcal{M}_n^{(6)} =  \mathcal{M}_n^{(1)}\bigg|_{0<\text{Im}(\Omega_{22})\leq2\text{Im}(\Omega_{12})\leq \text{Im}(\Omega_{11})}\,.\nonumber
\end{align}
The sum in \eqref{equ:patches_ampl} now combines  to the single expression, where the integration domain for the modular parameters is simply defined by
\begin{align}\label{equ:fund-ambi}
(i)& \quad -\frac{1}{2}\leq \text{Re}(\Omega_{11}),\text{Re}(\Omega_{12}),\text{Re}(\Omega_{22}) \leq \frac{1}{2}\,, \nonumber \\
(ii)&\quad 0<\text{Im}(\Omega_{IJ})\quad \forall_{I,J}\,,\\
(iii)& \quad \left|\text{det}(c\,\Omega+d)\right| >1 \quad \forall
\begin{pmatrix}a&b\\c&d\end{pmatrix} \in \text{Sp}(4,\mathbb{Z}) \,.\nonumber
\end{align}
Since this space plays an important role for the ambitwistor string, let us denote it by $\mathfrak{M}'_2$, where the notation is chosen to reflect its close relation to the moduli space $\mathfrak{M}_g$ of Riemann surfaces. The definition \eqref{equ:fund-ambi} naturally extends to the surface with vertex operators insertions, in which case the space is $\mathfrak{M}'_{2,n}$. We conclude that the amplitude is  given by \eqref{equ:Z_presplit} with $P=P^{(1)}=\ell^I_\mu \omega_I +\sum_{i=1}^n k_{i\,\mu}\omega_{i,*}$, but integrated over $\mathfrak{M}'_{2,n}$ and the loop momenta.\\

Before proceeding to calculate this amplitude, let us briefly comment on two aspects of the moduli space $\mathfrak{M}'_{g,n}$. Due to its close relation  to $\mathfrak{M}_{g,n}$, the compactification of  $\mathfrak{M}'_{g,n}$ can be defined in full analogy to the Deligne-Mumford compactification. However, note that a new feature emerges: instead of a single non-separating boundary divisor, $\widehat{\mathfrak{M}}'_{g,n}$ contains three distinct non-separating boundary divisors, each corresponding to a different degeneration of the Riemann surface; see figure \ref{fig:1stresthm} in \cref{sec:nodalRS_SE} for illustration. In contrast to string theory, where all of these degenerations would be the same after a modular transformation, they represent here genuinely different degenerations, with different loop momenta associated to each homology cycle. 

An important conclusion from our discussion here is that it would be both interesting and fruitful to study the  ambitwistor string moduli space more deeply. Important work in this context has been done by refs.~\cite{Casali:2017zkz} and \cite{Ohmori:2015sha}, but especially extensions to higher genus  remain largely an open problem, on which the treatment given here could shed some light. We postpone this topic for future investigation.

\subsection{The chiral partition function}\label{sec:partition-function}
\paragraph{Defining the chiral partition function}
As observed in \cref{sec:review}, a crucial property of the ambitwistor string action is that, after gauge fixing, it is free and linear. In particular, this means that we can decompose the correlator in \cref{equ:Z_presplit} into the correlators over the different tilded and untilded fields,
\begin{equation}\label{equ:Z_split}
 \mathcal{M}_n=\int \d^{20}\ell \,
 \d^3\Omega\prod_{I\leq J}\bar\delta\big(u^{IJ}\big)\,\sum_{\delta,\tilde\delta}\eta_\delta\eta_{\tilde\delta}\,\mathcal{Z}^{\text{chi}}[\delta]\,\widetilde{\mathcal{Z}}^{\text{chi}}[\tilde\delta]\left\langle \prod_{\alpha=1}^2 P\cdot\psi(x_\alpha)\,P\cdot\tilde\psi(x_\alpha)\,\prod_i\mathcal{V}_i \right\rangle_{\delta,\tilde\delta}
\end{equation}
where we have defined `chiral' partition functions (in analogy to the RNS superstring) by
\begin{subequations}
\begin{align}
 \mathcal{Z}^{\text{chi}}[\delta]&=\frac{1}{(\det \omega_I\omega_J(y_r))\,Z^{5}}\left\langle\prod_{r=1}^3 b(y_r)\,\prod_{\alpha=1}^2 \delta\Big(\beta(x_\alpha) \Big) \right\rangle_\delta\,,\\
 \widetilde{\mathcal{Z}}^{\text{chi}}[\tilde\delta] &= \frac{1}{(\det \omega_I\omega_J(y_r))\,Z^{5}}\left\langle\prod_{r=1}^3 \tilde b(y_r)\,\prod_{\alpha=1}^2 \delta\Big(\tilde\beta(x_\alpha) \Big) \right\rangle_{\tilde\delta}\,.
\end{align}
\end{subequations}
In particular, since both tilded and untilded fields obey the same OPEs, it is sufficient to evaluate $\mathcal{Z}^{\text{chi}}[\delta]$ -- the result will extend straighforwardly to $\widetilde{\mathcal{Z}}^{\text{chi}}[\tilde\delta]$. It is worth highlighting at this point a major difference with respect to the conventional RNS string: the ambitwistor string is inherently chiral, and there is no sense of chiral splitting into left- and right-moving sectors, since the latter sector does not exist. However, in analogy with the chiral splitting in the RNS string \cite{DHoker:1989cxq}, the ambitwistor string correlator exhibits a `chiral contribution squared' ({\it not} absolute squared), as we have seen in \cref{equ:Z_split}. Indeed, we have
\begin{equation}
 \widetilde{\mathcal{Z}}^{\text{chi}}[\tilde\delta] =  {\mathcal{Z}}^{\text{chi}}[\tilde\delta]\,.
\end{equation}
{}

Since all fields in $\mathcal{Z}^{\text{chi}}[\delta]$ are $\beta\gamma$ systems, the chiral partition function is easily constructed using the results of Verlinde \& Verlinde \cite{Verlinde:1986kw} reviewed in \cref{sec:Verlinde}.  We read off $\lambda_b=2$,  $\lambda_\beta=3/2$ and  $\lambda_\psi=1/2$ from the field definitions, and therefore
\begin{equation}
 \mathcal{Z}^{\text{chi}}[\delta]=\frac{\det' \dbar_{1-\lambda_b}}{(\det \omega_I\omega_J(y_r))\,Z^{5}}\frac{\left(\det \dbar_{1-\lambda_\psi;[\delta]}\right)^5}{\det' \dbar_{1-\lambda_\beta;[\delta]}} = \frac{\widehat{Z}_2\,(\widehat{Z}_{1/2}[\delta])^5}{(\det \omega_I\omega_J(y_r))\,Z^{15/2}\,\widehat{Z}_{3/2}[\delta]}\,.
 \label{equ:Zchidets}
\end{equation}
Inserting explicitly \cref{equ:Verlinde}, the full chiral partition function is thus
\begin{equation}\label{equ:Zchi_unsimpl}
 \mathcal{Z}^{\text{chi}}[\delta]=\frac{\vartheta[\delta](0)^5\,\vartheta(D_b)\,\prod_{r<s}E(y_r,y_s)\prod_r \sigma(y_r)^3}{Z^{15/2}\,\vartheta[\delta](D_\beta)\,E(x_1,x_2)\,\prod_\alpha\sigma(x_\alpha)\,\det \omega_I\omega_J(y_r)}\,,
\end{equation}
where we have abbreviated the ghost divisors for readability,
\begin{equation}
 D_b=\sum_{r=1}^3 y_r-3\Delta\,,\qquad\qquad D_\beta=\sum_{\alpha=1}^2 x_\alpha-2\Delta\,,
\end{equation}
and $\Delta$ is the vector of Riemann constants \eqref{eq:Riemannvector}. We reiterate here that the final amplitude is independent of the choice of $x_\alpha$ and $y_r$, although this is not manifest at this stage. In fact, $\mathcal{Z}^{\text{chi}}[\delta]$ by itself has to be independent of $y_r$ since the rest of the partition function is manifestly independent of these punctures. We will see this explicitly below.

\paragraph{Simplifying the chiral partition function.} For dealing with the chiral partition function, we will rely on the simplifications achieved in \cite{DHoker:2001jaf,DHoker:2005vch} for conventional superstring theory, where the same object appears. In \cite{DHoker:2001jaf}, it was shown that the chiral partition function \eqref{equ:Zchi_unsimpl} can be written as \begin{equation}\label{equ:Zchi_simpl}
 \mathcal{Z}^{\text{chi}}[\delta]=\frac{\mathcal{Z}_0 \,E(x_1,x_2)\,\vartheta[\delta](0)^5}{\vartheta[\delta](D_\beta)}\,.
\end{equation}
Here, $\mathcal{Z}_0$ is a $(-1,0)$ form in both $x_1$ and $x_2$, and is proportional to the bosonic string partition function $\mathcal{Z_B}$,
\begin{equation}
 \mathcal{Z}_0=\frac{\mathcal{Z_B} \,Z^{6}}{E(x_1,x_2)^2\,\sigma(x_1)^2\sigma(x_2)^2}\,,\qquad\qquad \mathcal{Z_B}=\frac{1}{\pi^{12}\Psi_{10}}\,,\label{equ:Z0}
\end{equation}
where $\Psi_{10} \equiv \prod_\delta \vartheta[\delta](0)^2$ is a modular form of weight 10. While \cref{equ:Zchi_simpl} still depends on $x_\alpha$ (as it must -- after all, the remaining correlator in \cref{equ:Z_split} depends on $x_\alpha$ as well), the above formula is indeed manifestly independent of $y_r$, as advertised above.

Following \cite{DHoker:2005vch}, and in view of the calculations involved in simplifying the scattering amplitude, it will moreover be useful to make a special choice for the two marked points $x_\alpha$: we will take them to be the zeros of a holomorphic $(1,0)$ form $\varpi$,
\begin{equation}\label{equ:varpi}
 \varpi(z) \equiv \omega_I(z)\partial_I\vartheta(x_1-\Delta)e^{2i\pi\kappa'\cdot(x_1-\Delta)} = -\,\omega_I(z)\partial_I\vartheta(x_2-\Delta)e^{2i\pi\kappa'\cdot(x_2-\Delta)} \,.
\end{equation}
As before, $\Delta$ is the vector of Riemann constants \eqref{eq:Riemannvector}, and $2\kappa$ is an arbitrary full period, i.e., $2\kappa\in {\mathbb Z^2}\oplus \Omega {\mathbb Z^2}$. In fact, the condition that the marked points $x_\alpha$ are the zeroes of a holomorphic differential is, in terms of the Abel map, that $x_1+x_2-2\Delta=2\kappa$. Notice that, at genus two, a holomorphic differential is defined up to a constant multiple by the location of its two zeros, and the normalisation used here is chosen for convenience. With this choice for the marked points $x_\alpha$, the chiral partition function can be simplified to
\begin{equation}
 \mathcal{Z}^{\text{chi}}[\delta]=\mathcal{Z}_0 \,E(x_1,x_2)\, e^{4i\pi \kappa'\Omega\kappa'} \langle \kappa |\delta\rangle\,\vartheta[\delta](0)^4\,.\label{equ:Zchi_kappa}
\end{equation}
Using these parameters, requiring the amplitude to be independent of the choice of $x_\alpha$ is equivalent to requiring independence of $x_1$ and $\kappa$. We will see in the calculation of the four-point amplitude how $\varpi$ simplifies various calculations; see e.g. \cref{sec:identities}.

\paragraph{The GSO projection and the cosmological constant:}
To conclude the derivation of the partition function of the type II ambitwistor string, we still need to impose the GSO projection. As at genus one, the requirement of modular invariance fixes all relative phases among even spin structures. Indeed, ref.~\cite{DHoker:2001jaf} proved that all relative phases in the GSO projection at genus two are unique, and equal. This amounts to
\begin{equation}\label{equ:GSO}
\eta_\delta=\eta_{\tilde\delta}=1\,.
\end{equation}
We will derive these phases explicitly for the ambitwistor string in \cref{sec:mod-inv}. 
The vacuum amplitude can then be evaluated easily: the remaining correlator in \eqref{equ:Z_split} over the $\psi$ system leads to a factor of $S_\delta(x_1,x_2)P(x_1)\cdot P(x_2)$, and the result for $\tilde\psi$ is analogous (with $\tilde\delta$). The GSO sum in \eqref{equ:Z_split} therefore vanishes as a consequence of the identity
\begin{equation}
 \sum_\delta \mathcal{Z}^{\text{chi}}[\delta]\,S_\delta(x_1,x_2)=0\,,
\end{equation}
which is one of the identities proven in \cite{DHoker:2005vch} and listed in our \cref{equ:I1}. Thus, the cosmological constant vanishes in the ambitwistor string.  A very similar argument, and the corresponding vanishing identities \cref{equ:I2} through \cref{equ:I7}, imply that all $n$-point amplitudes with $n<4$ vanish as well.

\subsection{The amplitude}
We have obtained the following expression for the amplitude:
\begin{equation}
 \mathcal{M}_n=\delta\Big(\sum_i k_i\Big)\int
 \d^{20}\ell\,\d^3\Omega\prod_{I\leq J}\bar\delta\big(u^{IJ}\big)\,\sum_{\delta, \tilde\delta} \mathcal{Z}^{\text{chi}}[\delta]\,{\mathcal{Z}}^{\text{chi}}[\tilde\delta]\left\langle \prod_{\alpha=1}^2 P\cdot\psi(x_\alpha)\,P\cdot\tilde\psi(x_\alpha)\,\prod_{i=1}^n \mathcal{V}_i \right\rangle_{\delta,\tilde\delta} \,
\end{equation}
where we have already included the GSO projection, and condensed the partition functions of various fields into the chiral partition functions $\mathcal{Z}^{\text{chi}}[\delta]$ and $\mathcal{Z}^{\text{chi}}[\tilde\delta]$. For readability, let us furthermore introduce the loop integrand $\mathfrak{I}$, such as
\begin{equation}\label{equ:amplitude_g=2}
 \mathcal{M}_n=\delta\Big(\sum_{i=1}^n k_i\Big)\,\int \d^{10}\ell_1\,\d^{10}\ell_2 \,\,\mathfrak{I}_n\,.
\end{equation}
Just as we have observed for the partition function, the free ambitwistor string action guarantees that the remaining correlator in the loop integrand $\mathfrak{I}$ splits into tilded and untilded systems. We can therefore evaluate each contribution independently,
\begin{equation}\label{equ:integrand_g=2}
 \mathfrak{I}_n=\int
 \d^3\Omega\,\prod_{I\leq J}\bar\delta\big(u^{IJ}\big)\,\prod_{i=1}^n\bar\delta\big(\big\langle \mu_i\,P^2\big\rangle\big)\,\,\mathcal{I}^{\text{chi}}_n\,\widetilde{\mathcal{I}}^{\text{chi}}_n\,,
\end{equation}
where we defined the `chiral' (untilded) integrand by
\begin{equation}\label{equ:int_n-point}
  \mathcal{I}^{\text{chi}}_n=\sum_{\delta}  \mathcal{Z}^{\text{chi}}[\delta]\,\left\langle \prod_{\alpha=1}^2 P\cdot\psi(x_\alpha)\,\prod_{i=1}^n (\epsilon_i\cdot P+k_i\cdot \psi \,\epsilon\cdot\psi)(z_i)  \right\rangle^{\text{nzms}}_{\delta}
\end{equation}
and similarly for the tilded integrand with $\tilde\delta$. We can evaluate the correlator to find
\begin{equation}\label{equ:intchi}
 \mathcal{I}^{\text{chi}}_n=\sum_{\delta}  \mathcal{Z}^{\text{chi}}[\delta]\,\pf\big(M_\delta^{(2)}\big)\,,
\end{equation}
where the $(2n+2)\times(2n+2)$ matrix $M^{(2)}_\delta$ is given by
\begin{subequations}
\begin{align}
 & &&M^{(2)}_\delta=\begin{pmatrix}A &-C^T\\C&B\end{pmatrix}\,,&&\\
 &A_{x_1x_2}=P(x_1)\cdot P(x_2) S_\delta(x_1,x_2)\,,&& A_{x_\alpha,j}=P(x_\alpha)\cdot k_j S_\delta(x_\alpha,z_j)\,,&& A_{ij}=k_i\cdot k_j S_\delta(z_i,z_j)\,,\\
 & && C_{x_\alpha,j}=P(x_\alpha)\cdot \epsilon_j S_\delta(x_\alpha,z_j)\,,&& C_{ij}=\epsilon_i\cdot k_j S_\delta(z_i,z_j)\,,\\
 & && C_{ii}\hspace{9.5pt}= P(z_i)\cdot\epsilon_i\,, && B_{ij}=\epsilon_i\cdot\epsilon_j  S_\delta(z_i,z_j)\,.
\end{align}
\end{subequations}

The formula for the genus-two $n$-point scattering amplitude presented here is the main result of \cref{sec:g=2}. In the remainder of this section, we will discuss the simplifications occurring for $n=4$, in \cref{sec:4pt}, and then we will show explicitly that the $n$-point formula satisfies the stringent constraint of modular invariance, in \cref{sec:mod-inv}.

As we have already mentioned, the genus-two scattering equations, which in principle localise the integration over the moduli space of the Riemann surface in the formula above, are too hard to solve explicitly. We cannot, therefore, evaluate directly the genus-two formula given in this section. We will show in \cref{sec:nodalRS} how to turn it into a much simpler formula on the Riemann sphere.

\subsection{The four-particle amplitude}\label{sec:4pt}

The formula for the genus-two $n$-point scattering amplitude presented above includes a sum over spin structures. This sum builds up the contributions of particle states running in the loops, both bosons and fermions, as we discussed in \eqref{eq:R-NS}. Since we are dealing with a supersymmetric theory, we expect that this sum provides a significant simplification. Indeed, this is what happens in the conventional RNS superstring.

Fortunately, for our purposes, we can rely on the superstring work \cite{DHoker:2005vch}, where the important identities for the four-particle amplitude were proven. These identities for sums over spin structures involving the chiral partition functions and the Szeg\H{o} kernels are listed in the \cref{sec:identities} for convenience. As a consequence of the identities, the only non-vanishing contributions to the four-point amplitude come from terms where the two picture changing operators $P\cdot\psi(x_1)$ and $P\cdot\psi(x_2)$ are contracted, leading to a factor of $P(x_1)\cdot P(x_2)\,S_\delta(x_1,x_2)$. The identities actually allow for an even stronger conclusion: since all terms with less than five Szeg\H{o} kernels vanish, the diagonal terms $C_{ii}=\epsilon_i\cdot P(z_i)$ do {\it not} contribute at four points. This leaves us with only two contraction cycles that both evaluate to the same permutation-invariant result \eqref{equ:Idres},
\begin{equation}
 I_{11}=I_{12}=-2\mathcal{Z}_0\prod_{i=1}^4\varpi(z_i)\,.
\end{equation}
Since all  worldsheet contractions contribute the same factor, we can  extract a kinematic prefactor $\mathcal{K}$ from each Pfaffian Pf$\big(M_\delta^{(2)}\big)$, given by 
\begin{align}
 \mathcal{K}=\tr\big(F_1F_2\big)&\tr\big(F_3F_4\big)+\tr\big(F_1F_3\big)\tr\big(F_2F_4\big)+\tr\big(F_1F_4\big)\tr\big(F_2F_3\big)\\
 &-4\,\tr\big(F_1F_2F_3F_4\big)-4\,\tr\big(F_1F_3F_2F_4\big)-4\,\tr\big(F_1F_2F_4F_3\big)\,,\nonumber
\end{align}
where $F_i^{\mu\nu} = k_i^{[\mu}\epsilon_i^{\nu]}$, and similarly $\widetilde{\mathcal{K}} = \mathcal{K}(\epsilon\rightarrow\tilde\epsilon)$ for the tilded system. Using this, the four-point loop integrand becomes
\begin{equation}\label{equ:4ptintegrand_not_simpl}
 \mathfrak{I}_4=\,\mathcal{K\tilde{K}}\int \prod_{I\leq J}\d\Omega_{IJ}\,\prod_{I\leq J}\bar\delta\big(u^{IJ}\big)\,\prod_{i=1}^n\bar\delta\big(\big\langle \mu_i\,P^2\big\rangle\big)\,\left(2 P(x_1)\cdot P(x_2)\,\, \mathcal{Z}_0\prod_{i=1}^4\varpi(z_i) \right)^2\,.
\end{equation}

We can further simplify this loop integrand by adding a judicious choice of terms that vanish on the support of the scattering equations. Notice that  this gives us considerable freedom in the representation of the integrand: since the Beltrami differentials $\{\mu_r\}_{r=1}^{n+3}$ form a basis of $H^{0,1}(\Sigma, T_\Sigma(-z_1-...-z_n))$, we are free to add any term containing
\begin{equation}\label{equ:anybelt}
 \big\langle \mu \,P^2\big\rangle =0\,,
\end{equation}
where  $\mu\in H^{0,1}(\Sigma, T_\Sigma(-z_1-...-z_n))$ is a linear combination of the Beltrami differentials used in the gauge fixing. Taking inspiration from the superstring \cite{DHoker:2005vch}, a particularly convenient choice for this differential is
\begin{equation}
 \mu_x(z)  =\frac{1}{2}\left(\frac{c_1}{c_2}\delta(z, x_1)+\frac{c_2}{c_1}\delta(z, x_2)\right)\,.\label{equ:mux}
\end{equation}
The factors $c_\alpha$ in this definition are given by $\varpi(z)=c_1 \Delta(x_1,z)=c_2 \Delta(x_2,z)$, where we used the (standard, though unfortunate) notation
\begin{equation}
 \Delta(z_i,z_j)=\Delta_{ij}=\epsilon^{IJ}\omega_I(z_i)\omega_J(z_j)\,,
\end{equation}
and we stress that $\Delta_{ij}$ is unrelated to the vector of Riemann constants $\Delta$.

In the RNS string at genus two \cite{DHoker:2005vch}, the Beltrami differential $\mu_x$ plays a role in proving that the amplitude is independent of the PCO gauge slice, i.e., the choice of $\chi_\alpha$ and $\tilde\chi_\alpha$ which determines the marked points $x_\alpha$. While the details of the calculations are quite different in the ambitwistor string -- where $\mu_x$ is associated to terms that vanish on the support of the scattering equations  --  we will see below that $\mu_x$  effectively leads to similar simplifications of the amplitude. Many useful identities involving the factors $c_\alpha$ have been derived in \cite{DHoker:2005vch}, and we have listed the relevant equations in  \cref{sec:other_ids} for convenience. In particular, let us highlight the relations to the holomorphic differentials and the partition function,
\begin{subequations}
 \begin{align}
  c_1 \omega_I(x_1)&=c_2 \omega_I(x_2)\,,\qquad I=1,2\,,\label{cw1=cw2}\\
  \mathcal{Z}_0 c_1c_2 \partial\varpi&(x_1)\partial\varpi(x_2)=1\label{equ:Zcc=1}\,,\\
  c_1\omega_{i*}(x_1)-c_2\omega_{i*}(x_2) &= -c_1^2\partial\varpi(x_1)\frac{\Delta_{i*}}{\varpi(z_i)\varpi(z_*)}\,,\label{equ:c1c2Delta}
 \end{align}
\end{subequations}
where $z_\ast$ is an arbitrary marked point. Moreover, we find that  $\mu_x$ obeys
\begin{subequations}
\begin{align}
 2\big\langle \mu_x\,\omega_I\omega_J\big\rangle &= \frac{c_1}{c_2}\omega_I(x_1)\omega_J(x_1)+\frac{c_2}{c_1}\omega_I(x_2)\omega_J(x_2)=\omega_I(x_1)\omega_J(x_2)+\omega_I(x_2)\omega_J(x_1)\,,\\
 2\big\langle \mu_x\,\omega_I\omega_{i*}\big\rangle &= \omega_I(x_1)\omega_{i*}(x_2)+\omega_I(x_2)\omega_{i*}(x_1)\,,
\end{align}
\end{subequations}
where we made use of the identity \eqref{cw1=cw2}. This implies that, on the support of the scattering equations, we are free to add the following to the loop integrand:
\begin{subequations}
\begin{align}\label{equ:mux_identity}
 0=\big\langle \mu_x \,P^2\big\rangle &= \ell_1^2\omega_1(x_1)\omega_1(x_2)+\ell_2^2\omega_2(x_1)\omega_2(x_2)+\ell_1\cdot\ell_2\,\big(\omega_1(x_1)\omega_2(x_2)+\omega_1(x_2)\omega_2(x_1)\big)\nonumber\\
 &\qquad +\sum_{I,i}\ell^I\cdot k_i\,\big(\omega_I(x_1)\omega_{i*}(x_2)+\omega_I(x_2)\omega_{i*}(x_1)\big)\\
 &\qquad +\frac{1}{2}\sum_{i,j}k_i\cdot k_j\,\left(\frac{c_1}{c_2}\omega_{i*}(x_1)\omega_{j*}(x_2)+\frac{c_2}{c_1}\omega_{i*}(x_2)\omega_{j*}(x_2)\right)\,.\nonumber
\end{align}
\end{subequations}
We can now define a quantity that agrees with $P(x_1)\cdot P(x_2)$ on the support of the scattering equations,
\begin{equation}\label{equ:Px1Px2}
 \px(x_1,x_2):=P(x_1)\cdot P(x_2)-\big\langle \mu_x \,P^2\big\rangle\,,
\end{equation}
with the very convenient property that all dependence on the zero modes $\ell^{I}_\mu$ of $P_\mu$ has been eliminated. We can simplify $\px(x_1,x_2)$ by using the  definition of $c_{1,2}$ as well as the identity \eqref{equ:c1c2Delta}  involving the  holomorphic one-form $\varpi(z)$,
\begin{align}\label{equ:Px1Px2_simpl}
 \px(x_1,x_2) 
 &=-\frac{1}{2}\sum_{i,j}\frac{k_i\cdot k_j}{c_1 c_2}\,\Big(c_1\omega_{i*}(x_1)-c_2\omega_{i*}(x_2)\Big)\Big(c_1\omega_{j*}(x_1)-c_2\omega_{j*}(x_2)\Big) \nonumber\\
 &=-\frac{1}{2}c_1c_2\partial\varpi(x_1)\partial\varpi(x_2)\sum_{i,j} k_i\cdot k_j\,\frac{\Delta_{i*}\Delta_{j*}}{\varpi(z_i)\varpi(z_j)\varpi(z_*)^2}\,.
\end{align}

Let us now revisit the  integrand  \eqref{equ:4ptintegrand_not_simpl} and simply substitute $P(x_1)\cdot P(x_2)$ by $\px(x_1,x_2)$, since these objects agree on the support of the scattering equations. The factor $c_1c_2\partial\varpi(x_1)\partial\varpi(x_2)$ from $\px(x_1,x_2)$ then combines with the partition function $\mathcal{Z}_0$ as in the identity \eqref{equ:Zcc=1}, and we can further choose the arbitrary marked point to coincide with one of the vertex operators,  $z_*=z_4$. This leads to the integrand 
\begin{subequations}\label{equ:4pt_genus2}
\begin{align}
 \mathfrak{I}_4&= \,\mathcal{K\tilde{K}}\int \d^3\Omega\,\prod_{I\leq J}\bar\delta\big(u^{IJ}\big)\,\prod_{i=1}^n\bar\delta\big(\big\langle \mu_i\,P^2\big\rangle\big)\,\mathcal{Y}^2\\
 \text{where} \qquad \mathcal{Y}&=\sum_{i,j} k_i\cdot k_j\,\frac{\Delta_{i4}\Delta_{j4}}{\varpi(z_i)\varpi(z_j)\varpi(z_4)^2}\prod_{k=1}^4\varpi(z_k)\,.
\end{align}
\end{subequations}
As a last step, we can mirror again the superstring calculation  \cite{DHoker:2005vch} and simplify $\mathcal{Y}$  by using the Jacobi-like relations
\begin{equation}
 \varpi(z_b)\Delta_{ac}-\varpi(z_a)\Delta_{bc}=\varpi(z_c)\Delta_{ab}\,.
\end{equation}
Repeated application of these identities leads to the following compact expression for the integrand:
\begin{equation}\label{equ:4ptintegrand-final}
 \mathcal{Y}=s\Delta_{14}\Delta_{23}-t\Delta_{12}\Delta_{34}\,,
\end{equation}
which is manifestly independent of the marked points $x_\alpha$ and the associated holomorphic differential $\varpi$.

This concludes the derivation of the four-point amplitude from the RNS ambitwistor string. Luckily, it has revealed some manipulations that will be useful later on, namely the introduction of the object $\px(x_1,x_2)$ in substitution of $P(x_1)\cdot P(x_2)$, which is valid on the support of the genus-two scattering equations.

The type II supergravity four-point amplitude was previously derived from the pure spinor ambitwistor string in \cite{Adamo:2015hoa}, following earlier results from the (non-minimal) pure spinor superstring \cite{Berkovits:2006bk,Mafra:2008ar,Gomez:2010ad}. As such, it formed the basis of preliminary work on using global residue theorems to localise genus-two supergravity amplitudes on bi-nodal Riemann spheres \cite{Geyer:2016wjx}, where the four-point loop integrand was also matched to a known for of the integrand, thereby checking its validity. While \cite{Geyer:2016wjx} relied on factorisation arguments to account for certain factors, we give a full derivation of the global residue theorems and the resulting $n$-point amplitudes on the bi-nodal Riemann sphere in \cref{sec:nodalRS} and \cref{sec:contour_argument}.

\subsection{Modular and gauge invariance}\label{sec:mod-inv}
In this section, we discuss two essential checks on the amplitude: independence of the PCO gauge slice $\chi_\alpha$ and $\tilde\chi_\alpha$ (i.e., of the marked points $x_\alpha$), and modular invariance. Both will play a crucial role for the residue theorem applied  in the next section, where only a judicious choice of representation of the integrand and scattering equations will lead to a localisation on the maximal non-separating degeneration.

\subsubsection{Independence of $x_\alpha$}
The amplitude -- and therefore  the chiral integrand $\mathcal{I}^{\text{chi}}_n$ -- must be independent of the PCO gauge slice defined by the marked points  $x_\alpha$. This constitutes an important check of our results and is easily proven  using Liouville's theorem. In the following, we verify the absence of poles in $x_\alpha$ on the support of the scattering equations, and hence that the integrand is bounded. Liouville's theorem then guarantees that  the chiral integrand $\mathcal{I}^{\text{chi}}_n$ is constant in $x_\alpha$.\footnote{Notice that the chiral integrand $\mathcal{I}^{\text{chi}}_n$ is a function of  $x_\alpha$, i.e., it has form degree zero in $\d x_\alpha$. This can be checked from the definitions of the ingredients in \eqref{equ:intchi}. The amplitude would not be well defined otherwise.}

By inspection of \eqref{equ:int_n-point}, it is evident that there are only two types of potential poles involving $x_\alpha$: 
\begin{enumerate}[(A)]
 \item when the insertions of the two PCOs coincide, $x_1- x_2= \varepsilon$,
 \item when one of the PCOs collides with a vertex operator, $x_\alpha -z_i =\varepsilon$. \footnote{Throughout this section, we use $\varepsilon$ to denote a small separation of marked points on the Riemann surface.}
\end{enumerate}

\paragraph{Case A.} Let us first consider the coefficients of poles in  $x_1- x_2= \varepsilon$. The only sources of these poles are the measure $\mathcal{Z}^{\text{chi}}[\delta]\sim\varepsilon^{-1}$ and the component $A_{x_1\,x_2}=\varepsilon^{-1}(P(x_1)^2+\mathcal{O}(\varepsilon))$ of the Pfaffian.  At order $\mathcal{O}(\varepsilon^{-2})$, the integrand therefore contains only terms proportional to $P^2(x_1)$, which vanish on  the support of the scattering equations. The subleading term at order $\mathcal{O}(\varepsilon^{-1})$ is also trivial, since the rows and columns corresponding to $x_1$ and $x_2$ are identical,
\begin{equation}
 M^{(2)}_{x_1,a}\Big|_{\varepsilon^{-1}}=M^{(2)}_{x_2,a}\Big|_{\varepsilon^{-1}}\,\qquad \text{for any } a\neq x_1,x_2\,,
\end{equation}
and therefore the matrix $M^{(2)}\big|_{\varepsilon^{-1}}$ is degenerate.

\paragraph{Case B.} Consider now poles in  $x_\alpha -z_i =\varepsilon$, from one of the PCOs contracting with an integrated vertex operator. The partition functions do not contribute to this pole, and the leading term at  $\mathcal{O}(\varepsilon^{-2})$ originating from the $A_{i,x_\alpha}$ vanishes trivially due to $k_i^2=0$. The subleading term $\mathcal{O}(\varepsilon^{-1})$ is again given by $\pf(M^{(2)})\big|_{\varepsilon^{-1}}$ with
\begin{subequations}
\begin{align}
 &A_{x_\beta\,x_\alpha}=A_{x_\beta\,i}, &&A_{i,x_\alpha}=\varepsilon^{-1}P(z_i)\cdot k_i, && A_{j,x_\alpha}=A_{ji},\\
 & && C_{i,x_\alpha}=C_{ii}, && C_{j,x_\alpha}=C_{ji}\,,
\end{align}
\end{subequations}
to leading order  $\mathcal{O}(\varepsilon^{-1})$ . Similarly to the case A  above, the matrix $M^{(2)}\big|_{\varepsilon^{-1}}$ becomes degenerate, with identical  rows and columns for $x_\alpha$ and $i$,
\begin{equation}
 M^{(2)}_{a\, x_\alpha}\Big|_{\varepsilon^{-1}}=M^{(2)}_{a\,i}\Big|_{\varepsilon^{-1}}\,\qquad \text{for any } a\,,
\end{equation}
and so  the coefficient of the potential pole vanishes.\\

Using Liouville's theorem, the chiral integrand $\mathcal{I}^{\text{chi}}_n$ is thus independent of the choice of the insertion points $x_\alpha$ of the picture changing operators. \\

Looking ahead to the degeneration to the nodal Riemann sphere in \cref{sec:nodalRS}, it is worth highlighting a fundamental difference between case A and case B. While the coefficient of $(x_\alpha-z_i)^{-1}$ vanishes on the support of the vertex scattering equations $\langle \mu_i P^2\rangle=0$ alone (even off the support of the moduli scattering equations, $u^{IJ}=0$), the absence of the pole in $x_1-x_2$ relies on the support of all scattering equations to guarantee that $P^2=0$. This distinction will play an important role when applying the global residue theorem to localise on the non-separating degeneration: applying the residue theorem relaxes two of the constraints  $u^{IJ}=0$, and therefore $P^2\neq 0$ on the resulting lower-genus Riemann surface.

\subsubsection{Modular invariance} \label{sec:mod_inv_sub}
The GSO projection plays a crucial role in the ambitwistor string by restricting to the correct degrees of freedom for type II supergravity. In the path integral formalism, the GSO projection is implemented via the sum over spin structures, and modular invariance restricts the relative phases in that sum, as asserted at the end of \cref{sec:partition-function}. After that section, we postponed a more detailed discussion in favour of calculating the full amplitude \eqref{equ:amplitude_g=2}. Here, we return to the question of modular invariance by deriving the phase factors explicitly, and provide a direct  proof of modular invariance for the $n$-point amplitude. We conclude with a discussion contrasting the modular invariance of the full amplitudes with the loop-momenta-fixed integrands considered in \cite{Ohmori:2015sha}, which explicitly break modular invariance.

Further details on the modular properties of theta functions and chiral partition functions can be found in refs.~\cite{fay1973theta, DHoker:1988ta, Verlinde:1986kw, DHoker:2001jaf}.

\paragraph{The basics.} 
Recall from \cref{sec:rev_basics} that the modular group Sp$(4,\mathbb{Z})$ acts on the period matrix $\Omega_{IJ}$ and on the holomorphic Abelian differentials $\omega_I$ as
\begin{equation}\label{equ:mod_transform_omega}
 \Omega\rightarrow\widetilde\Omega= \big(a\Omega +b\big)\,\big(c\Omega+d\big)^{-1}\,,\qquad\qquad \omega\rightarrow\widetilde\omega=\omega\big(c\Omega+d\big)^{-1}\,.
\end{equation}
 For the integrand and the scattering equations to be well-defined, the  one-form $P_\mu$ given by \eqref{equ:ell} must transform homogeneously under modular transformations.  We will see soon that the meromorphic differentials in $P_\mu$ are invariant under modular transformations, and therefore the holomorphic part, $\ell^I_\mu\omega_I$, must also be invariant. This implies in turn that  the loop momenta $\ell^I_\mu$ must absorb the transformation of the Abelian differentials $\omega_I$:
\begin{equation}
 \ell\rightarrow\widetilde\ell=\big(c\Omega+d\big)\ell\,.\label{equ:mod_transform_ell}
\end{equation}
The integration over the loop momenta therefore has modular weight $+10$, and the integration over the modular parameters $\Omega$ and the localisation on the scattering equations have weight $-3$ each,
\begin{subequations}
\begin{align}
 &\prod_{I\leq J}\d\Omega_{IJ} &&\rightarrow &&\prod_{I\leq J}\d\widetilde\Omega_{IJ} = \det\big(c\Omega+d\big)^{-3} \prod_{I\leq J}\d\Omega_{IJ}\,, \\
 &\prod_{I\leq J}\bar\delta\big(u^{IJ}\big)&&\rightarrow &&\prod_{I\leq J}\bar\delta\big(\widetilde{u}^{IJ}\big)= \det\big(c\Omega+d\big)^{-3}\prod_{I\leq J}\bar\delta\big(u^{IJ}\big)\,, \\
&\d^{20}\ell &&\rightarrow && \d^{20}\widetilde{\ell}=\det(c\Omega+d)^{10}\d^{20}\ell \,.\label{eq:mod_transform_ell_measure}
\end{align}
\end{subequations}
For fields of half integer weight, the transformation of the partition function further depends on the  action of the modular group on the spin structures. Following Ref.~\cite{DHoker:1988ta, DHoker:2001jaf},  this is most conveniently expressed when the spin structures are assembled into a single column vector, 
\begin{align}
 \begin{pmatrix}\kappa' \\ \kappa''\end{pmatrix}&\rightarrow \begin{pmatrix}\widetilde\kappa' \\ \widetilde\kappa''\end{pmatrix} =\underbrace{\begin{pmatrix}d & -c \\ -b & a\end{pmatrix} \begin{pmatrix}\kappa' \\ \kappa''\end{pmatrix}}+ \underbrace{\frac{1}{2}\begin{pmatrix}\text{diag}\big( cd^T\big) \\\text{diag}\big(ab^T\big)\end{pmatrix}}\,.\\
 \kappa \hspace{10pt}& \rightarrow\hspace{10pt} \widetilde\kappa\hspace{10pt} =\hspace{32pt}\hat\kappa\hspace{32pt} +\hspace{30pt}\alpha\nonumber
\end{align}
Here, diag$(m)$ denotes the column vector containing the diagonal  elements of the matrix $m$. The  theta functions then  transform as  \cite{fay1973theta, DHoker:1988ta, Verlinde:1986kw}
\begin{equation}
 \vartheta[\widetilde\kappa]\big(\widetilde\zeta,\widetilde\Omega\big) = \epsilon(M)\,e^{i\pi\varphi(\kappa)}\det\big(c\Omega+d\big)^{1/2}e^{i\pi\zeta^T(c\Omega+d)^{-1}c\,\zeta}\vartheta[\kappa]\big(\zeta,\Omega) \label{equ:mod_transform_theta}\,,
\end{equation}
with the argument $\widetilde\zeta$  of the theta function and the phase $\varphi$ defined by
\begin{equation}
 \widetilde\zeta= \left(\big(c\Omega+d\big)^{T}\right)^{-1} \zeta\,,\qquad\qquad\varphi(\kappa)=\hat\kappa'\cdot\hat\kappa'' - \kappa'\cdot\kappa''+2\hat\kappa'\cdot\alpha''\,.
\end{equation}
Moreover,  $\epsilon(M)=\epsilon(a,b,c,d)$ denotes a transformation-dependent phase factor satisfying $\epsilon(M)^8=1$, whose specific form is not important since it will cancel out in the chiral partition function, as we shall see below. For more details on $\epsilon(M)$, including detailed tables for the generators of the modular group, the interested reader is referred to \cite{DHoker:2001jaf,Verlinde:1986kw}.

\paragraph{Prime form, Szeg\H{o} kernels and the partition function.} Proceeding in analogy to \cref{sec:review}, we are now in a position to review the modular properties of the key objects relevant for CFTs on higher-genus Riemann surfaces: the propagators and the partition functions. In particular, the modular behaviour of the prime form follows directly from  \cref{equ:mod_transform_theta}, 
\begin{equation}\label{equ:mod_transform_prime}
 E(z,w)\rightarrow E(z,w)\,\,e^{i\pi\zeta^T(c\Omega+d)^{-1}c\,\zeta}\,,\qquad \text{where }\zeta = \int_w^z\omega\,.
\end{equation}
This property ensures  that the Abelian differentials of the third kind $\omega_{w_1,w_2}$ are invariant under the action of the modular group Sp$(2g,\mathbb{Z})$. As for the Szeg\H{o} kernels $S_\delta$, we have a relation among different (even) spin structures,
\begin{equation}
 S_\delta(z,w|\Omega)\rightarrow S_{\widetilde\delta}(z,w|\widetilde\Omega) = S_\delta(z,w|\Omega) \,.
\end{equation}
This extends immediately to the Pfaffians $\pf\big(M^{(2)}_\delta\big)$  after taking into account the invariance of $P_\mu$ discussed above,
\begin{equation}
\pf\big(M^{(2)}_\delta|\Omega\big)\rightarrow \pf\big(M^{(2)}_{\widetilde\delta}|\widetilde\Omega\big) = \pf\big(M^{(2)}_\delta|\Omega\big) \,.
\end{equation}
Since modular transformations interpolate between different even spin structures, the Pfaffians are {\it not} modular forms, despite having trivial modular weight. Only the full amplitude, when summed over spin structures with appropriate phase factors, will be modular invariant.

The action of the modular group on a chiral determinant $\det \dbar_{1-\lambda}$ associated to partition functions of chiral $bc$ and $\beta\gamma$ systems was derived in Ref.~\cite{Verlinde:1986kw},
\begin{subequations}\label{equ:mod_partition_functions}
\begin{align}
 \det \dbar_{1-\lambda}(\widetilde\Omega) &=  \epsilon(M)^{2/3}\det(c\Omega +d)^{-\lambda}\,\det\dbar_{1-\lambda}(\Omega) && \lambda\in\mathbb{Z}\,, \\
   \det \dbar_{1-\lambda;[\widetilde\kappa]} (\widetilde\Omega) & = \epsilon(M)^{2/3}e^{i\pi\widetilde\varphi(\kappa,\lambda)}\det(c\Omega +d)^{-\lambda}\, \det \dbar_{1-\lambda;[\kappa]}(\Omega) && \lambda\in\mathbb{Z}+\frac{1}{2}\,.
\end{align}
\end{subequations}
where the phase factor $\widetilde\varphi$ depending on the weight $\lambda$ is given by
\begin{equation}
 \widetilde\varphi(\kappa,\lambda) = \varphi(\kappa)+2(2\lambda-1)\big(\hat\kappa'\cdot\alpha''+\hat\kappa''\cdot\alpha'\big)\,.
\end{equation}
We can now assemble these ingredients to study the action of modular transformations on the ambitwistor chiral partition function  \eqref{equ:Zchidets},
\begin{equation}
 \mathcal{Z}^{\text{chi}}[\delta]=\frac{1}{\det\big(\omega_I\omega_J(y_r)\big)}\frac{(\det' \dbar_{1-2}) (\det\dbar_{1-1/2;[\delta]})^5}{(\det'\dbar_{1-3/2;[\delta]})(\det\dbar_{1-1})^5}
\end{equation}
From \cref{equ:mod_partition_functions}, we see that $\mathcal{Z}^{\text{chi}}$ has modular weight $2+5\cdot\frac{1}{2}-\frac{3}{2}-5\cdot 1=-2$.
The factor $\epsilon(M)$ cancels between the fermionic and the bosonic systems, and the remaining phase simplifies to $e^{i\pi\delta_\varphi}$ with
\begin{equation}
 i\pi\delta_\varphi =
 4i\pi\,\big(\hat\kappa'\cdot\hat\kappa'' - \kappa'\cdot\kappa''+\hat\kappa'\cdot\alpha''-\hat\kappa''\cdot\alpha'\big)\,.
\end{equation}
This is easily confirmed to be a multiple of $2i\pi$ for all generators of the modular group Sp$(4,\mathbb{Z})$. The full chiral partition function thus has modular weight $-2$ and {\it no} relative  sign factors between different spin structures,
\begin{equation}\label{equ:mod_partition_Z}
 \mathcal{Z}^{\text{chi}}[\widetilde\delta](\widetilde\Omega)=\det\big(c\Omega+d\big)^{-2} \mathcal{Z}^{\text{chi}}[\delta](\Omega)\,.
\end{equation}
It is worth highlighting again that the contribution $\mathcal{Z}^{\text{chi}}[\delta] \pf\big(M^{(2)}_\delta\big)$ to the chiral integrand for any single spin structure is not a modular form because modular transformations involve different spin structures. Since each contribution carries no modular weight, however, it is straighforward to construct a modular invariant object by summing over spin structures with appropriate coefficients to absorb any relative phases, leading to the chiral integrand $\sum_\delta \mathcal{Z}^{\text{chi}}[\delta] \pf\big(M^{(2)}_\delta\big)$. Since all phases in \cref{equ:mod_partition_Z} are trivial, this concludes our proof of $\eta_\delta=1$ for all even spin structures.

To summarise, the amplitude transforms under modular transformations as follows:
\begin{equation*}
 \mathcal{M}_n=\int\underbrace{\vphantom{\prod_{I\leq J}}\d^{10}\ell_1\,\d^{10}\ell_2}_{\textcolor{blue}{+10}} \,\underbrace{\prod_{I\leq J}\d\Omega_{IJ}}_{\textcolor{blue}{-3}} \underbrace{\prod_{I\leq J}\bar\delta\big(u^{IJ}\big)}_{\textcolor{blue}{-3}} \,\underbrace{\prod_{i=1}^n \bar\delta\big(\langle\mu_i P^2\rangle\big)}_{\textcolor{blue}{0}}\,\sum_{\delta,\tilde\delta}\underbrace{\vphantom{\prod_{I\leq J}}\mathcal{Z}^{\text{chi}}[\delta]}_{\textcolor{blue}{-2}} \underbrace{\vphantom{\prod_{I\leq J}}\mathcal{Z}^{\text{chi}}[\tilde\delta]}_{\textcolor{blue}{-2}}\,\underbrace{\vphantom{\prod_{I\leq J}}\pf\big(M^{(2)}_\delta\big) \pf\big(M^{(2)}_{\tilde\delta}\big)}_{\textcolor{blue}{0}}\,.
\end{equation*}
Modular weights of each factor are indicated in blue, and evidently sum to zero. All phase factors cancel as discussed above, and so the full expression is modular invariant.

\paragraph{Modular invariance for four particles.} For amplitudes with  four external particles, the analysis of modular invariance simpifies considerably. It is sufficient to observe that due to \cref{equ:mod_transform_omega},
\begin{subequations}
\begin{align}
 \qquad\qquad\qquad\qquad\qquad &\Delta_{ij} &&\rightarrow && \widetilde\Delta_{ij}=\det\big(C\Omega+D\big)^{-1}\Delta_{ij}\,,  \qquad\qquad\qquad\qquad\qquad\\
 \qquad\qquad\qquad\qquad\qquad &\mathcal{Y}&&\rightarrow &&\widetilde{\mathcal{Y}}^2=\det\big(C\Omega+D\big)^{-4}\mathcal{Y}^2\,.  \qquad\qquad\qquad\qquad\qquad
\end{align}
\end{subequations}
The chiral integrand is therefore a modular form of weight $-4$, and combines with the modular measure and the scattering equations to a modular form of weight $-10$, balancing the modular weight $+10$ from the loop integration:
\begin{equation}\label{eq:mod_inv_4pt}
 \mathcal{M}_4= \int \underbrace{\vphantom{\prod_{I\leq J}}\d^{10}\ell_1\,\d^{10}\ell_2}_{\textcolor{blue}{+10}} \,\underbrace{\prod_{I\leq J}\d\Omega_{IJ}}_{\textcolor{blue}{-3}} \underbrace{\prod_{I\leq J}\bar\delta\big(u^{IJ}\big)}_{\textcolor{blue}{-3}} \,\underbrace{\vphantom{\prod_{I\leq J}} \mathcal{Y}^4\,.}_{\textcolor{blue}{-4}}
\end{equation}

\paragraph{Modular transformation of the loop momenta.} 
The non-trivial transformation property \eqref{equ:mod_transform_ell} of the loop momenta $\ell_\mu^I$,
\begin{equation}\label{equ:mod_transform_ell_2}
 \ell^I\rightarrow\widetilde\ell^I={\big(c\Omega+d\big){}^I{_J}}\;\ell^J\,,
\end{equation}
plays a crucial role in the modular invariance of ambitwistor string amplitudes. It ensures that $P_\mu$ has homogeneous (vanishing) modular weight, and is consequently responsible for the nice transformation properties of the scattering equations,
\begin{subequations}
\begin{align}
 &\prod_{I\leq J}\bar\delta\big(u^{IJ}\big)&&\rightarrow &&\prod_{I\leq J}\bar\delta\big(\widetilde{u}^{IJ}\big)= \det\big(c\Omega+d\big)^{-3}\prod_{I\leq J}\bar\delta\big(u^{IJ}\big)\,,\\
 & \prod_{i=1}^n\bar\delta\left(\left\langle \mu_i P^2\right\rangle\right)&&\rightarrow &&\prod_{i=1}^n\bar\delta\left(\left\langle \mu_i P^2\right\rangle\right)\,.
\end{align}
\end{subequations}
The importance of \eqref{equ:mod_transform_ell_2} mirrors the situation at one loop \cite{Adamo:2013tsa}, where the transformation property $\ell_\mu\rightarrow (c\tau +d)\ell_\mu$ was essential in proving modular invariance. Our discussion above demonstrates that this feature  -- also observed in a different guise in the null string \cite{Casali:2017zkz} --  persists at higher loops. To summarise, the modular invariance of the amplitude relies on the modular transformation properties \eqref{equ:mod_transform_ell_2} of the loop momenta.

To highlight this point, let us contrast the above results with the loop-momenta-fixed integrands considered in \cite{Ohmori:2015sha}. The loop-momenta-fixed integrands $ \hat{\mathfrak{I}}_n^{\text{fixed}}$ are defined by 
inserting a set of delta-functions into the correlator to localise  the loop integration,
\begin{equation}
 \hat{\mathfrak{I}}_n^{\text{fixed}}(\hat\ell)=\delta\Big(\sum_{i=1}^n k_i\Big)\,\int \d^{10}\ell_1\,\d^{10}\ell_2\; \prod_{\mu,I} \delta\left(\hat\ell_\mu^I-\oint_{A_I}P_\mu\right)\,\,\mathfrak{I}_n\,.
\end{equation}
Here, we distinguish between the zero-mode coefficients $\ell_\mu^I$ of $P_\mu$, which transform under \eqref{equ:mod_transform_ell_2} so that $P_\mu$ is invariant, and the loop momenta $\hat\ell_\mu^I$, which do \emph{not} transform under the modular group. If we choose to work with the loop-momenta-fixed integrands, then the delta-functions integrand explicitly break modular invariance; notice that the cycles $A_I$ transform. Equivalently, this can also be observed in the scattering equations: since  $\hat\ell_\mu^I$ do not transform under modular transformations, different terms in the scattering equations transform with different modular weights, and thus break the modular invariance of the integrand. 

Of course, these two approaches of understanding the amplitude are compatible. If we take $\hat\ell_\mu^I$ to transform as  \eqref{equ:mod_transform_ell_2} under the action of the modular group (instead of considering fixed loop momenta), then $  \hat{\mathfrak{I}}_n^{\text{fixed}}(\hat\ell)$ is a modular form of weight $-10$.  The fixed integrand $  \hat{\mathfrak{I}}_n^{\text{fixed}}(\hat\ell)$ can then be integrated against a measure $\d^{20}\hat\ell$ to recover the modular invariant amplitude
\begin{equation}
 \mathcal{M}_n = \int \d^{20}\hat\ell \,\,  \hat{\mathfrak{I}}_n^{\text{fixed}}(\hat\ell)\,.
\end{equation}

\paragraph{Modular  invariance vs finiteness.} To conclude, let us briefly comment on an important aspect of modular invariance in the ambitwistor string. In contrast to standard string theory, modular invariance does not restrict the ambitwistor string correlators to a compact integration domain. This is due to the very distinct relation between the loop momenta (which can also be introduced in conventional string theory) and the modular parameters imposed by the scattering equations. Importantly, this means that the amplitudes are not expected to be finite, but  contain in fact the ultraviolet divergence of the loop integration expected for maximal supergravity in ten dimensions. In our work, we only deal with the loop integrand, for which finiteness is not an issue.

%%%%%%%%%%%%%%%%%%%%%%%%%%%%%%
%%%%%%%%%%%%%%%%%%%%%%%%%%%%%%
\section{From genus two to the nodal Riemann sphere}\label{sec:nodalRS}
As a correlator in the genus expansion of the ambitwistor string, the supergravity amplitude \eqref{equ:integrand_g=2} is the natural generalisation of lower loop orders.
This higher-genus representation has many desirable aspects: it manifests both modular invariance and the localisation on the scattering equations, and it makes a wide array of string theory techniques available due to the close similarity of the amplitudes.  The underlying mathematical structure, however, becomes increasingly challenging at higher genus, and obscures the relation to known rational field theory integrands.  While these are expected consequences of working with a `stringy' representation, they raise the question of how a manifestly rational loop integrand can possibly appear from the higher-genus formalism of the ambitwistor string --  especially in the absence of the string parameter $\alpha'$ governing the degeneration of the string moduli space in the field theory limit.

At one loop, a resolution of this problem was offered in \cite{Geyer:2015bja,Geyer:2015jch}. The ambitwistor string amplitudes can be localised on the non-separating boundary divisor,  rather than the higher-genus scattering equations, via the residue theorem in the moduli space $\mathfrak{M}_{1,n}$. This residue theorem moves the integration contour from a pole defined by the scattering equations to the only other simple pole in the integrand, the boundary divisor $\mathfrak{D}_{1,n}^{\text{non-sep}}\cong\widehat{\mathfrak{M}}_{0,n+2}$. The resulting integrand -- localised on the non-separating boundary of the moduli space -- is naturally formulated over  a nodal Riemann sphere, with the loop momentum running through the node. This representation of the amplitude has the advantage of reducing the computationally challenging ambitwistor higher-genus expressions to simple formulae on nodal Riemann spheres that are manifestly rational, and thus easier to match to known field theory integrands. Moreover, integrands are known not only for supergravity, but also for super-Yang-Mills theory \cite{Geyer:2015bja}, bi-adjoint scalar theory \cite{He:2015yua,Cachazo:2015aol}, pure Yang-Mills theory and gravity \cite{Geyer:2015jch}, are valid in any dimension $d$ and can be obtained directly from the nodal Riemann sphere \cite{Roehrig:2017gbt}.

In this section, we extend this argument to genus two. We have seen above that the two crucial properties necessary for applying a residue theorem  -- modular invariance and localisation on the scattering equations -- persist at two loops. Our goal is therefore  to localise the amplitudes  on the maximal non-separating boundary divisor  $\mathfrak{D}_{2,n}^{\text{max}}\cong\widehat{\mathfrak{M}}_{0,n+4}$ by using the residue theorem in the  moduli space $\widehat{\mathfrak{M}}'_{2,n}$, reducing the genus-two surface to a bi-nodal Riemann sphere as proposed in \cite{Geyer:2016wjx}. While higher-genus residue theorems are in general subtle to implement, the degeneration can be achieved iteratively by two uses of the residue theorem, each collapsing a single $A$-cycle. The first step is to move the integration contour to the boundary divisor corresponding to a nodal torus $\mathfrak{D}_{2,n}^{\text{non-sep}}\cong\widehat{\mathfrak{M}}_{1,n+2}$, while the second step localises the amplitude on  the bi-nodal Riemann sphere $\mathfrak{D}_{1,n+2}^{\text{non-sep}}\cong\widehat{\mathfrak{M}}_{0,n+4}$. 

\begin{figure}[ht]
	\centering 
	\begin{subfigure}[t]{0.29\textwidth} 
        \centering
        \includegraphics[width=\textwidth]{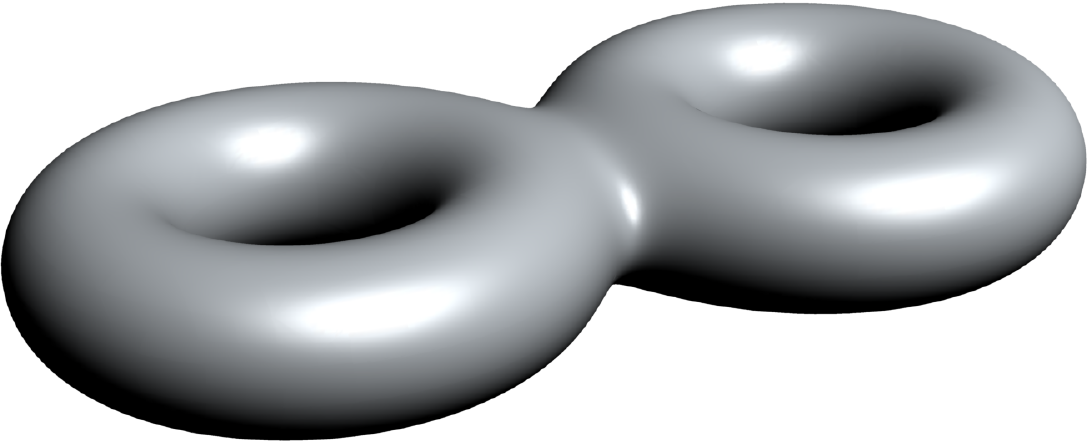}
    \end{subfigure}\hfill
    \begin{subfigure}[c]{0.05\textwidth} \vspace{-50pt}
    \centering
        $\leadsto$
    \end{subfigure}\hfill
    \begin{subfigure}[t]{0.29\textwidth}
       \includegraphics[width=\textwidth]{lattice_genus2_degen1.pdf}
        \centering
    \end{subfigure}\hfill
    \begin{subfigure}[c]{0.05\textwidth} \vspace{-50pt}
    \centering
        $\leadsto$
    \end{subfigure}\hfill
    \begin{subfigure}[t]{0.29\textwidth}
       \includegraphics[width=\textwidth]{lattice_genus2_degen2.pdf}
        \centering
    \end{subfigure}%
	\caption{The effect of the residue theorem. The amplitude, initially localised on a genus-two Riemann surface, localises on a nodal torus at an intermediate stage and finally on a bi-nodal sphere.}
	\label{fig:resthm}
\end{figure}

The residue theorem approach outlined above assumes directly that the amplitude only has poles at the scattering equations -- constituting the pole at which the amplitude is formulated on the higher-genus surface -- and the non-separating boundary divisor. This is indeed true for   the genus-two integrand, so the first application of the residue theorem is straightforward. The resulting expression on the nodal torus, however, contains in general many poles besides the divisor $\mathfrak{D}_{1,n+2}^{\text{non-sep}}$ and the scattering equations, leading to a variety of unwanted terms from the second application of the residue theorem. While this may  seem like an obstruction to obtaining an amplitude on a bi-nodal Riemann sphere, there is considerable freedom in the choice of  the integrand: any basis of Beltrami differentials may be chosen to define the scattering equations (see \cref{sec:SE}), and the  integrand is only defined modulo terms proportional to the scattering equations.\footnote{In fact, we have used this freedom to define the compact integrand \eqref{equ:4ptintegrand-final} at four points.} In \cref{sec:nodalRS_SE} and \cref{sec:nodalRS_integrand}, we demonstrate that we can use this freedom to construct an integrand containing only  poles at the maximal non-separating boundary divisor and the scattering equations. Using this representation, we can finally apply the residue theorem to localise the amplitude on the boundary divisor $\mathfrak{D}_{2,n}^{\text{max}}$.

In \cref{sec:map_to_RS}, we discuss the mapping of the remaining modular parameter on $\mathfrak{D}_{2,n}^{\text{max}}$ to the bi-nodal Riemann sphere. While any appropriate map could be chosen, we use a convenient trick to simplify the calculation. By extending the domain of integration to the full complex plane, using modular invariance, the modulus  maps to a cross-ratio of the marked points parametrising the nodes. The extension of the integral to the full complex plane is unique if we require  $\mathfrak{D}_{2,n}^{\text{max}}$ and the scattering equations  to remain the only poles of the integrand. Using the global residue theorem, the ambitwistor string correlator then localises straightforwardly  on the bi-nodal Riemann sphere. We conclude with a discussion of the resulting formula for $n$-point two-loop amplitudes in \cref{sec:contour_argument}.

Given the particularly technical nature of the discussion in this section, and for the benefit of the time-constrained reader who may want to skip on the details, we will present a brief summary of the results at the beginning of \cref{sec:contour_argument}.

\subsection{The scattering equations}\label{sec:nodalRS_SE}
The core idea of this section is to use the global residue theorem to localise ambitwistor string amplitudes on the non-separating boundary divisor,  rather than the higher-genus scattering equations.  A necessary prerequisite for this to work is that the {\it only} simple pole of the integrand -- besides the scattering equations, of course -- is the maximal non-separating boundary divisor $\mathfrak{D}_{2,n}^{\text{max}}\subset \widehat{\mathfrak{M}}'_{2,n}$. While this is certainly not true for all representations of the integrand, we will construct a representation of the integrand for which it holds.  Since  $\mathfrak{D}_{2,n}^{\text{max}}$ is a divisor of co-dimension two, we will analyse the global residue theorem iteratively, considering at each step a co-dimension one  divisor, $\mathfrak{D}_{2,n}^{\text{max}}\cong \mathfrak{D}_{1,n+2}^{\text{non-sep}}\subset\mathfrak{D}_{2,n}^{\text{non-sep}}\subset \widehat{\mathfrak{M}}'_{2,n}$, where the intermediate stage $\mathfrak{D}_{2,n}^{\text{non-sep}}$ corresponds to the (compactified) moduli space of the nodal torus. Fortunately, we can split the task of finding an appropriate representation of the integrand  into two parts:
\begin{enumerate}
 \item Finding a basis of Beltrami differentials (or equivalently a linear combination of the genus-two scattering equations $u^{IJ}=0$) such that  $\mathfrak{D}_{2,n}^{\text{max}}$ is the unique pole of the measure 
 \begin{equation}
  \prod_{I\leq J}\d\Omega_{IJ}\,\prod_{I\leq J}\bar\delta\big(u^{IJ}\big)\,,
 \end{equation}
 apart from the obvious pole $u^{IJ}=0$ in which the measure is originally defined.\footnote{We recall that, given the definition $2\pi i \bar\delta(z)=\bar\partial(1/z)$, Stokes theorem implies that the localisation on the delta functions can be seen as a multi-dimensional residue.}
 \item Finding a representation of  the  integrand $\mathcal{I}^{\text{chi}}_n=\mathcal{Z}^{\text{chi}}[\delta]\,\pf\big(M_\delta^{(2)}\big)$ that does not contain any  poles on the support of the scattering equations, both on $\widehat{\mathfrak{M}}'_{2,n}$ and on the nodal torus $\mathfrak{D}_{2,n}^{\text{non-sep}}$.
\end{enumerate}
Here, we will focus on part 1 -- finding the basis of Beltrami differentials -- while \cref{sec:nodalRS_integrand} tackles  constructing the integrand. Since both parts are interlinked, we will assume the existence of such an integrand for the remainder of this section.\\

Let us work in the parametrisation \eqref{equ:paramet_Omega} of the period matrix adapted to studying non-separating degenerations,
\begin{equation}\label{equ:param_omega_v2}
 \Omega = \begin{pmatrix} \tau_1+\tau_3 & \tau_3 \\ \tau_3 & \tau_2+\tau_3 \end{pmatrix}\,.
\end{equation}
All non-separating degenerations are represented by $\tau_r=i\infty$ for some $r\in\{1,2,3\}$, and so \eqref{equ:param_omega_v2} parametrises the  moduli space near all non-separating boundary divisors  $\mathfrak{D}_{2,n}^{\text{non-sep}}$ of $\widehat{\mathfrak{M}}'_{2,n}$. 
Of particular convenience are the exponentiated variables  $q_r$, defined in analogy with \cref{equ:q_II_def} to be
\begin{equation}\label{equ:def_q_r}
 q_1 = e^{i\pi\tau_1}\,,\qquad q_2 = e^{i\pi\tau_2}\,,\qquad q_3 = e^{2i\pi\tau_3}\,.
\end{equation}
This leads to the following integration measure for the period matrix;
\begin{equation}
 \frac{\d^3\Omega}{2(i\pi)^3} = \frac{\d q_1\,\d q_2\,\d q_3}{q_1 q_2 q_3}\,,
\end{equation}
where the poles at the non-separating boundary divisors, now given by $q_r=0$, are manifest. We will verify in \cref{sec:nodalRS_integrand} and \cref{sec:int_nodalRS}  explicitly that $q_r=0$ and the scattering equations are the only simple poles of the integrand.

To capture the freedom we have in representing the amplitude, let us use a generic basis $u_r = \big\langle \mu_r P^2\big\rangle$ for the moduli scattering equations. In the amplitude \cref{equ:amplitude_g=2}, this comes in general at the cost of a  Jacobian factor associated to the change of basis for the Beltrami differentials. (In our case that Jacobian will turn out to be trivial.) For the  purpose of discussing the residue theorem,  it will be useful to  introduce the following short-hand notation for the amplitude:
\begin{equation}\label{equ:def_R_notation}
 \mathcal{M}_n\equiv \mathfrak{R}(u_{1},u_{2},u_{3})\,.
\end{equation}
This compact notation is designed to exhibit only the moduli scattering equations $u_r$, while the remaining scattering equations as well as all other dependences remain  implicit. The original representation \cref{equ:amplitude_g=2} of the two-loop amplitude corresponds to $\mathcal{M}_n=\mathfrak{R}(u_{11},u_{22},u_{12})$ in this new notation. The goal of the remainder of \cref{sec:nodalRS_SE} is to express the $u_r$ in terms of the $u_{IJ}$ such that the amplitude localises on the bi-nodal Riemann sphere. The result will be given in \cref{equ:SE_final_choice}.\\

\begin{figure}[ht]
	\centering 
	\begin{subfigure}[t]{0.29\textwidth} 
        \centering
        \includegraphics[width=\textwidth]{lattice_genus2_degen1.pdf}
    \end{subfigure}\hfill
    \begin{subfigure}[c]{0.05\textwidth} \vspace{-50pt}
    \centering
        $+$
    \end{subfigure}\hfill
    \begin{subfigure}[t]{0.29\textwidth}
       \includegraphics[width=\textwidth]{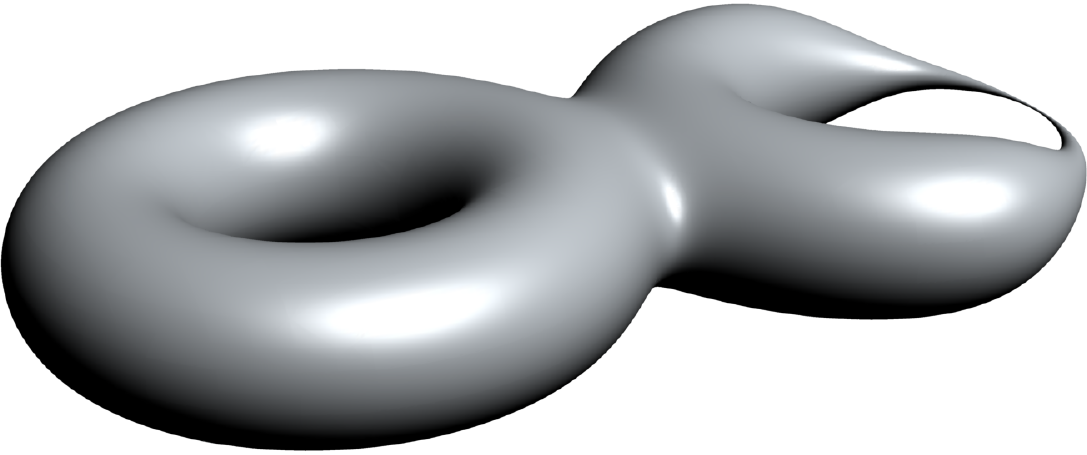}
        \centering
    \end{subfigure}\hfill
    \begin{subfigure}[c]{0.05\textwidth} \vspace{-50pt}
    \centering
        $+$
    \end{subfigure}\hfill
    \begin{subfigure}[t]{0.29\textwidth}
       \includegraphics[width=\textwidth]{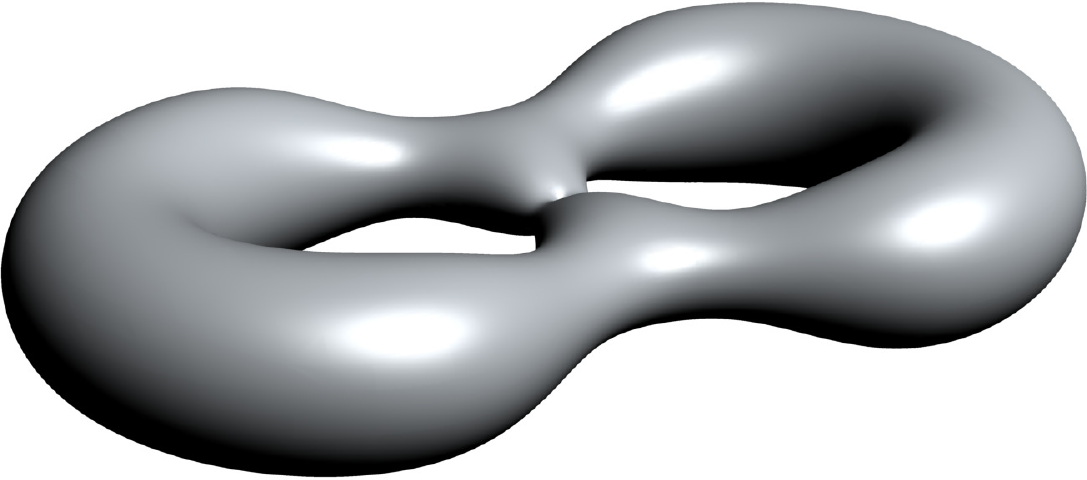}
        \centering
    \end{subfigure}%
	\caption{After applying the global residue theorem, the amplitude localises on three different nodal tori, corresponding to the boundary divisors $q_r=0$ for $r\in\{1,2,3\}$.}
	\label{fig:1stresthm}
\end{figure}

 Let us first explore how the global residue theorem plays out with this generic set of scattering equations, $u_r=0$. Since the two-loop amplitude  is fully localised over the moduli space $\widehat{\mathfrak{M}}'_{2,n}$, we can use the residue theorem to move the integration contour away from one of the poles defined by the scattering equations, say $u_2=0$. Since $q_r=0$ are the only other poles, this leads to three contributions, each localised on a non-separating boundary divisor,
\begin{equation}\label{equ:1st-res-thm}
 \mathfrak{R}(u_1,u_2,u_3) = - \mathfrak{R}(u_1,q_1,u_3)-\mathfrak{R}(u_1,q_2,u_3)-\mathfrak{R}(u_1,q_3,u_3)\,,
\end{equation}
as illustrated in figure \ref{fig:1stresthm}. The novel feature, compared to standard worldsheet theories, is that {\it all three} terms on the right hand side contribute on the moduli space $\mathfrak{M}'_{2,n}$ of the ambitwistor string. They represent genuinely distinct degenerations due to the different loop momenta associated to each homology cycle, and cannot be related by modular transformations without relabelling the loop momenta.\footnote{Contrast this with conventional worldsheet theories formulated over $\mathfrak{M}_{2,n}$, for whom the non-separating degeneration is unique on the fundamental domain. Concretely, this implies that, for the bosonic string, only the second term is relevant due to the definition of the fundamental domain \eqref{equ:fund-domain}: $q_2=0$ automatically implies $q_1=0$, and similarly for $q_3$. Both $\mathfrak{R}(u_1,q_1,u_3)$ and $\mathfrak{R}(u_1,q_3,u_3)$ vanish for generic momenta after an appropriate blow-up procedure to regulate the limit.}

All three terms in \eqref{equ:1st-res-thm} are formulated over a nodal torus, and since the amplitude localises over $\mathfrak{D}^{\text{non-sep}}_{2,n}\cong\widehat{\mathfrak{M}}_{1,n+2}$, we are free to apply another residue theorem. However, applying this second residue theorem does {\it not} localise the amplitude on the bi-nodal Riemann sphere. A short calculation shows that unless two of the above terms vanish, the amplitude receives contributions from nodal tori as well as the bi-nodal Riemann sphere,
\begin{equation}
 \mathfrak{R}(u_1,u_2,u_3) =\underbrace{ \mathfrak{R}(q_2,q_1,u_3)+\mathfrak{R}(q_3,q_1,u_3)+\dots}_{\text{on bi-nodal sphere}}+\underbrace{\mathfrak{R}(u_2,q_1,u_3)+\dots}_{\text{on nodal torus}}\,,
\end{equation}
where we represented explicitly only the contributions coming from the first term on the right-hand side of \cref{equ:1st-res-thm}.
This not only demonstrates that the amplitude, formulated using a generic basis for the scattering equations, fails to localise on the bi-nodal sphere after applying the residue theorem on moduli space, but also suggests a resolution: choose a basis of scattering equations such that two of the terms contributing to each residue theorem vanish. \\

\begin{figure}[ht]
	\centering 
	\includegraphics[width=6cm]{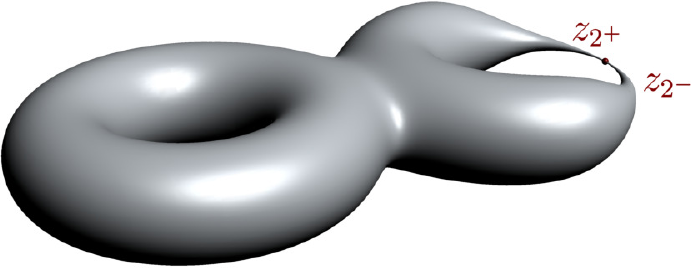}
	\caption{The non-separating boundary divisor $q_2=0$, corresponding to a nodal torus. The node is parametrised by $z_{2^+}$ and $z_{2^-}$.}
	\label{fig:degen3_labels}
\end{figure}

To find this basis, let us investigate the asymptotics of the scattering equations on the nodal tori $q_r=0$.  Clearly, it depends on the behaviour of the holomorphic differentials and the period matrix in the non-separating degeneration limit, reviewed in \cref{sec:degen-review}. For concreteness, let us focus on $\mathfrak{R}(q_2,u_1,u_3)$. In the limit $q_2\to0$, the holomorphic differential $\omega_2$ associated to the degenerating $A_2$-cycle develops simple poles at the node, while $\omega_1$ descends to the holomorphic differential $\d z$ on the torus.  In this case, Fay's degeneration formula \cref{equ:Fay-hol-g} gives the asymptotics\footnote{It is clear from \cref{equ:Fay-hol-g} that the first correction to \eqref{equ:Fay-hol_a} is of order $\mathcal{O}(q_2^4)$, since we have $\omega_1^{(1)}(z_{g^\pm})=\d z_{g^\pm}$ for the torus. This is not important in our analysis, and a correction of order $\mathcal{O}(q_2^2)$, which occurs at higher genus, would suffice.}
\begin{subequations}\label{equ:Fay-hol}
\begin{align}
 \omega_1 (z)& = \d z +\mathcal{O}(q_2^4)\,, \label{equ:Fay-hol_a}\\
 \omega_2 (z)& = \frac1{2\pi i}\,\omega_{2^+\!,2^-}(z)  +\mathcal{O}(q_2^2)\,,
\end{align}
\end{subequations}
where the node is parametrised by $z_{2^+}$ and $z_{2^-}$, and $\omega_{2^+\!,2^-}(z) $ denotes the Abelian differential of the third kind with simple poles at the node; see \cref{fig:degen3_labels}. The subleading term at $\mathcal{O}(q_2^2)$ vanishes due to the translation invariance of the torus. The asymptotics of the period matrix, given in \cref{equ:Fay_period_matrix_g}, imply that the component $\Omega_{11}$ of the period matrix descends to the modular parameter $\tau$ of the torus, while the off-diagonal entries $\Omega_{12}= \int_{z_{2^-}}^{z_{2^+}}\d z=z_{2^+}-z_{2^-} $ encode the moduli associated to the  node,
\begin{equation}\label{equ:Fay_period_matrix}
 \Omega = \begin{pmatrix}\tau &z_{2^+}-z_{2^-} \\z_{2^+}-z_{2^-} & \;\frac{1}{i\pi}\ln q_2+\text{const} \end{pmatrix}+\mathcal{O}(q_2^2)\,.
\end{equation}
By fixing the translation invariance of the torus, we can  align the modulus $\Omega_{12}$ directly with the location of the node by fixing one of the nodal points, for example $\Omega_{12}=z_{2^+}$ using $z_{2^-}=0$.  Note that $\tau$ is indeed integrated over the fundamental domain, since $\tau=\Omega_{11}$ with Re$(\tau)\in [-\frac{1}{2}, \frac{1}{2}]$ by $(i)$ of \cref{equ:fund-domain}, while the condition $(iii)$  implies that $|\Omega_{11}|>1$.\footnote{Using the modular transformation $a=d=\begin{pmatrix}  0 & 0 \\ 0 & 1\end{pmatrix}$ and 
$ c=-b=\begin{pmatrix} 1 & 0 \\ 0 & 0\end{pmatrix}$} We will discuss the range of the remaining modulus $\Omega_{12}=z_{2^+}-z_{2^-}$ in detail in \cref{sec:map_to_RS}.\\

Fay's degeneration formula now  allows us to calculate the asymptotics of the scattering equations on the nodal torus. Since $\omega_2$ develops simple poles at the node, $u_{22}$ and $u_{12}$ can be identified as  the coefficients of the second  and first order pole at $z_{2^+}$ of $P^2$, respectively. Moreover, with $\omega_1=\d z$ to leading order, the remaining scattering equation $u_{11}=u^{(1)}$ becomes the coefficient of the (single) quadratic holomorphic differential $\d z^2$ on the torus. The scattering equations on 
$\mathfrak{R}(q_2,u_1,u_3)$  thus take the following form: 
\begin{subequations}\label{equ:degen_q2}
\begin{align}
 &u_{11} = u^{(1)}\big(q=q_{11}^2\big) +\mathcal{O}(q_2) &&\text{as }q_2\rightarrow 0\,,\\
 &u_{12} = \mathcal{E}^{(1)}_{2^+}\big(q=q_{11}^2\big) +\mathcal{O}(q_2)   &&\text{as }q_2\rightarrow 0\,,\\
 &u_{22} = \ell_2^2+\mathcal{O}(q_2) &&\text{as }q_2\rightarrow 0\,.
\end{align}
\end{subequations}
This means that $u_{22}=\ell_2^2$ is associated to the pole at the node, while the other scattering equations descend to the nodal torus: $u_{11}=u^{(1)}$ becomes the modular scattering equation associated to $q=e^{2i\pi\tau}$, while $u_{12}$ turns into a vertex scattering equation $ \mathcal{E}^{(1)}_{2^+}$ for the nodal point $z_{2^+}$ with momentum $\ell_2$.\footnote{Notice that both loop momenta $\ell_1$ and $\ell_2$ appear linearly in $ \mathcal{E}^{(1)}_{2^+}$.} Moreover, let us highlight that the amplitude on the nodal torus $\mathfrak{R}(u_1,q_2,u_3)$ does not localise on $u_{2}=0$ -- this is precisely the constraint relaxed by applying the residue theorem -- so, if we take $u_{2}=u_{22}$, \eqref{equ:degen_q2} does not imply a cut of the loop momentum $\ell_2$.

Of course, the exact same arguments can be applied to the amplitude $\mathfrak{R}(u_1,q_1,u_3)$ on the nodal torus $q_1=0$, but with reversed roles for the holomorphic differentials $\omega_I$. Let us denote the node resulting from the degeneration of the $A_1$-cycle by $z_{1^+}$ and $z_{1^-}$. Then Fay's degeneration formula implies that
\begin{subequations}
\begin{align}
 \omega_1 (z)& =  \frac1{2\pi i}\,\omega_{1^+\!,1^-}(z)  +\mathcal{O}(q_1^2)\,,\\
 \omega_2 (z)& = \d z +\mathcal{O}(q_1^4)\,,
\end{align}
\end{subequations}
and likewise $\Omega_{22}=\tau$, $\Omega_{12}=z_{1^+}-z_{1^-}$ to leading order. By identifying the coefficients of the single and double pole of $P^2$ at the nodal point $z_{1^+}$, we find the following asymptotics for the scattering equations:
\begin{subequations}\label{equ:degen_q1}
\begin{align}
 &u_{11} = \ell_1^2+\mathcal{O}(q_1) &&\text{as }q_1\rightarrow 0\,,\label{equ:degen_q1_u11}\\
 &u_{12} = \mathcal{E}^{(1)}_{1^+}\big(q=q_{22}^2\big) +\mathcal{O}(q_1)   &&\text{as }q_1\rightarrow 0\,,\\
 &u_{22} = u^{(1)}\big(q=q_{22}^2\big) +\mathcal{O}(q_1) &&\text{as }q_1\rightarrow 0\,.
\end{align}
\end{subequations}
Again, the roles are reversed with respect to \cref{equ:degen_q2},  so that $u_{11}=\ell_1^2$ becomes the momentum squared flowing through the node, while $u_{22}$ and $u_{12}$ descend to the scattering equations on the nodal torus. \\

Given these asymptotics,  let us return to our objective of constructing a basis of scattering equations such that  only a single term contributes to the residue theorem. From \eqref{equ:degen_q1_u11}, it is evident how to choose the scattering equation $u_1$ in order to make $\mathfrak{R}(u_1,q_1,u_3)$ vanish for generic loop momenta,
\begin{equation}
  \mathfrak{R}(u_1,q_1,u_3) \equiv \mathfrak{R}(u_{11},q_1,u_3)=\mathfrak{R}(\ell_1^2,q_1,u_3)=0\,.
\end{equation}
Strictly speaking, we have not yet seen that the choice $u_1=u_{11}$ precludes contributions  of the form $\delta\big(\ell_1^2\big)$ to the cut of the amplitude. We will revisit this question  at the end of this section, where we show that no additional terms contribute to the amplitude on a cut.

Having fixed $u_1=u_{11}$,  we also have to choose $u_2=u_{22}$ in order to preserve the symmetry between the degenerations of the cycles $A_1$ and $A_2$. This is a natural requirement, because the amplitude should be unaffected by our choice of relaxing the scattering equation $u_1$ or $u_2$ first. However, this implies that we require $\mathfrak{R}(q_3,u_1,u_3)=0$ to obtain a formulation of the amplitude on the bi-nodal sphere.\\

\begin{figure}[ht]
	\centering 
	\begin{subfigure}[t]{0.29\textwidth} 
        \centering
        \includegraphics[width=\textwidth]{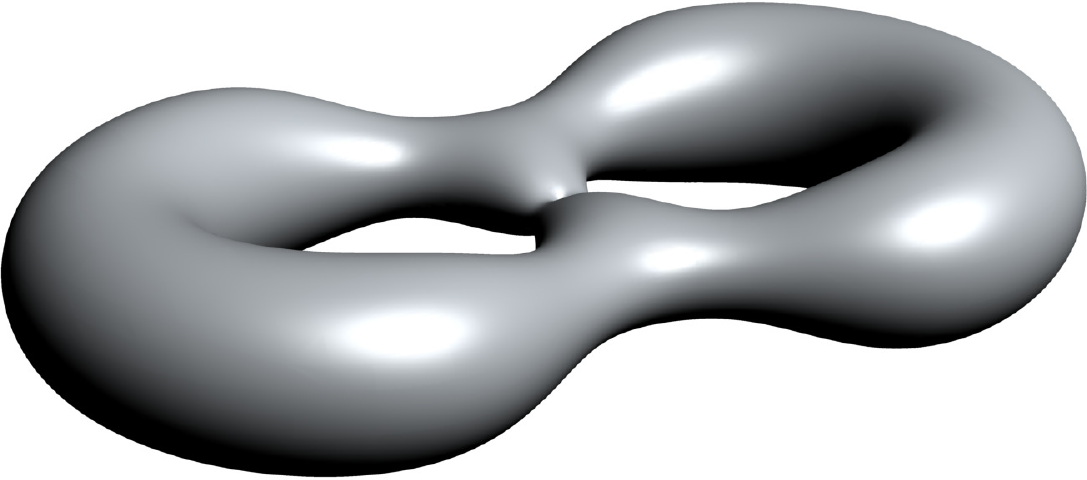}
    \end{subfigure}\hspace{-7pt}
    \begin{subfigure}[c]{0.2\textwidth} \vspace{-50pt}
    \centering
        $\scalebox{2}{\Bigg|}_{\big(\ell_1,\ell_2\big)}\hspace{10pt}{ \longleftrightarrow}\hspace{10pt}$
    \end{subfigure}
    \begin{subfigure}[t]{0.29\textwidth}
       \includegraphics[width=\textwidth]{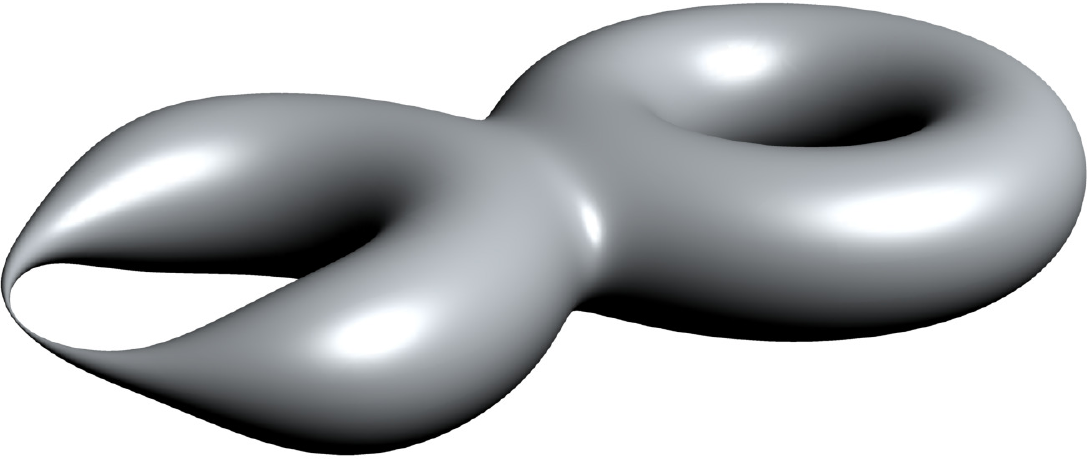}
        \centering
    \end{subfigure}\hspace{1pt}
     \begin{subfigure}[c]{0.1\textwidth} \vspace{-50pt}
    \centering
        $\scalebox{2}{\Bigg|}_{\big(\ell_1+\ell_2,\ell_2\big)}$
    \end{subfigure}\hfill
	\caption{The form of the scattering equations at $q_3=0$ can be determined from the scattering equations at $q_1=0$ by using a modular transformation to relate the degenerations. In particular, the loop momenta transform as $\big(\ell_1,\ell_2\big) \leftrightarrow \big(\ell_1+\ell_2,\ell_2\big)$.}
	\label{fig:modrel}
\end{figure}

\begin{figure}[ht]
	\centering 
	  \includegraphics[width=7cm]{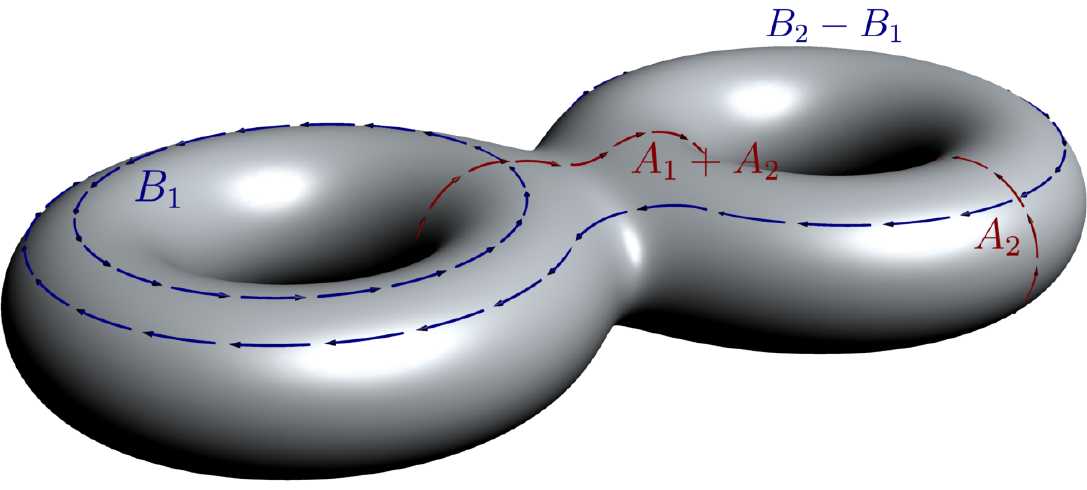}
	\caption{The homology basis $\big(\widetilde{A}_1, \widetilde{B}_1\big) = \big(A_1+A_2,B_1\big)$ and $\big(\widetilde{A}_2, \widetilde{B}_2\big) = \big(A_2,B_2-B_1\big)$, after the modular transformation \eqref{equ:mod-transform_scatt}. The loop momentum flowing through the cycle $\widetilde{A}_1$ is given by $\widetilde{\ell}_{1}=\oint_{A_1+A_2}P=\ell_1+\ell_2$, as can be seen intuitively from the intersection of $\widetilde{A}_1$ with the original cycles $B_1$ and $B_2$.}
	\label{fig:homol_after_mod}
\end{figure}

Notice  that calculating the asymptotics of the scattering equations for the contribution $\mathfrak{R}(q_3,u_1,u_3)$ requires an additional step, compared to \eqref{equ:degen_q2} and \eqref{equ:degen_q1}: to use Fay's degeneration formula, we need to exchange the role of $\tau_3$  with e.g. $\tau_1$ via a modular transform. We will do this below, but let us first take a look at the result. At $q_3=0$, the scattering equations become
\begin{subequations}\label{equ:degen_q3}
\begin{align}
 &u_{11} = (\ell_1+\ell_2)^2+\mathcal{F}_{11}(q_1,q_2)+\mathcal{O}(q_3) &&\text{as }q_3\rightarrow 0\,,\\
 &u_{22} = \mathcal{F}_{22}(q_1,q_2)+\mathcal{O}(q_3)  &&\text{as }q_3\rightarrow 0\,,\\
 &u_{12} = \mathcal{F}_{12}(q_1,q_2)+\mathcal{O}(q_3) &&\text{as }q_3\rightarrow 0\,,
\end{align}
\end{subequations}
with $\mathcal{F}_{IJ}(q_1,q_2)\neq 0$ but $\mathcal{F}_{11}(q_1,q_2)+\mathcal{F}_{22}(q_1,q_2)+\mathcal{F}_{12}(q_1,q_2)=0$; see also \cref{fig:homol_after_mod}. This implies that we cannot choose  $u_3=u_{12}$, as we might have guessed, because $\mathfrak{R}(u_{22},q_3,u_3)=\mathfrak{R}(u_{22},q_3,u_{12})\neq 0$,\footnote{This term appears from a second use of the residue theorem on the last term of \cref{equ:1st-res-thm}.} and thus the amplitude would receive contributions from a nodal torus. To see what linear combination of the $u_{IJ}$ we should choose instead for $u_3$, we will need to  prove the degeneration \eqref{equ:degen_q3} using modular invariance.

\paragraph{Proof.} The main idea is to exchange the roles of $\tau_1$ and $\tau_3$ using a modular transformation.  Consider therefore the modular transformation $M$ with
\begin{equation}\label{equ:mod-transform_scatt}
 a=\begin{pmatrix}1 & 0 \\ -1 & 1\end{pmatrix}\,,\qquad d=\begin{pmatrix} 1 & 1\\0 & 1\end{pmatrix}\,,\qquad b=c=0\,.
\end{equation}
In terms of the basis of homology cycles, this corresponds to 
\begin{equation}
 \big(\widetilde{A}_1, \widetilde{B}_1\big) = \big(A_1+A_2,B_1\big)\,,\qquad \big(\widetilde{A}_2, \widetilde{B}_2\big) = \big(A_2,B_2-B_1\big)\,,
\end{equation}
so this transformation indeed exchanges the cycle $A_1$ with $A_1+A_2$, see figure \ref{fig:homol_after_mod}. Using \cref{equ:mod_transform_omega} for the modular transformations of the period matrix, we also confirm directly that this exchanges $\tau_1$ and $\tau_3$,
\begin{equation}
 \widetilde{\Omega} = \begin{pmatrix} \tau_3+\tau_1 & -\tau_1 \\ -\tau_1 & \tau_2+\tau_1 \end{pmatrix}\,.
\end{equation}
Recalling the discussion of modular invariance,  \cref{equ:mod_transform_omega} and \cref{equ:mod_transform_ell} describe the behaviour of the holomorphic differentials, as well as the loop momenta,
\begin{equation}\label{equ:ell1=ell1+ell2}
 \big(\widetilde{\omega}_1, \widetilde{\omega}_2\big) = \big(\omega_1,\omega_2-\omega_1\big)\,,\qquad \big(\widetilde{\ell}_1, \widetilde{\ell}_2\big) = \big(\ell_1+\ell_2,\ell_2\big)\,,
\end{equation}
Note that $\widetilde{\ell}_1=\ell_1+\ell_2$ is the loop momentum flowing through the cycle $\widetilde{A}_1=A_1+A_2$ as illustrated in figure  \ref{fig:homol_after_mod}.

\noindent With the above properties, the modular transformation maps $P_\mu$ to
\begin{equation}
 P_\mu(z) = \big(\ell_1+\ell_2\big)_\mu\widetilde{\omega}_1+\ell_{2,\mu}\,\widetilde{\omega}_2+\sum_i k_{i,\mu} \omega_{i,*}\,.
\end{equation}
As before, we can use the support of the particle scattering equations to write $P^2$ as a holomorphic quadratic differential, now in the basis $\widetilde{\omega}_I$ of holomorphic differentials,
\begin{equation}\label{equ:P2_mod}
 P^2 = u_{IJ}\,\omega_I\omega_J= \underbrace{\big(u_{11}+u_{12}+u_{22}\big)}_{\textcolor{blue}{\equiv\widetilde{u}_{11}}}\widetilde{\omega}_1^2 + \underbrace{\big(u_{12}+2u_{22}\big)}_{\textcolor{blue}{\equiv\widetilde{u}_{12}}}\widetilde{\omega}_1 \widetilde{\omega}_2 +\underbrace{\phantom{\big(}u_{22}}_{\textcolor{blue}{\equiv\widetilde{u}_{22}}}\widetilde{\omega}_2^2\,.
\end{equation}
At this stage, we are able to use Fay's degeneration formula for the differentials $\big(\widetilde{\omega}_1, \widetilde{\omega}_2\big)$ to investigate the limit $q_3=\tilde{q}_1\rightarrow 0$ of the scattering equations. This now mirrors exactly the procedure from above. Denoting the locations of the node  resulting from the degeneration of the $\widetilde{A}_1$-cycle by $z_{3^+}$ and $z_{3^-}$, we find that
\begin{subequations}
\begin{align}
 \widetilde{\omega}_1 (z)& =  \frac1{2\pi i}\,\omega_{3^+,3^-}(z)  +\mathcal{O}(q_3^2)\,,\\
 \widetilde{\omega}_2 (z)& = \d z +\mathcal{O}(q_3^4)\,,
\end{align}
\end{subequations}
By identifying the coefficients of the poles of $P^2$ at the nodal point $z_{3^+}$, we can again extract the asymptotics for the scattering equations;
\begin{subequations}
\begin{align}
 &\widetilde{u}_{11} = \big(\ell_1+\ell_2\big)^2+\mathcal{O}(q_3) &&\text{as }\widetilde{q}_1=q_3\rightarrow 0\,,\\
 &\widetilde{u}_{12} = \mathcal{E}^{(1)}_{3^+}\big(q=\widetilde{q}_{22}^2\big) +\mathcal{O}(q_3)   &&\text{as }\widetilde{q}_1=q_3\rightarrow 0\,,\\
&\widetilde{u}_{22} = u^{(1)}\big(q=\widetilde{q}_{22}^2\big) +\mathcal{O}(q_3) &&\text{as }\widetilde{q}_1=q_3\rightarrow 0\,.
\end{align}
\end{subequations}
We stress that this has now the same interpretation as for the nodal tori $q_1=0$ or $q_2=0$: the scattering equation $\widetilde{u}_{11}=\widetilde{\ell}_1^2=(\ell_1+\ell_2)^2$ becomes the momentum squared flowing through the node, while $\widetilde{u}_{22}$ and $\widetilde{u}_{12}$ descend to the scattering equations on the nodal torus. In particular, these  scattering equations evidently depend on the modular parameter $\widetilde{q}_{22}^2=q$ and the modulus of the node, $\widetilde{\tau_3}=-\tau_1=z_{3^-}-z_{3^+}$. In turn, this implies that all of
\begin{equation*}
 u_{22}=\underbrace{\widetilde{u}_{22}(q_1,q_2)}_{\equiv \mathcal{F}_{22}}\,,\qquad u_{12}=\underbrace{\widetilde{u}_{12}(q_1,q_2)-2\widetilde{u}_{22}(q_1,q_2)}_{\equiv \mathcal{F}_{12}}\,,\qquad u_{11}= \big(\ell_1+\ell_2\big)^2+\underbrace{\widetilde{u}_{22}(q_1,q_2)-\widetilde{u}_{12}(q_1,q_2)}_{\equiv  \mathcal{F}_{11}}\,,
\end{equation*}
depend on the modular parameters $q_1$ and $q_2$. We arrive therefore at  \eqref{equ:degen_q3}, upon identifying $ \mathcal{F}_{IJ}$ as given above, and comparing to \eqref{equ:P2_mod} to show that $\mathcal{F}_{11}(q_1,q_2)+\mathcal{F}_{22}(q_1,q_2)+\mathcal{F}_{12}(q_1,q_2)=0$. \hfill $\square$
\newline

This discussion  now allows us to construct $u_3$ such that $\mathfrak{R}(q_3,u_2,u_3)$ vanishes for generic loop momenta.  From \eqref{equ:P2_mod}  and in analogy with the argument for the torus $q_1=0$,  we must take $u_3 = \widetilde{u}_{11} = u_{11}+u_{12}+u_{22}$, because 
\begin{equation}
 u_{3} = \big(\ell_1+\ell_2\big)^2+\mathcal{O}(q_3) \qquad\qquad\text{as }q_3\rightarrow 0\,,
\end{equation}
so that $\mathfrak{R}(u_2,q_3,u_3)=0$ for generic loop momenta. We will  discuss the cut $(\ell_1+\ell_2)^2=0$ shortly.\\

To summarise, we conclude that the residue formula \eqref{equ:1st-res-thm} only results in a single contribution from the non-separating boundary divisor $\mathfrak{D}_{2,n}^{\text{non-sep}}$ if we choose the scattering equations
\begin{equation}\label{equ:SE_final_choice}
 u_1 = u_{11}\,,\qquad u_2 = u_{22}\,,\qquad u_3 = u_{12}+u_{11}+u_{22}\,.
\end{equation}
Note that the Jacobian associated to this basis choice is trivial, so the  integrand $\mathcal{I}^{\text{chi}}_n$ of the amplitude is unaffected.  In the next section, we construct a representation of this integrand that does not contain poles on support of the scattering equations  -- part 2 in our roadmap outlined at the beginning of \cref{sec:nodalRS_SE}. Once proven on the nodal torus, we can again apply the residue theorem to localise the amplitude on the bi-nodal Riemann sphere,
\begin{equation}\label{equ:full-residue-theorem}
 \mathfrak{R}(u_1,u_2,u_3) = - \mathfrak{R}(u_1,q_2,u_3)=\mathfrak{R}(q_1,q_2,u_3)\,.
\end{equation}
All other terms  vanish due to our choice of scattering equations,
\begin{equation}\label{equ:res_vanishing}
 \mathfrak{R}(q_1,u_1,u) =  \mathfrak{R}(q_2,u_2,u)=\mathfrak{R}(q_3,u_3,u)=0\,,\qquad\text{ for any }u\,.
\end{equation}
Before proceeding, let us highlight briefly an interpretation of the relations \eqref{equ:res_vanishing}. Since the maximal non-separating boundary divisor $\mathfrak{D}^{\text{max}}_{2,n}$ has co-dimension two in the $(n+3)$-dimensional moduli space $\widehat{\mathfrak{M}}'_{2,n}$, the full residue theorem localising on $\mathfrak{D}^{\text{max}}_{2,n}$ is two-dimensional as well. However, the relations \eqref{equ:res_vanishing} effectively diagonalise this two-dimensional residue theorem, reducing it to two consecutive residue theorems in separate variables.

\paragraph{Contribution on a cut.} While the above discussion seems to suggest that contributions from cuts are subtle and need to be treated with care, they actually represent the simplest scenario. To see this, let us investigate the genus-two amplitude $\mathfrak{R}(u_1,u_2,u_3)$  on the cut $\ell_2^2=0$. From \eqref{equ:degen_q2}, we see that this cut forces $u_2\propto q_2$, and thus trivially
\begin{equation}
 \mathfrak{R}(u_1,u_2,u_3)\Big|_{\ell_2^2=0}=\mathfrak{R}(u_1,q_2,u_3)\Big|_{\ell_2^2=0}\,.
\end{equation}
This is indeed the same result we obtained for generic loop momenta after the first use of the residue theorem. If furthermore $\ell_1^2\neq0$ and $(\ell_1+\ell_2)^2\neq 0$, applying a single residue theorem is sufficient to localise the full amplitude on $\mathfrak{R}(q_1,q_2,u_3)\big|_{\ell_2^2=0}$. Note that this exactly matches the result obtained from a cut of the amplitude \eqref{equ:full-residue-theorem} on the bi-nodal sphere $\mathfrak{D}^{\text{max}}_{2,n}$, so the cut commutes with the residue theorem.

Similarly, for a cut in one of the other loop momenta, $\ell_1^2=0$ or $(\ell_1+\ell_2)^2= 0$, we find respectively that the amplitude is given by
\begin{subequations}
 \begin{align}
  \mathfrak{R}(u_1,u_2,u_3)\Big|_{\ell_1^2=0}\hspace{23pt}&=\mathfrak{R}(q_1,u_2,u_3)\Big|_{\ell_1^2=0}\,,\\
  \mathfrak{R}(u_1,u_2,u_3)\Big|_{(\ell_1+\ell_2)^2=0}&=\mathfrak{R}(u_1,u_2,q_3)\Big|_{(\ell_1+\ell_2)^2=0}\,.
 \end{align}
\end{subequations}
After using the residue theorem (twice for $(\ell_1+\ell_2)^2= 0$), the result is again the same as taking a cut of the amplitude \eqref{equ:full-residue-theorem} on the bi-nodal sphere $\mathfrak{D}^{\text{max}}_{2,n}$. This analysis extends straightforwardly to multiple cuts, and thus the two-loop amplitude localises on the maximal non-separating degeneration irrespective of the loop momentum configuration.

\subsection{The integrand}\label{sec:nodalRS_integrand}
Throughout the last section, we assumed  the existence of a representation of the integrand $\mathcal{I}_n^{\text{chi}}$ that does not contain  poles on the support of the scattering equations. Let us now return to this point and explicitly construct this representation.

It is easily checked that the integrand of \eqref{equ:integrand_g=2} does not have poles on the genus-two Riemann surface for generic kinematics, because poles in the location of vertex operators $z_i-z_j$ correspond -- via the scattering equations -- to factorisation channels of the amplitude. Moreover, recall from \cref{sec:mod-inv} that there are no poles associated to the PCO gauge slice $x_{\alpha}$, so no additional poles contribute in the first residue theorem. However, since this PCO gauge invariance relies on the support of {\it all} scattering equations, we expect the integrand to develop poles in $x_{\alpha}$ on the nodal torus $\mathfrak{D}^{\text{non-sep}}_{2,n}$. 

This can be made explicit. Once on the nodal torus, one of the genus-two holomorphic differentials becomes meromorphic,  $\omega_2=\omega_{2^+\!,2^-}(z)$,  and the modular parameter $\tau_3=z_{2^+}-z_{2^-}$ encodes the location of the node. The terms in the Pfaffian containing a factor of $P(x_\alpha)$, 
\begin{equation*}
 A_{x_1\,x_2}=P(x_1)\cdot P(x_2)\,S_\delta (x_1,x_2)\,,\qquad A_{x_\alpha,j}=P(x_\alpha)\cdot k_j S_\delta(x_\alpha,z_j)\,,\qquad C_{x_\alpha,j}=P(x_\alpha)\cdot \epsilon_j S_\delta(x_\alpha,z_j)\,,
\end{equation*}
thus develop simple poles in the modular parameter $q_3$. While the coefficients of $A_{x_\alpha,j}$ and $C_{x_\alpha,j}$ still vanish on the support of the scattering equations,\footnote{The argument is completely analogous to the one presented in \cref{sec:mod-inv}.} the coefficient of $ A_{x_1\,x_2}$ is non-zero on the nodal torus. If the amplitude is represented using the integrand \eqref{equ:intchi}, the second application of the residue theorem thus leads to a contribution from the poles $q_3=e^{2i\pi x_\alpha}$ of $P(x_1)\cdot P(x_2)$. Of course, the full amplitude remains invariant under different PCO gauge choices, but fails to localise on the bi-nodal sphere.

Luckily, we have already seen how to eliminate these poles  when discussing the four-particle amplitude in \cref{sec:4pt}. In that case, the full amplitude is proportional to $ A_{x_1\,x_2}$, and we used a linear combination of the scattering equations determined by the Beltrami differential \eqref{equ:mux},
\begin{equation}
\mu_x=\frac{1}{2}\left(\frac{c_1}{c_2}\delta(z,x_1)+\frac{c_2}{c_1}\delta(z,x_2)\right)\,,
\end{equation}
to simplify  $P(x_1)\cdot P(x_2)$ to $\px(x_1,x_2)$. This procedure characteristically removes the terms in $P(x_1)\cdot P(x_2)$ proportional to the holomorphic differentials, and will thus eliminate poles in $q_3$ on the nodal torus. Generalising from four to $n$ particles, we indeed find
\begin{align}\label{equ:px_version2}
 \px(x_1,x_2)&=P(x_1)\cdot P(x_2) - \big\langle\mu_x\,P^2\big\rangle\,
 =-\frac{1}{2}c_1c_2\partial\varpi(x_1)\partial\varpi(x_2)\sum_{i,j} k_i\cdot k_j\,\frac{\Delta_{i*}\Delta_{j*}}{\varpi(z_i)\varpi(z_j)\varpi(z_*)^2}\,.
\end{align}
Note that while the right hand side of \eqref{equ:px_version2} still depends on $x_1$ and $x_2$, the coefficient of the pole in $x_1-x_2$ vanishes manifestly when multiplied by the partition function due to \cref{equ:Zcc=1}. Moreover, the coefficients of the poles in $x_\alpha-z_i$ vanish because  the  Pfaffian matrix becomes degenerate at this order; the relevant calculation proceeds  in close analogy to the discussion in \cref{sec:mod-inv}.

We can thus use $\px(x_1,x_2)$ to define a representation of the  integrand free of poles in the modular parameters.\\

In summary, we will use the following representation of the amplitude for the residue theorem:
\begin{equation}\label{equ:ampl_final}
 \mathcal{M}_n=\delta\Big(\sum_{i=1}^n k_i\Big)\,\int \d^{10}\ell_1\,\d^{10}\ell_2 \,\,\int_{\mathfrak{M}_{2,n}}\hspace{-15pt}
 \d^3\tau\,\prod_{r=1}^3\bar\delta\big(u_r\big)\,\prod_{i=1}^n\bar\delta\big(\big\langle \mu_i\,P^2\big\rangle\big)\,\,\mathcal{I}^{\text{chi}}_n\,\widetilde{\mathcal{I}}^{\text{chi}}_n\,.
\end{equation}
Just as in \cref{equ:intchi}, the chiral integrand is defined by
\begin{equation}\label{equ:chiral-int_final}
 \mathcal{I}^{\text{chi}}_n=\sum_{\delta}  \mathcal{Z}^{\text{chi}}[\delta]\,\pf\big(M_\delta^{(2)}\big)\,,
\end{equation}
but in contrast to the original representation,  the $(2n+2)\times(2n+2)$ matrix $M^{(2)}_\delta$ is now given by
\begin{subequations}
\begin{align}
 & &&M^{(2)}_\delta=\begin{pmatrix}A &-C^T\\C&B\end{pmatrix}\,,&&\\
 &A_{x_1x_2}=\px(x_1,x_2) S_\delta(x_1,x_2)\,,&& A_{x_\alpha,j}=P(x_\alpha)\cdot k_j S_\delta(x_\alpha,z_j)\,,&& A_{ij}=k_i\cdot k_j S_\delta(z_i,z_j)\,,\\
 & && C_{x_\alpha,j}=P(x_\alpha)\cdot \epsilon_j S_\delta(x_\alpha,z_j)\,,&& C_{ij}=\epsilon_i\cdot k_j S_\delta(z_i,z_j)\,,\\
 & && C_{ii} = P(z_i)\cdot \epsilon_i\,, && B_{ij}=\epsilon_i\cdot\epsilon_j  S_\delta(z_i,z_j)\,.
\end{align}
\end{subequations}
As discussed above, $A_{x_\alpha,j}$ and $ C_{x_\alpha,j}$ do not give rise to  poles in $q_3$ because the respective coefficients still vanish on the nodal torus on the support of the vertex scattering equations. The integrand therefore meets our requirement of not containing  poles on  the nodal torus $\mathfrak{D}^{\text{non-sep}}_{2,n}$, and we can proceed with the second application of the residue theorem,\eqref{equ:full-residue-theorem}
\begin{equation}
 \mathfrak{R}(u_1,u_2,u_3) = - \mathfrak{R}(u_1,q_2,u_3)=\mathfrak{R}(q_1,q_2,u_3)\,,
\end{equation}
as indicated in the previous section. Note that the discussion presented here bears a close resemblance to the issues arising from the scattering equations in the last section: while the degeneration to the nodal torus was straightforward, only a specific representation of the integrand allows for a further application of the residue theorem to fully localise on the maximal non-separating degeneration. These strong requirements on the representation of the integrand to localise on higher non-separating degenerations seem to be a general feature that we strongly expect  to extend to higher genus.

\subsection{Integration over the moduli}\label{sec:map_to_RS}
As discussed over the course of the last two sections, a suitable representation of the two-loop amplitude localises on  the maximal non-separating boundary divisor $\mathfrak{D}^{\text{max}}_{2,n}$ after applying a global residue theorem. One last subtlety remains to be resolved: the isomorphism $\mathfrak{D}^{\text{max}}_{2,n}\cong \widehat{\mathfrak{M}}_{0,n+4}$ with the bi-nodal Riemann sphere. While Fay's degeneration formulae \eqref{equ:Fay-hol} already incorporates this map to the nodal Riemann sphere, more care is needed with the remaining modulus $\tau_3$ of the period matrix. One way to see this is as follows. Recall first that, due to modular invariance, the integration over the moduli runs over the fundamental domain, so that $|q_r|< 1$. On the other hand, on the bi-nodal Riemann sphere, $\tau_3$ is expected to correspond to the location of one of the nodes  upon fixing the other nodal points using M\"obius invariance (see Fays' degeneration formula for the period matrix \eqref{equ:Fay_period_matrix}), and $q_3=e^{2\pi i\tau_3}$ should thus be unconstrained.\footnote{In fact, the same argument already applies for the nodal torus. We will see below that the resolution offered here does not impact the arguments of the preceding sections.} Evidently, this implies that the isomorphism $\mathfrak{D}^{\text{max}}_{2,n}\cong \widehat{\mathfrak{M}}_{0,n+4}$ is non-trivial.

There exists however a nice way to trivialise $\mathfrak{D}^{\text{max}}_{2,n}\cong \widehat{\mathfrak{M}}_{0,n+4}$ using modular invariance: extend the integration domain for the modular parameter $q_3$ to the full complex plane. This method also has the advantage of considerably simplifying the degeneration because it obviates the construction of an explicit map from $\mathfrak{D}^{\text{max}}_{2,n}$ to $\widehat{\mathfrak{M}}_{0,n+4}$. Let us see how this trivialisation works in more detail.

Since we are interested in extending the domain of integration to the full complex plane, the natural modular transformation to consider is $\widetilde{q}_{12}= 1/q_{12}$, where we recall that $q_3=q_{12}=e^{2\pi i \Omega_{12}}$. This suggests that the extension of $q_3$ to the full complex plane is best seen in the  parametrisation \eqref{equ:q_II_def} of the period matrix, which we will use in what follows. Explicitly, the modular transformation $\widetilde{q}_{12}= 1/q_{12}$ is given by
\begin{equation}\label{equ:mod-transform_1/q12}
 a=d=\begin{pmatrix}1 & 0 \\0 & -1\end{pmatrix}\,,\qquad b=c=0\,.
\end{equation}
At the level of the homology cycles, this means
\begin{equation}
  \big(\widetilde{A}_1,\,\widetilde{B}_1\big)=\big(A_1,\,B_1\big)\,,\qquad  \big(\widetilde{A}_2,\,\widetilde{B}_2\big)=\big(-A_2,\,-B_2\big)\,.
\end{equation}
Of course, we could have chosen to reverse the orientations of $(A_1,\,B_1)$  instead, while keeping  $(A_2,\,B_2)$  invariant. The period matrix  transforms under this modular transformation as
\begin{equation}
 \widetilde{\Omega}=\begin{pmatrix}\Omega_{11} & -\Omega_{12} \\ -\Omega_{12} &\Omega_{22}\end{pmatrix}\,,
\end{equation}
confirming that the modular transformation \eqref{equ:mod-transform_1/q12} indeed corresponds to $\widetilde{q}_{12}= 1/q_{12}$. The behaviour of the holomorphic differentials and the zero modes $\ell^I_\mu$  can be read off from \cref{equ:mod_transform_omega} and \cref{equ:mod_transform_ell},
\begin{equation}
  \big(\widetilde{\omega}_1,\,\widetilde{\omega}_2\big)=\big(\omega_1,\,-\omega_2\big)\,,\qquad \big(\widetilde{\ell}_1,\,\widetilde{\ell}_2\big)=\big(\ell_1,\,-\ell_2\big)\,.
\end{equation}
Let us schematically write the two-loop amplitude \eqref{equ:ampl_final} as
\begin{equation}\label{equ:schematic-ampl}
 \mathcal{M}_n \equiv \int_{|q_{12}|< 1}\d\mu_{2,n}\,\mathcal{I}_n =  \int_{|q_{12}|> 1}\d\mu_{2,n}\,\mathcal{I}_n\,,
\end{equation}
where the second equality holds due to modular invariance, as just discussed. The amplitude can thus be expressed as 
\begin{equation}\label{equ:condition_f}
 \mathcal{M}_n = \int\d\mu_{2,n}\,\mathcal{I}_n\, f\big(q_{12}\big)\,,\qquad\text{ where }  f\big(q_{12}\big)+ f\big(q_{12}^{-1}\big)=1\,,
\end{equation}
and where the integration is unconstrained and runs over full complex plane. Of course, there are  many possible choices for $f(q_{12})$ if we only require it to be subject to $f(q_{12})+ f(1/q_{12})=1$. In the context of the residue theorem, however, there is an additional natural requirement: $f(q_{12})$ should not introduce poles into the integrand. Equivalently, we can demand  that the maximal non-separating boundary divisor $\mathfrak{D}^{\text{max}}_{2,n}$ remains the only global residue (apart from the one that defines the original amplitude). The simplest example $f(q_{12})=1/2$  fails this additional requirement, because it introduces a pole as $q_{12}\rightarrow\infty$. In fact, requiring the integrand to remain holomorphic on the support of the scattering equations implies uniquely that
\begin{equation}\label{equ:choice_f}
  f\big(q_{12}\big) = \frac{1}{1-q_{12}}\,.
\end{equation}
While this naively introduces a pole at $q_{12}=1$, the amplitudes of type II supergravity \eqref{equ:ampl_final} vanish on the separating degeneration. This can be understood intuitively in analogy with  the superstring, where only massive poles contribute to this channel \cite{DHoker:2005jhf}, which are absent in the ambitwistor string. An explicit proof for four particles  was  given in \cite{Adamo:2015hoa}, and we  extend this argument to all $n$ in \cref{sec:f}.

Having extended the domain of integration to trivialise the map to the bi-nodal sphere, let us return  to the parametrisation \eqref{equ:paramet_Omega} of the period matrix. With $q_{12}=q_3$,  the amplitude is given by 
\begin{equation}\label{equ:ampl_schematic}
   \mathcal{M}_n = \int\d\mu_{2,n}\,\mathcal{I}_n\,\frac{1}{1-q_{3}}\,.
\end{equation}
Since no new poles are introduced by $f(q_{12})=f(q_3)$,  the amplitude\footnote{Here, $\mathcal{I}_n$ is chosen in the representation established in the last two sections.} localises on the maximal non-separating divisor $\mathfrak{D}^{\text{max}}_{2,n,}$ after applying the global residue theorem. At this stage, the isomorphism $\mathfrak{D}^{\text{max}}_{2,n}\cong \widehat{\mathfrak{M}}_{0,n+4}$ is  trivial, and $\tau_3$ is determined by  Fay's degeneration formula \eqref{equ:Fay_period_matrix}.

\section{The amplitude on the bi-nodal Riemann sphere}\label{sec:contour_argument}
Let us briefly summarise our conclusions from \cref{sec:nodalRS}. In \cref{sec:map_to_RS}, we established that we can trivialise the isomorphism $\mathfrak{D}^{\text{max}}_{2,n}\cong\widehat{\mathfrak{M}}_{0,n+4}$ by extending the domain of integration for the modular parameter $q_3$ from the fundamental domain to the full complex plane, using modular invariance. This introduces a factor of $f(q_3)=(1-q_3)^{-1}$ into the integrand. We will work with the representation given in \cref{equ:chiral-int_final} of the integrand, as well as the basis \eqref{equ:SE_final_choice} for the scattering equations, 
\begin{equation}
  \mathcal{M}_n = \int\d\mu_{2,n}\mathcal{I}_n\,\frac{1}{1-q_{3}}\equiv \mathfrak{R}(u_{1},u_{2},u_{3})\,.
\end{equation}
In a slight abuse of notation, we denote the amplitude again by $\mathfrak{R}(u_{1},u_{2},u_{3})$, which denotes the residue at $u_r=0$. Since the only simple poles of the integrand are the scattering equations and  $q_r=0$,\footnote{In \cref{sec:nodalRS_integrand} and \cref{sec:map_to_RS}, we have seen that these are indeed the only poles. See \cref{sec:int_nodalRS} for the calculation showing that the poles $q_r=0$ are  simple.} the amplitude can be localised on the non-separating degenerations $\mathfrak{D}^{\text{non-sep}}_{2,n}\cong\widehat{\mathfrak{M}}_{1,n+2}$ using the residue theorem. This residue theorem moves the contour from  one of the scattering equations, e.g. $u_2$,\footnote{Note that, in the application of the residue theorem, nothing forced us to relax the scattering equation $u_2$. An equivalent result would have been obtained if we chose to relax $u_1$ or $u_3$ instead.}  to circle the poles at $q_r=0$. With the choice \eqref{equ:SE_final_choice} for $u_r$,
\begin{equation}
 u_1 = u_{11}\,,\qquad u_2 = u_{22}\,,\qquad u_3 = u_{12}+u_{11}+u_{22}\,,
\end{equation}
only one pole contributes to the residue theorem since 
\begin{equation}
 \mathfrak{R}(u_1,q_1,u_{3})=\mathfrak{R}(u_1,q_3,u_{3})=0	\,.
\end{equation}
The residue theorem thus results in a single contribution $\mathfrak{D}^{\text{non-sep}}_{2,n}\cong\widehat{\mathfrak{M}}_{1,n+2}$ on a nodal torus, 
\begin{equation}
 \mathfrak{R}(u_{1},u_{2},u_{3})=-\mathfrak{R}(u_1,q_{2},u_{3})\,.
\end{equation}
Using the representation \eqref{equ:chiral-int_final} of the integrand, the only simple pole on the torus (apart from the one where the amplitude is defined) sits at the non-separating boundary divisor $\mathfrak{D}_{1,n+2}^{\text{non-sep}}\cong\widehat{\mathfrak{M}}_{0,n+4}$. Therefore, a second application of the residue theorem localises the amplitude onto the nodal Riemann sphere $\mathfrak{D}^{\text{max}}_{2,n}$,
\begin{equation}
 \mathfrak{R}(u_{1},u_{2},u_{3})=-\mathfrak{R}(u_1,q_{2},u_{3}) = \mathfrak{R}(q_1,q_2,u_{3})\,,
\end{equation}
again using $\mathfrak{R}(u_2,q_2,u_{3})=\mathfrak{R}(q_3,q_2,u_{3})=0$. Since we trivialised the isomorphism $\mathfrak{D}^{\text{max}}_{2,n}\cong\widehat{\mathfrak{M}}_{0,n+4}$ by extending the domain of integration for $q_3$, the resulting amplitude is formulated directly over the nodal Riemann sphere.

\subsection{The measure on the bi-nodal Riemann sphere} 
\begin{figure}[ht]
	\centering 
	  \includegraphics[width=7cm]{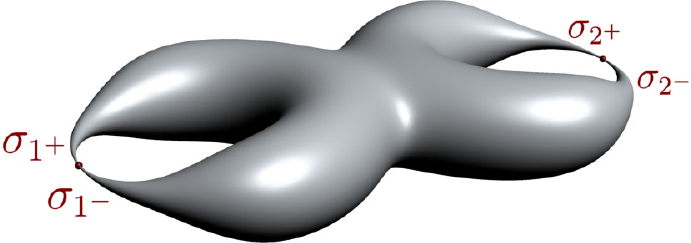}
	\caption{The bi-nodal Riemann sphere, with nodes parametrised by $\sigma_{1^\pm}$ and $\sigma_{2^\pm}$. }
	\label{fig:bindalRS_labelled}
\end{figure}

Let us introduce the coordinate $\sigma \in\mathbb{CP}^1$ on the (bi-nodal) Riemann sphere, to distinguish it from the previous coordinate $z$ at higher genus. Then the $\sigma_i$ denote the locations of the $n$ marked points associated to the external particles, while  $\sigma_{1^\pm}$ and $\sigma_{2^\pm}$ denote the location of the nodes; see  figure \ref{fig:bindalRS_labelled} for illustration. 

\noindent In this notation, Fay's degeneration formula for the holomorphic differentials \eqref{equ:Fay-hol-g} gives the following asymptotics to order $\mathcal{O}(q_1^2,q_2^2)$,
\begin{subequations}\label{equ:hol_nodalRS}
\begin{align}
 \omega_1(\sigma) &= \frac1{2\pi i}\, \omega_{1^+\!,1^-}(\sigma)=\frac1{2\pi i}\,\frac{(\sigma_{1^+}-\sigma_{1^-})}{(\sigma-\sigma_{1^+})(\sigma-\sigma_{1^-})}\,\d \sigma\,,\\
  \omega_2(\sigma) &=  \frac1{2\pi i}\,\omega_{2^+\!,2^-}(\sigma)=\frac1{2\pi i}\,\frac{(\sigma_{2^+}-\sigma_{2^-})}{(\sigma-\sigma_{2^+})(\sigma-\sigma_{2^-})}\,\d \sigma\,.
\end{align}
\end{subequations}
Since the isomorphism $\mathfrak{D}^{\text{max}}_{2,n}\cong\widehat{\mathfrak{M}}_{0,n+4}$ is trivial after extending the domain of integration for $q_3$,  Fay's degeneration formula for the period matrix  \eqref{equ:Fay_period_matrix_g} straighforwardly determines the asymptotics for the remaining modular parameter,
\begin{equation}
  \Omega_{12}=\oint_{B_2}\omega_1 = \frac{1}{2\pi i}\int_{\sigma_{2^-}}^{\sigma_{2^+}}\hspace{-5pt}\frac{\d \sigma\,(\sigma_{1^+}-\sigma_{1^-})}{(\sigma-\sigma_{1^+})(\sigma-\sigma_{1^-})}=\frac{1}{2\pi i}\ln \frac{(1^+2^+)\,(1^-2^-)}{(1^+2^-)\,(1^-2^+)}\,,
\end{equation}
where we introduced the notation $(ij)\equiv\sigma_i-\sigma_j$. The exponentiated parameter $q_3=e^{2\pi i \Omega_{12}}$ thus becomes the cross-ratio of the location of the nodes,
\begin{equation}\label{equ:q3}
 q_3 = \frac{(1^+2^+)\,(1^-2^-)}{(1^+2^-)\,(1^-2^+)}\,.
\end{equation}
The M\"obius symmetry of the Riemann sphere allows us to fix three of the marked points. For practical calculations, a convenient choice is given by the gauge $\sigma_{1^-}=1$, $\sigma_{2^+}=0$ and $\sigma_{2^-}=\infty$, leading to $q_3=\sigma_{1^+}$ encoding the location of the remaining node. To arrive at an SL$(2,\mathbb{C})$-invariant representation of the amplitude, however, we will not choose a specific gauge and instead quotient by the volume of the symmetry group,
\begin{equation}
 \frac{\d q_3}{q_3} = \frac{J}{\text{vol SL}(2,\mathbb{C})}\,,\quad\qquad\text{where } \,\, J =\frac{\d\sigma_{1^+}\d\sigma_{1^-}\d\sigma_{2^+}\d\sigma_{2^-}}{(1^+2^+)(1^+2^-)(1^-2^+)(1^-2^-)}\,.
\end{equation}
Using the asymptotics for the holomorphic differentials and the moduli, we can now  take a closer look at the remaining ingredients for the amplitude.  The expression \eqref{equ:q3} for $q_3$ leads directly to
\begin{equation}
 f(q_3) = \frac{1}{1-q_3} = \frac{(1^+2^-)(1^-2^+)}{(1^+1^-)(2^+2^-)}\,.
\end{equation}
Moreover, we already  established the degeneration of the two scattering equations relaxed by the residue theorems  in \cref{sec:nodalRS_SE}: the limit $q_1, q_2\rightarrow 0$ forces $u_1=\ell_1^2$ and $u_2=\ell_2^2$. Therefore, the  SL$(2,\mathbb{C})$-invariant representation of the amplitude on the bi-nodal Riemann sphere is given by
\begin{equation}
 \mathcal{M}_n=\int \!\frac{\d^{10}\ell_1\,\d^{10}\ell_2}{\ell_1^2\ell_2^2} \int  \hspace{-15pt}{\phantom{\Bigg(}}_{\mathfrak{M}_{0,n+4}}\hspace{-10pt}
\frac{1}{\text{vol SL}(2,\mathbb{C})}\,
\bar\delta\big(u_3\big)\,\prod_{i=1}^n\bar\delta\big(\mathcal{E}_i)\,\,\,J\,\mathcal{I}^{\text{chi}}_n\,\widetilde{\mathcal{I}}^{\text{chi}}_n\, \frac{(1^+2^-)(1^-2^+)}{(1^+1^-)(2^+2^-)}\,,
\end{equation}
where $\sigma_A\in\{\sigma_{1^\pm},\sigma_{2^\pm},\sigma_i\}$. In this expression, the scattering equations $u_3$ and $\mathcal{E}_i\equiv\big\langle \mu_i\,P^2\big\rangle$ as well as the chiral integrands $\mathcal{I}^{\text{chi}}_n$ are evaluated implicitly on the maximal non-separating divisor $q_1=q_2=0$. We will derive the explicit form of the scattering equations in the next section, and discuss the asymptotics of the chiral integrand in \cref{sec:int_nodalRS}. Notice also that the pre-factor $(\ell_1^2\ell_2^2)^{-1}$ of the loop integrand arises from the poles  that were relaxed in the residue theorem, $(u_1u_2)^{-1}$; see also \cref{equ:u1u2binodal} below.

\subsection{The scattering equations on the bi-nodal Riemann sphere} \label{sec:SE_on_nodalRS}
Let us first focus on the asymptotics of the scattering equations on the bi-nodal Riemann sphere. From the degeneration of the holomorphic differentials \eqref{equ:hol_nodalRS}, we obtain  $P_\mu$ as\footnote{To avoid the proliferation of $2\pi i$ factors in the pairing $\ell^I\omega_I$, we perform the redefinition $\ell^I_{\mu}\to 2\pi i\, \ell^I_{\mu}$, which cancels the $(2\pi i)^{-1}$ factors in \eqref{equ:hol_nodalRS}. We also redefine the normalisation of the loop integrand so that no such factor remains.}
\begin{equation}
 P_\mu(\sigma)=\ell_{1\,\mu}\,\omega_{1^+\!,1^-}(\sigma)+\ell_{2\,\mu}\,\omega_{2^+\!,2^-}(\sigma)+\sum_i k_{i\,\mu}\,\omega_{i,*}(\sigma)\,,
\end{equation}
where $\omega_{i,*}$ are the meromorphic differentials on the Riemann sphere. The form of $P_\mu$ strongly resembles the forward limit of the tree-level solution $P_\mu^{(0)}$, but with $n+4$ legs, two of which have been identified pairwise with equal-opposite loop momenta $\ell_{I\,\mu}$. Notice that,  as expected from a forward limit, the loop momenta are off-shell, $\ell_I^2\neq 0$.

The particle scattering equations $\mathcal{E}_i\equiv\big\langle \mu_i\,P^2\big\rangle$, calculated as the residue of $P^2$ at the vertex operator insertions $\sigma_i$, degenerate straightforwardly to the bi-nodal sphere. Moreover,   $u_1=u_{11}$ and $u_2=u_{22}$ are given respectively by the coefficients of the quadratic differentials $\omega_{1^+\!,1^-}^2$ and $\omega_{2^+\!,2^-}^2$ in $P^2$. As discussed in \cref{sec:nodalRS_SE}, this simply implies that
\begin{equation}
\label{equ:u1u2binodal}
 u_1=\text{Res}_{\sigma_{1^+}}(\sigma-\sigma_{1^+})P^2(\sigma)=\ell_1^2\,,\qquad\qquad u_2=\text{Res}_{\sigma_{2^+}}(\sigma-\sigma_{2^+})P^2(\sigma)=\ell_2^2\,.
\end{equation}
Following the discussion in \cref{sec:nodalRS_SE}, we choose the remaining scattering equation to take the form  $u_3=u_{11}+u_{22}+u_{12}=\ell_1^2+\ell_2^2+u_{12}$, so only $u_{12}$ remains to be determined on the nodal Riemann sphere. A convenient way to do so is to calculate the  residue at $\sigma_{1^+}$ of the quadratic differential $P^2-\ell_1^2\omega_{1^+\!,1^-}^2\!-\ell_2^2\omega_{2^+\!,2^-}^2$,
\begin{equation}\label{equ:degen_u12}
 u_{12}\,\,\omega_2(\sigma_{1^+}) = \text{Res}_{\sigma_{1^+}}\left(P^2-\ell_1^2\omega_{1^+\!,1^-}^2\!-\ell_2^2\omega_{2^+\!,2^-}^2\right)=2\ell_1\cdot\ell_2\,\, \omega_{2^+\!,2^-}(\sigma_{1^\pm})+\sum_{j} 2\ell_1\cdot k_j\,\,\omega_{j,*}(\sigma_{1^\pm})\,.
\end{equation}
Of course, we could have chosen alternatively to calculate $u_{12}$ as the residue at any of the other three nodal points, consistent with the SL$(2,\mathbb{C})$ M\"obius symmetry of the Riemann sphere. The full set of scattering equations $\{u_3,\mathcal{E}_i\}$ can thus be expressed in a manifestly  SL$(2,\mathbb{C})$-invariant form,
\begin{subequations}\label{equ:SE_nodalRS_v1}
\begin{align}
 \mathcal{E}_i&=k_i\cdot\ell_1\,\omega_{1^+\!,1^-}(\sigma_i)+k_i\cdot\ell_2\,\omega_{2^+\!,2^-}(\sigma_i)+\sum_{j\neq i} k_i\cdot k_j \,\omega_{j,*}(\sigma_i)\,,\\
\pm \mathcal{E}_{1^\pm}&=\frac{1}{2}\left(\ell_1+\ell_2\right)^2 \omega_{2^+\!,2^-}(\sigma_{1^\pm})+\sum_{j} \ell_1\cdot k_j\,\omega_{j,*}(\sigma_{1^\pm})\,,\\
\pm \mathcal{E}_{2^\pm}&=\frac{1}{2}\left(\ell_1+\ell_2\right)^2\omega_{1^+\!,1^-}(\sigma_{2^\pm})+\sum_{j} \ell_2\cdot k_j\,\omega_{j,*}(\sigma_{2^\pm})\,.
\end{align}
\end{subequations}
We recover $u_3$, dressed by a factor of $\omega_{I^+,I^-}(\sigma_{J^\pm})$, upon fixing the SL$(2,\mathbb{C})$ symmetry.\footnote{For example when gauge fixing the constraints $\mathcal{E}_{1^-}$, $\mathcal{E}_{2^+}$ and $\mathcal{E}_{2^-}$, the remaining scattering equation  $\mathcal{E}_{1^+}$ becomes $u_3\,\omega_2(\sigma_{1^+})$, while the integrand picks up the usual SL$(2,\mathbb{C})$ Fadeev-Popov factor $(1^-2^+)(2^+2^-)(2^-1^-)$.}
The scattering equations \eqref{equ:SE_nodalRS_v1} can be written more compactly by introducing an auxiliary quadratic differential
\begin{equation}
 \mathfrak{P}_2(\sigma)=P^2(\sigma)-\ell_1^2\,\omega_{1^+\!,1^-}^2(\sigma)-\ell_2^2\,\omega_{2^+\!,2^-}^2(\sigma) +\left(\ell_1^2+\ell_2^2\right)\omega_{1^+\!,1^-}(\sigma)\,\omega_{2^+\!,2^-}(\sigma)\,.
\end{equation}
Both the nodal and the particle scatttering equations   are then given by the residues of $\mathfrak{P}_2$ at the marked points,
\begin{equation}\label{equ:SE_nodalRS_v2}
 \mathcal{E}_A = \text{Res}_{\sigma_A}\mathfrak{P}_2(\sigma)\,,\qquad\qquad\text{for }\sigma_A\in\{\sigma_{1^\pm},\sigma_{2^\pm},\sigma_i\}\,.
\end{equation}
The three linear relations among these $n+4$ scattering equations -- encoding the M\"obius invariance of the Riemann sphere --  are given by
\begin{equation}
  \sum_A \sigma_A^q \mathcal{E}_A=0\,,\qquad\qquad\text{for }q=0,1,2\,,
\end{equation}
in this notation. Both the form of these relations and the construction of the scattering equations \eqref{equ:SE_nodalRS_v2}  are strongly reminiscent of the tree-level case, where the SL$(2,\mathbb{C})$-invariant form of the scattering equations has been studied in \cite{Dolan:2013isa}. Note, however, that while the scattering equations \eqref{equ:SE_nodalRS_v2} bear a close structural resemblance  with the tree-level scattering equations $\mathcal{E}_i=\text{Res}_{\sigma_i}P^2$ in the forward limit, the defining quadratic differential has to be modified from $P^2$ (whose vanishing we relaxed in the residue theorem) to $\mathfrak{P}_2$ at two loops. A nice interpretation of the analogous feature at one loop was given recently in \cite{Roehrig:2017gbt}: the full amplitude can be constructed directly from the Riemann sphere by introducing a `gluing operator' that effectively creates the node. BRST invariance requires this operator to contain a non-local term  compensating for the off-shell state running through the node, which in turn leads to an effective BRST operator $Q\supset\oint\frac{\tilde{c}}{2}\big(P^2-\ell^2\omega_{+,-}^2\big)\equiv\oint\frac{\tilde{c}}{2}\,\mathfrak{P}_2^{(1)}$. It would be interesting to give a similar interpretation  to the quadratic differential $\mathfrak{P}_2$ at two loops.

Using the M\"obius-invariant form of the scattering equations introduces an additional factor of the Jacobian $J$ into the amplitude. The full two-loop  integrand $\mathfrak{I}_n$ is then given by the CHY-type formula, 
\begin{equation}
 \mathfrak{I}_n=
 \int  \hspace{-15pt}{\phantom{\Bigg(}}_{\mathfrak{M}_{0,n+4}}\hspace{-10pt}
\frac{1}{\text{vol SL}(2,\mathbb{C})^2}\,
\,\,\prod_{A}\,\bar\delta\big( \mathcal{E}_A\big)\,\,\mathcal{I}^{(2)}_n\,\widetilde{\mathcal{I}}^{(2)}_n\, \frac{(1^+2^-)(1^-2^+)}{(1^+1^-)(2^+2^-)}\,,
 \label{equ:integrandsugra}
\end{equation}
The additional quotient by SL$(2,\mathbb{C})$ refers to the choice of  scattering equations, and leads to the usual Fadeev-Popov factor. Moreover, we rescaled the chiral integrands $\mathcal{I}_n^{\text{chi}}$ by a factor of the Jacobian
\begin{equation}
 \mathcal{I}_n^{(2)}=J\,\mathcal{I}_n^{\text{chi}}=\frac{\d\sigma_{1^+}\d\sigma_{1^-}\d\sigma_{2^+}\d\sigma_{2^-}}{(1^+2^+)(1^+2^-)(1^-2^+)(1^-2^-)}\,\mathcal{I}_n^{\text{chi}}\,,
\end{equation}
and analogously $\widetilde{\mathcal{I}}_n^{(2)}=J\,\widetilde{\mathcal{I}}_n^{\text{chi}}$. The new integrand factor $\mathcal{I}_n^{(2)}$ defined in this manner has form degree one and vanishing SL$(2,\mathbb{C})$-weight in each of the marked points, including the nodes $\sigma_{1^+}$, $\sigma_{1^-}$, $\sigma_{2^+}$ and $\sigma_{2^-}$.
The full expression is thus manifestly invariant under both the SL$(2,\mathbb{C})$ fixing any three of the marked points, and the SL$(2,\mathbb{C})$ associated to the choice of $n+1$ out of the $n+4$ scattering equations. Finally, notice that the product of $\bar\delta$ delta functions has $(1,0)$-form degree $-1$ and $(0,1)$-form degree $1$ in each of the punctures, since $ \mathcal{E}_A = \text{Res}_{\sigma_A}\mathfrak{P}_2(\sigma)$ has $(1,0)$-form degree 1. The total form degree of the expression under the integral is therefore of the appropriate type for the integration over the moduli space: a $(1,1)$ form in each of the punctures.\footnote{We have abused notation slightly at several points regarding the form degrees, so here we just want to clarify that the formula is consistent.}

The formula above, with $\mathcal{I}^{(2)}_n$ further simplified as below in \eqref{equ:int_nodal_final}, is the main result of this paper for supergravity.

\subsection{The chiral integrand on the bi-nodal Riemann sphere} \label{sec:int_nodalRS}
The last missing ingredient is the chiral integrand on the maximal non-separating boundary divisor $\mathfrak{D}^{\text{max}}_{2,n}$. In particular, we are interested in the asymptotics of the Szeg\H{o} kernels as well as the partition functions $\mathcal{Z}^{\text{chi}}[\delta]$ around $q_1=q_2=0$. 

By the definition \eqref{equ:szegodef}, the behaviour of the Szeg\H{o} kernels near the boundary divisor depends on the theta functions and the prime form $E(z,w)$. While the degeneration of the theta functions can be obtained straightforwardly from \cref{equ:theta_expansion}, the expansion of the prime form needs more care.  First note that Fay's degeneration formula \eqref{equ:Fay-hol-g} ensures that the subleading $\mathcal{O}(q_1,q_2)$ contribution  to the holomorphic differentials $\omega_I$ vanishes, so subleading terms  in the prime form $E$ can only originate from the theta functions. A short calculation shows that these  terms cancel, and so\footnote{We use the standard convention that $\mathcal{O}(\epsilon)$ denotes a contribution at order $\epsilon$, whereas $o(\epsilon)$ denotes a contribution at order strictly lower than $\epsilon$.}
\begin{equation}\label{equ:degen_prime-form}
 E(z,w) = \frac{z-w}{\sqrt{dz}\sqrt{dw}} +\,o(q_1,q_2)\,.
\end{equation}
Therefore, the subleading asymptotics of the Szeg\H{o} kernels depend only on the behaviour of the theta function near the non-separating boundary divisor. With the expansion  \eqref{equ:theta_expansion} of the theta function, the Szeg\H{o} kernels can be grouped into NS-NS, NS-R, R-NS and R-R  Szeg\H{o} kernels  as follows:
\begin{subequations}\label{equ:Szego-expansion}
\begin{align}
  S_{\delta}(z,w) &= \!\!\!\sum_{n_1,n_2\in\{0,1\}}\!\! (-1)^{2(n_1 \delta''_1+n_2\delta''_2)}\,q_1^{n_1}q_2^{n_2}\,S^{(n_1,n_2)}_{\text{NS}}(z,w) &&\delta\in\{\delta_1,\,\delta_2,\,\delta_3,,\delta_4\}\,,\\
  S_{\delta}(z,w) &= \sum_{n_1\in\{0,1\}}\, (-1)^{2n_1 \delta''_1}\,q_1^{n_1}\,S^{(n_1,n_2)}_{\text{R}2}(z,w)&&\delta\in\{\delta_5,\,\delta_6\}\,,\\
   S_{\delta}(z,w) &= \sum_{n_2\in\{0,1\}}\, (-1)^{2n_2\delta''_2}\,q_2^{n_2}\,S^{(n_1,n_2)}_{\text{R}1}(z,w) &&\delta\in\{\delta_7,\,\delta_8\}\,,\\
 S_{\delta}(z,w) &= S^{(0,0)}_{\text{RR}_i}(z,w)  &&\delta\in\{\delta_9,\,\delta_0\}\,,\,\,i=9,0\,.
\end{align}
\end{subequations}
All expansions are given to order $o(q_1q_2)$, which suffices for our purposes, and the notation for the Szeg\H{o} kernels is chosen to reflect the sector of the spin structure according to \eqref{eq:R-NS}. The explicit form of the respective orders can be found in \cref{sec:Szego}. Below, we will primarily make use of  the relative signs in the expansion \eqref{equ:Szego-expansion}. The other ingredient in the integrand are the partition functions, whose  asymptotics  are completely determined by the degeneration of the prime form and the theta functions. To order $o(1)$ in the degeneration parameters, we find 
\begin{subequations}\label{equ:partition-expansion}
\begin{align}
 \mathcal{Z}^{\text{chi}}[\delta]&= \!\!\!\sum_{n_1,n_2\in\{0,1\}} \!\!(-1)^{2(n_1 \delta''_1+n_2\delta''_2)}\,q_1^{-n_1}q_2^{-n_2}\,\mathcal{Z}^{(-n_1,-n_2)}_{\text{NS}} &&\delta\in\{\delta_1,\,\delta_2,\,\delta_3,,\delta_4\}\,,\\
  \mathcal{Z}^{\text{chi}}[\delta] &= \sum_{n_1\in\{0,1\}}\, (-1)^{2n_1 \delta''_1}\,q_1^{-n_1}\,\mathcal{Z}^{(-n_1,0)}_{\text{R}2}&&\delta\in\{\delta_5,\,\delta_6\}\,,\\
 \mathcal{Z}^{\text{chi}}[\delta] &=\sum_{n_2\in\{0,1\}}\, (-1)^{2n_2\delta''_2}\,q_2^{-n_2}\,\mathcal{Z}^{(0,-n_2)}_{\text{R}1} &&\delta\in\{\delta_7,\,\delta_8\}\,,\\
 \mathcal{Z}^{\text{chi}}[\delta] &= \mathcal{Z}_{\text{RR}_i}^{(0,0)}&&\delta\in\{\delta_9,\,\delta_0\}\,,\,\,i=9,0\,.
\end{align}
\end{subequations}
All details and the explicit form of the NS-NS, NS-R, R-NS and R-R partition functions can be found in \cref{sec:partition}. Note in particular that the highest pole in the partition functions is of order $q_1^{-1}q_2^{-1}$, so it is indeed sufficient to expand  \eqref{equ:Szego-expansion} only to order $o(q_1q_2)$.  With the asymptotics  \eqref{equ:Szego-expansion} and \eqref{equ:partition-expansion}, we can  proceed to study the  chiral integrand on the bi-nodal Riemann sphere,
\begin{equation}
 \mathcal{I}^{(2)}_n=J\,\sum_{\delta}  \mathcal{Z}^{\text{chi}}[\delta]\,\pf\Big(M_\delta^{(2)}\Big)\Bigg|_{\substack{q_1\rightarrow 0\\ q_2\rightarrow 0}}\,.
\end{equation}
Let us first check explicitly that all terms of order $q_1^{-1}$, $q_2^{-1}$ and $q_1^{-1}q_2^{-1}$ cancel.\footnote{Recall that we assumed this in the degeneration to the nodal Riemann sphere; if not, $q_1=0$ and $q_2=0$ would not have been simple poles.} As a first check, this is easily verified for four external particles from the degeneration of the relations in \cref{sec:identities}. To generalise this cancellation to $n$ points, we will rely on the relative signs in the partition functions and the Szeg\H{o} kernels for the different spin structures. Explicitly, we find that the contribution to the chiral integrand at order $q_1^{-1}q_2^{-1}$ vanishes due to the sign differences in the NS-NS spin structures,
\begin{align}\label{equ:chiral_oq1q2-1}
 \mathcal{I}^{(2)}_n\Big|_{q_1^{-1}q_2^{-1}} &= J\,\underbrace{\sum_{i=1}^4\,(-1)^{2(\delta''_{i,1}+\delta''_{i,2})}}_{=1-1-1+1} \mathcal{Z}^{(-1,-1)}_{\text{NS}} \,\left( \pf\big(M_{\text{NS}}\big)\big|_{q_1^0q_2^0}\right)=0\,.
\end{align}
To improve readability, we introduced the notation $ \pf\big(M_{\text{NS}}\big)\big|_{q_1^aq_2^b}$ to indicate that the matrix $M$ is defined using the NS-NS Szeg\H{o} kernels introduced in \eqref{equ:Szego-expansion}, and evaluated at order $q_1^aq_2^b$ for $a,b=0,1$.\footnote{
That is, to order $o(q_1q_2)$, we have \;$
\pf\Big(M_\delta^{(2)}\Big)=\sum_{n_1,n_2\geq0} \,(-1)^{2(n_1 \delta''_1+n_2\delta''_2)}\,q_1^{-n_1}q_2^{-n_2}\,\pf\big(M_{\text{NS}}\big)\big|_{q_1^0q_2^0}
$ \; for $\delta\in\{\delta_1,\,\delta_2,\,\delta_3,,\delta_4\}$, with $M_{\text{NS}}=M_\delta^{(2)}(S_\delta\to S_{\text{NS}})$ and $S_{\text{NS}}\equiv \sum_{n_1,n_2\geq0}\,q_1^{n_1}q_2^{n_2}\,S^{(n_1,n_2)}_{\text{NS}}$.}
In particular, all signs due to different spin structures have been extracted, and only contribute an  overall factor -- which of course vanishes in \eqref{equ:chiral_oq1q2-1}.
The calculation for the chiral integrand at order $q_1^{-1}$ then proceeds in complete analogy, 
\begin{align}
  \mathcal{I}^{(2)}_n\Big|_{q_1^{-1}q_2^{0}}&=J\, \underbrace{\sum_{i=1}^4\,(-1)^{2 \delta''_{i,2}}}_{=1-1+1-1} \mathcal{Z}^{(-1,-1)}_{\text{NS}} \,\left( \pf\big(M_{\text{NS}}\big)\big|_{q_1^1q_2^0}\right)+J\,\underbrace{\sum_{i=1}^4\,(-1)^{2 \delta''_{i,1}}}_{=1+1-1-1} \mathcal{Z}^{(-1,0)}_{\text{NS}} \,\left( \pf\big(M_{\text{NS}}\big)\big|_{q_1^0q_2^0}\right)\nonumber\\
  &\qquad+ J\,  \underbrace{\sum_{i=5}^6\,(-1)^{2\delta''_{i,1}}}_{=1-1} \mathcal{Z}^{(-1,0)}_{\text{R}2} \,\left( \pf\big(M_{\text{R}2}\big)\big|_{q_1^0q_2^0}\right)=0\,.
\end{align}
The same argument holds for $ \mathcal{I}^{\text{chi}}_n\big|_{q_1^{0}q_2^{-1}}$, with the R1 spin structures contributing instead of  R2. The leading contribution of the chiral integrand is thus of order one, proving the assertion  that $q_1=0$ and $q_2=0$ are indeed simple poles of the integrand.\footnote{While we do not need the equivalent statement for $q_3$ here, note that it can be proven by the same methods after exchanging the roles of $\tau_1$ and $\tau_3$ using a modular transformation. }\\

\noindent
The  full chiral integrand is therefore given by the $\mathcal{O}(1)$ contribution on the bi-nodal sphere,
\begin{equation}\label{equ:int_nodal_final}
 \mathcal{I}_n^{(2)}=\mathcal{I}_n^{\text{NS}}+\mathcal{I}_n^{\text{R}2}+\mathcal{I}_n^{\text{R}1}+\mathcal{I}_n^{\text{RR}}\,,
\end{equation}
where we defined NS-NS, NS-R, R-NS and R-R  integrands
\begin{subequations}\label{equ:integrand_NS-R}
\begin{align}
  \mathcal{I}_n^{\text{NS}}&=4J\,\sum_{n_1,n_2\in\{0,1\}}\,\mathcal{Z}_{\text{NS}}^{(-n_1,-n_2)}\, \pf\big(M_{\text{NS}}\big)\big|_{q_1^{n_1}q_2^{n_2}}\,,\\
  \mathcal{I}_n^{\text{R}2}&=2J\,\Bigg(\mathcal{Z}_{\text{R}2}^{(0,0)}\, \pf\big(M_{\text{R}2}\big)\big|_{q_1^0q_2^0}+\mathcal{Z}_{\text{R}2}^{(-1,0)}\, \pf\big(M_{\text{R}2}\big)\big|_{q_1^1q_2^0}\Bigg)\,,\\
  \mathcal{I}_n^{\text{R}1}&=2J\,\Bigg(\mathcal{Z}_{\text{R}1}^{(0,0)}\, \pf\big(M_{\text{R}1}\big)\big|_{q_1^0q_2^0}+\mathcal{Z}_{\text{R}1}^{(0,-1)}\, \pf\big(M_{\text{R}1}\big)\big|_{q_1^0q_2^1}\Bigg)\,,\\
  \mathcal{I}_n^{\text{RR}}&=J\,\mathcal{Z}_{\text{RR}_9}^{(0,0)}\, \pf\big(M_{\text{RR}_9}\big)\big|_{q_1^0q_2^0}+J\,\mathcal{Z}_{\text{RR}_0}^{(0,0)}\, \pf\big(M_{\text{RR}_0}\big)\big|_{q_1^0q_2^0}\,.
\end{align}
\end{subequations}
Just as  on the genus two Riemann surface, ten different terms contribute to the amplitude, as can be easily seen from the expansions \eqref{equ:integrand_NS-R}. However, these terms are not aligned with the spin structures any more, but rather reflect the sector as well as the  asymptotics of the degeneration. The  $(2n+2)\times(2n+2)$ matrices $M$ are defined as before by
\begin{subequations}
\begin{align}
 & &&M^{(2)}_\text{S}=\begin{pmatrix}A &-C^T\\C&B\end{pmatrix}\,,&&\\
 &A_{x_1x_2}=\px(x_1,x_2) S_\text{S}(x_1,x_2)\,,&& A_{x_\alpha,j}=P(x_\alpha)\cdot k_j S_\text{S}(x_\alpha,z_j)\,,&& A_{ij}=k_i\cdot k_j S_\text{S}(z_i,z_j)\,,\\
 & && C_{x_\alpha,j}=P(x_\alpha)\cdot \epsilon_j S_\text{S}(x_\alpha,z_j)\,,&& C_{ij}=\epsilon_i\cdot k_j S_\text{S}(z_i,z_j)\,,\\
 & && C_{ii}=P(\sigma_i)\cdot\epsilon_i\,, && B_{ij}=\epsilon_i\cdot\epsilon_j  S_\text{S}(z_i,z_j)\,,
\end{align}
\end{subequations}
where S $\in\{$NS, R1, R2, RR$\}$ denotes the types of states propagating through the nodes. All expressions for the partition functions and the Szeg\H{o} kernels can be found in \cref{sec:degen}. To find the asymptotics of $\px(x_1,x_2)$ on the bi-nodal Riemann sphere, note that the holomorphic differential $\varpi$ degenerates to
 \begin{align}
  \varpi(\sigma) &= \sqrt{\frac{\omega_{1^+\!,1^-}(x_1)}{\omega_{2^+\!,2^-}(x_1)}}\,\,\omega_{2^+\!,2^-}(\sigma)-\sqrt{\frac{\omega_{2^+\!,2^-}(x_1)}{\omega_{1^+\!,1^-}(x_1)}}\,\,\omega_{1^+\!,1^-}(\sigma)\\
  & =\sqrt{\frac{\omega_{1^+\!,1^-}(x_2)}{\omega_{2^+\!,2^-}(x_2)}}\,\,\omega_{2^+\!,2^-}(\sigma)-\sqrt{\frac{\omega_{2^+\!,2^-}(x_2)}{\omega_{1^+\!,1^-}(x_2)}}\,\,\omega_{1^+\!,1^-}(\sigma)\,,\nonumber
 \end{align}
and thus, from the definition $\varpi(\sigma) = c_\alpha\Delta(x_\alpha,\sigma)$, 
\begin{equation}
 c_\alpha = \sqrt{\frac{(x_\alpha 1^+)(x_\alpha 1^-)(x_\alpha 2^+)(x_\alpha 2^-)}{(1^+1^-)(2^+2^-)}}\, \frac1{\d x_\alpha}= \big(\omega_{1^+,\!1^-}(x_\alpha)\,\omega_{2^+\!,2^-}(x_\alpha)\big)^{-1/2}\,.
\end{equation}
In particular, the expression for $\varpi$ and $c_\alpha$ is only valid for $x_1$ and $x_2$ related by
\begin{equation}\label{equ:relx1x2RS}
 \omega_{1^+\!,1^-}(x_1)\,\omega_{2^+\!,2^-}(x_2) = \omega_{1^+\!,1^-}(x_2)\,\omega_{2^+\!,2^-}(x_1)\,,
\end{equation}
as can be seen from \cref{cw1=cw2}. The two identities for $\varpi$ are related straightforwardly by \eqref{equ:relx1x2RS}, and it is easily checked that $x_1$ and $x_2$ are the unique zeroes of $\varpi$. 
%Using the simplified form of $\px(x_1,x_2)$ derived in \cref{equ:Px1Px2_simpl},  we thus arrive at
%\begin{equation}
% \px(x_1,x_2)  =-\frac{1}{2}\sum_{i,j}k_i\cdot k_j\,\sqrt{\frac{\omega_{1^+\!,1^-}(x_2)\,\omega_{2^+\!,2^-}(x_1)}{\omega_{1^+\!,1^-}(x_1)\,\omega_{2^+\!,2^-}(x_2)}} \,\,\frac{\omega_{x_1,x_2}(\sigma_i)}{\omega_{1^+\!,1^-}(\sigma_i)}\,\,\frac{\omega_{x_1,x_2}(\sigma_j)}{\omega_{2^+\!,2^-}(\sigma_j)}\,.
%\end{equation}
%\begin{equation}
% \px(x_1,x_2)  =-\frac{1}{2}\sum_{i,j}\frac{k_i\cdot k_j}{(ij)}\,\left( -2\,\frac{(ij)\,\d x_1\d x_2}{(x_1 i)(x_2 j)}
% +\frac{c_1}{c_2} \, \omega_{i,j}(x_1)\, \d x_1 + \frac{c_2}{c_1} \, \omega_{i,j}(x_2) \,\d x_2
% \right)\,.
%\end{equation}
We recall for convenience the form of $\px(x_1,x_2)$ from \cref{equ:Px1Px2_simpl},
\begin{equation}
 \px(x_1,x_2) 
 =-\frac{1}{2}\sum_{i,j}\frac{k_i\cdot k_j}{c_1 c_2}\,\Big(c_1\omega_{i,*}(x_1)-c_2\omega_{i,*}(x_2)\Big)\Big(c_1\omega_{j,*}(x_1)-c_2\omega_{j,*}(x_2)\Big)\,,
\end{equation}
which can be easily checked not to depend on the marked point $\sigma_\ast$.

The two-loop supergravity integrand $\mathfrak{I}_n$ from \eqref{equ:integrandsugra}, with $\mathcal{I}_n^{(2)}$ as described here, is the main result of this section. Below, we will briefly derive  the simplifications for four external particles, and discuss some basic checks for the formula.

\paragraph{The four-particle amplitude.} As expected from the genus two results, the four-particle amplitude simplifies considerably. Degenerating directly from the expression \eqref{equ:4pt_genus2} found in \cref{sec:4pt}, the integrand reduces to 
\begin{equation}\label{equ:4pt_nodalRS}
 \mathfrak{I}_4=
 \int  \hspace{-15pt}{\phantom{\Bigg(}}_{\mathfrak{M}_{0,4+4}}\hspace{-10pt}
\frac{1}{\text{vol SL}(2,\mathbb{C})^2}\,
\,\,\prod_{A}\,\bar\delta\big( \mathcal{E}_A\big)
\,\,\mathcal{K}\widetilde{\mathcal{K}}\,\,\widehat{\mathcal{Y}}^2\, \frac{(1^+2^-)(1^-2^+)}{(1^+1^-)(2^+2^-)}\,.
\end{equation}
where we rescaled the integrand again by the Jacobian form $J$, 
\begin{equation}
 \widehat{\mathcal{Y}}=J\,\mathcal{Y} =\frac{\d\sigma_{1^+}\d\sigma_{1^-}\d\sigma_{2^+}\d\sigma_{2^-}}{(1^+2^+)(1^+2^-)(1^-2^+)(1^-2^-)}\,\mathcal{Y}\,. 
\end{equation}
This agrees directly with the four particle supergravity integrand given in \cite{Geyer:2016wjx}. Instead, we could have chosen to work with the general form of the integrand as a sum over spin structures, and only simplified the formula after the residue theorem, when the amplitude is localised on the bi-nodal Riemann sphere -- the result  agrees with \cref{equ:4pt_nodalRS}. As shown in  \cite{Geyer:2016wjx}, this integrand indeed reproduces the known four-point integrand of supergravity \cite{Bern:1998ug}, if both the planar and the non-planar double boxes are written in the `mostly-linear' representation of the propagators. This representation can be achieved from the standard representation via the use of partial fraction identities and shifts in the loop momenta, and is related to the Q-cut construction \cite{Baadsgaard:2015twa}.

\subsection{Absence of unphysical poles} \label{sec:unphys-poles}
The two-loop integrand \eqref{equ:4pt_nodalRS} can be shown to match known supergravity integrands exactly for four external particles \cite{Geyer:2016wjx}. To generalise this to $n$ points, a proof of the formula \eqref{equ:int_nodal_final} could in principle be given by studying the behaviour of the amplitude near the boundary of the (sphere) moduli space to establish the standard field theory factorisation properties of the integrand. In practice, a full factorisation proof of \eqref{equ:int_nodal_final} is beyond the scope of this paper due to the Ramond states flowing through the nodes.\footnote{Closed $n$-point formulas involving Ramond states have been discussed previously in \cite{Haertl:2009yf, Hartl:2010ks}, indicating that these difficulties can be resolved.} As a first step towards factorisation, we show below that the amplitude  only contains physical poles. Since the absence of unphysical poles  relies on properties of the two-loop scattering equations  \eqref{equ:SE_nodalRS_v1}  established in previous work  \cite{Geyer:2016wjx}, we include a brief review for completeness.

\begin{figure}[ht]
	\centering 
	\begin{subfigure}[t]{0.45\textwidth} 
        \centering
        \includegraphics[width=6cm]{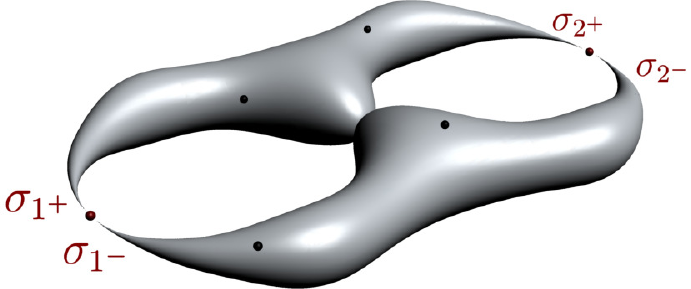}
        \caption{Physical factorisation channel, corresponding to poles of the form $\big(\ell_1+\ell_2+K_D\big)^2$.}
    \end{subfigure}\hfill
    \begin{subfigure}[t]{0.45\textwidth}
            \centering
       \includegraphics[width=6cm]{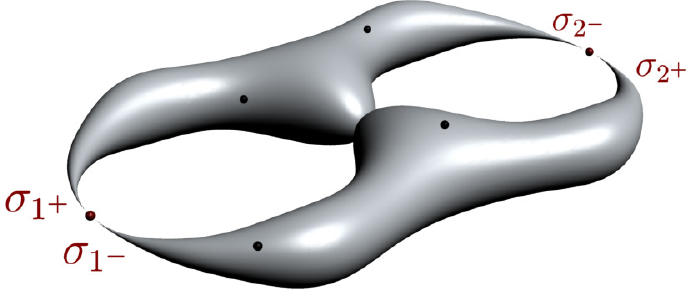}
       \caption{Unphysical factorisation channel, corresponding to poles of the form $\big(\ell_1+\ell_2\big)^2+2\big(\ell_1-\ell_2\big)\cdot K_D+K_D^2$}
    \end{subfigure}\hfill
    \caption{Different factorisation channels of the two-loop scattering equations.}
	\label{fig:unphys_pole}
\end{figure}

\paragraph{Separating degenerations.} The key feature of the scattering equations, at both tree and loop level, is that they relate factorisation channels of the amplitude to the boundary of the moduli space $\widehat{\mathfrak{M}}'_{g,n}$. This characteristic is preserved when degenerating to the bi-nodal Riemann sphere, and the potential poles are completely determined by the scattering equations via
\begin{equation}
 \sum_{A\in D}(\sigma_A-\sigma_D)\mathcal{E}_A=0\,,
\end{equation}
for some subset $D$ of the vertex operators coalescing to a point $\sigma_D$. The poles arising from such separating degenerations have been classified in \cite{Geyer:2016wjx}. With $K_{D\,\mu}=\sum_{i\in D}k_{i\,\mu}$ denoting the sum of external momenta in $D$, the scattering equations encode the  poles described in \cref{table:poles}.
\begin{table}[ht]
\begin{center}
\begin{tabular}{l l}
 subset $D$ & pole\\\hline
 $\{\sigma_i\}$ & $K_D^2$\\
 $\{\sigma_{I^\pm},\,\sigma_i\}$ & 2\,$\ell_I\cdot K_D\pm K_D^2$\\
 $\{\sigma_{I^+},\,\sigma_{I^-},\,\sigma_i\}$ & $K_D^2$\\
 $\{\sigma_{1^+},\,\sigma_{2^+},\,\sigma_i\}$ & $\big(\ell_1+\ell_2+K_D\big)^2$\\
 $\{\sigma_{1^-},\,\sigma_{2^-},\,\sigma_i\}$ & $\big(\ell_1+\ell_2-K_D\big)^2$\\
 $\{\sigma_{1^+},\,\sigma_{2^-},\,\sigma_i\}$ & $\big(\ell_1+\ell_2\big)^2+2\big(\ell_1-\ell_2\big)\cdot K_D +K_D^2$\\
 $\{\sigma_{1^-},\,\sigma_{2^+},\,\sigma_i\}$ & $\big(\ell_1+\ell_2\big)^2+2\big(\ell_2-\ell_1\big)\cdot K_D +K_D^2$
\end{tabular}
\end{center}
\caption{Separating degenerations and associated poles.}
\label{table:poles}
\end{table}
This highlights two important features of the amplitude \eqref{equ:integrandsugra}. 
\begin{enumerate}[(i)]
 \item Most notably, we observe that the poles containing only one of the loop momenta are linear. The loop integrand is therefore represented in the `mostly-linear' propagator representation (as opposed to Feynman propagators), related to the Q-cut construction of ref.~\cite{Baadsgaard:2015twa}. This representation of the integrand can be obtained from the standard representation by generalised `partial fraction identities' of the form
 \begin{equation}
  \frac{1}{\prod_i D_i} = \sum_{i} \frac{1}{D_i\prod_{j\neq i}\big(D_j-D_i\big)}\,,
 \end{equation}
 where $1/D_i$ denote standard Feynman propagators. In particular, the right-hand side of the above relation contains only one quadratic propagator, given by $\ell^2$ up to shifts in the loop momentum, while all other terms are linear in $\ell$.\\
 This result at two loops mirrors the amplitude representation at one loop, and is expected from the basis choice of Beltrami differentials: by extracting the residues of $P^2$ at the vertex operator insertions, the scattering equations $\mathcal{E}_i$ can only contain $\ell$ linearly.
\item The other important  aspect in \cref{table:poles} are the unphysical poles $\big(\ell_1+\ell_2\big)^2+2\big(\ell_1-\ell_2\big)\cdot K_D+K_D^2$. Since these poles do not correspond to factorisation channels of the loop integrand, they must be absent from $ \mathfrak{I}_n$, which serves  as an important check for our formula.degeneration
\end{enumerate}
From \cref{table:poles}, the scattering equations relate these unphysical poles to separating divisors $\mathfrak{D}^{\text{sep}}_{0,n+4}$   that retain $\sigma_{1^+}$ and $\sigma_{2^-}$  (together with some subset $\sigma_i$ for $i\in D$) on one component of the separating degeneration,  while $\sigma_{1^-}$ and $\sigma_{2^+}$ lie on the other sphere; see \cref{fig:unphys_pole}. Of course, the presence of this unphysical divisor on the moduli space does {\it not} imply that the loop integrand contains a pole there -- indeed, a CHY-type formula does not necessarily realise all factorisation channels encoded in the scattering equations. We can test for the presence or absence of the unphysical pole by probing the behaviour of the integrand close to the boundary divisor.

While it is possible to verify the absence of the unphysical poles explicitly from the form of the integrand \eqref{equ:int_nodal_final}, the calculation is quite involved. Luckily, there is a much more elegant solution relying exclusively on properties of the amplitude already discussed in the degeneration to the bi-nodal Riemann sphere. To see this, note that the separating boundary divisors containing the unphysical poles correspond to $q_3=\infty$ according to \cref{equ:q3}, since either $\sigma_{1^+}$ and $\sigma_{2^-}$ coalesce, leading to $(1^+2^-)=0$, or $\sigma_{1^-}$ and $\sigma_{2^+}$, giving $(1^-2^+)=0$. The limit $q_3\rightarrow\infty$,  on the other hand, has already been studied  in \cref{sec:map_to_RS}, where we established the absence of a pole in the amplitude. Let us briefly recall the argument here.

Modular invariance guarantees that the integrand $\mathcal{I}_n$ in
\begin{equation}
 \mathcal{M}_n= \int_{|q_{12}|< 1}\d\mu_{2,n}\,\mathcal{I}_n =  \int_{|q_{12}|> 1}\d\mu_{2,n}\,\mathcal{I}_n\,,
\end{equation}
has a simple pole at $q_3=q_{12}=\infty$, related to the pole at $q_3=0$ by the modular transformation $q_{12}\leftrightarrow 1/q_{12}$. After trivialising the isomorphism $\mathfrak{D}^{\text{max}}_{2,n}\cong \widehat{\mathfrak{M}}'_{0,n+4}$ by extending the domain of integration over $q_3$ however, the full amplitude contains a factor of $f(q_3)=(1-q_3)^{-1}$,
\begin{equation}
 \mathcal{M}_n  =\int\d\mu_{2,n}\,\mathcal{I}_n \,\frac{1}{1-q_3}\,,
\end{equation}
which cancels the pole at $q_3=\infty$. The final expression on the bi-nodal sphere is thus finite at $q_3\rightarrow \infty$, and does not contain the unphysical pole $\big(\ell_1+\ell_2\big)^2+2\big(\ell_1-\ell_2\big)\cdot K_D+K_D^2$. The argument presented here highlights the interplay between the form of the amplitude on the bi-nodal sphere and the residue theorem: the absence of a  pole at $q_3=\infty$ ensured both that the amplitude would localise on the sphere after applying the residue theorem, and that only physical factorisation channels are realised.

To gain some additional  intuition for this unphysical pole, let us briefly revisit the original formula \eqref{equ:integrand_g=2} on the genus-two Riemann surface. The $n+3$ genus-two scattering equations did not contain an equivalent of the unphysical pole -- modular invariance guarantees both that the integrand has a simple pole at $q_3=\infty$ and that the scattering equations relate this to the (physical) pole $\big(\ell_1+\ell_2\pm K_D\big)^2$. After the degeneration to the nodal Riemann sphere, however, the remaining (independent) $n+1$ scattering equations do not relate these poles, so the pole at $q_3=\infty$ assumes the unphysical form seen above -- which must of course be absent from the loop integrand, because the original loop integrand did not contain the unphysical pole. In the residue theorem, this is implemented concretely by the map $\mathfrak{D}^{\text{max}}_{2,n}\cong \widehat{\mathfrak{M}}'_{0,n+4}$, which provides the additional factor $f(q_3)$ that cancels the pole at infinity.

\paragraph{Degenerate solutions.} The scattering equations  \eqref{equ:SE_nodalRS_v1} on the nodal Riemann sphere encode an additional, more subtle unphysical pole: a Gram determinant of Mandelstam variables that can be localised on the  so-called  `degenerate solutions' to the loop scattering equations \cite{Cachazo:2015aol}.\footnote{We refer to the original paper \cite{Cachazo:2015aol}  for details on how to relate the Gram determinant pole to the degenerate solutions.} These degenerate solutions appear due to the form of the scattering equations for the nodal points:   the two constraints associated to the same node have the same functional form, 
\begin{equation}
 \pm \mathcal{E}_{1^\pm}(\sigma_{1^\pm})=0\,,\qquad \qquad\pm \mathcal{E}_{2^\pm}(\sigma_{2^\pm})=0\,.
\end{equation}
Solutions to the scattering equations thus fall into two classes, regular solutions  where $\sigma_{I^+}$ and $\sigma_{I^-}$ localise on different roots, and `degenerate solutions' with $\sigma_{I^+}=\sigma_{I^-}$ for at least one of $I=1,2$. The latter class accounts for the unphysical Gram determinant pole at
\begin{equation}
 \det (k_A\cdot k_B)=0\,,\qquad\qquad k_{A,B}\in\{ \ell,\,k_1,\,...k_n\}\,.
\end{equation}
Let us recall why poles of this type are absent at one loop, where their potential to appear was first observed. Note in this context that, while modular invariance at higher genus places strict requirements on the worldsheet theories, one of the main strengths of the representation on the nodal sphere is its versatility. Integrands have been proposed for a variety of theories in various dimensions, ranging from supergravity and super Yang-Mills \cite{Geyer:2015bja} to theories with less supersymmetry like pure Yang-Mills theory,  NS-gravity and the bi-adjoint scalar theory \cite{Geyer:2015jch,He:2015yua}. For these theories, two different strategies have been developed to establish the absence of the Gram determinant pole. The simplest case is that of supersymmetric theories, whose integrands vanish on the degenerate solutions, so no further analysis is necessary. For non-supersymmetric theories on the other hand, the contribution from the unphysical pole  can be shown to vanish after integration because it is homogeneous in the loop momentum. This distinction between the behaviour of supersymmetric and non-supersymmetric theories on degenerate solutions  is closely linked to the UV behaviour of the theory \cite{Geyer:2015jch}, because the only solutions that contribute as $\ell\rightarrow\infty$ become degenerate in that limit. In particular, the fact that the integrand vanishes on the degenerate solutions is associated to the absence of bubbles in a diagrammatic representation of the loop integrand. We may thus naturally expect some of the above discussion to carry over to two loops. Indeed, we will find below that, in analogy to one loop, the two-loop supergravity integrand vanishes on the degenerate solutions.

\begin{figure}[ht]
	\centering 
	\begin{subfigure}[t]{0.28\textwidth} 
        \centering
        \includegraphics[width=5.5cm]{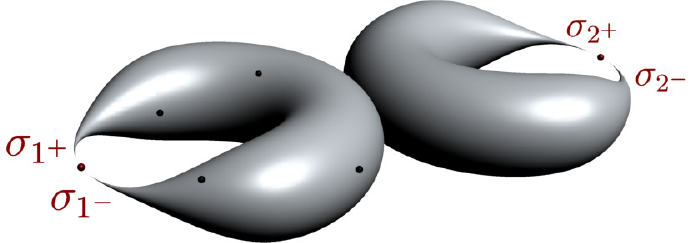}
        \caption{Type A.}
    \end{subfigure}\hfill
    \begin{subfigure}[t]{0.3\textwidth}
            \centering
       \includegraphics[width=5.7cm]{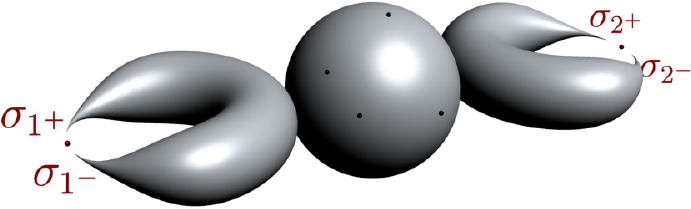}
       \caption{Type B.}
    \end{subfigure}\hfill
    \begin{subfigure}[t]{0.25\textwidth}
              \centering
       \includegraphics[width=3.2cm]{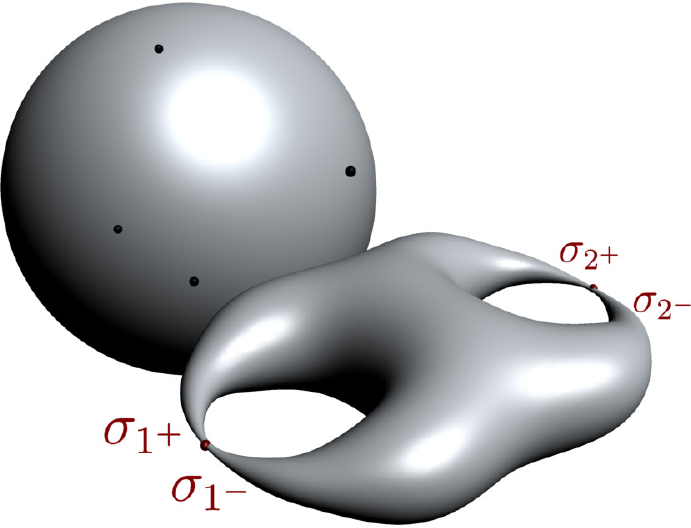}
       \caption{Type C.}
    \end{subfigure}%
    \caption{The three types of degenerate solutions to the scattering equations at two loops.}
	\label{fig:degen}
\end{figure}

With this background in mind, let us return to the scattering equations  on the bi-nodal Riemann sphere (analysed in detail in \cite{Geyer:2016wjx}). In generalising from one to two loops, a new feature appears: the degenerate solutions can be further split into three types, best summarised by \cref{fig:degen}. Note however that all three types of degenerate solutions satisfy   $\sigma_{I^+}=\sigma_{I^-}$ for at least one node $I=1,2$, so they all imply $q_3=1$, or equivalently $\tau_3=\tau_{12}=0$. 

Conveniently,  we have encountered this divisor before in \cref{sec:map_to_RS} on the genus-two Riemann surface, in the context of trivialising the isomorphism $\mathfrak{D}^{\text{max}}_{2,n}\cong \widehat{\mathfrak{M}}'_{0,n+4}$. As discussed there, the  amplitude at a given spin structure scales as  $\mathcal{O}(1)$ in $\tau_3\rightarrow 0$, even in the presence of additional punctures on both components of the degeneration. To show further that the supergravity integrand vanishes on the degenerate solutions, and thus that the Gram determinant pole is absent, consider the type A configuration, where no additional punctures are present on one of the tori. Clearly, this factorisation channel vanishes in the genus-two representation of the amplitude, because all $n$-point one-loop amplitudes for $n < 4$ vanish in type II supergravity. On the bi-nodal sphere, both the leading and the subleading contribution from the degenerate solutions thus vanish, and the chiral integrand scales as $\mathcal{O}(\tau_3)$. 

This argument is easily generalised to degenerate solutions of type B and type C, using respectively that one-loop and two-loop $n$-point amplitudes vanish for $n<4$. Note, moreover, that this argument for the absence of degenerate solutions relies on considering the full amplitude, including the sum over spin structures -- no such cancellations are expected for any individual spin structure, or the contribution from just the NS sector of the amplitude, for example. As mentioned above for one loop, this behaviour is expected from the known UV properties of those theories.

In conclusion, the chiral integrand for type II supergravity at two loops behaves as 
\begin{equation}
 \mathcal{I}_n^{(2)}=\mathcal{O}(\tau_3)\,,
\end{equation}
for all types of degenerate solutions. Including the  factor of $f(q_3)\sim\tau_3^{-1}$,  the full integrand thus vanishes on the degenerate solutions, 
\begin{equation}
 \mathcal{I}_n^{(2)}\,  \widetilde{\mathcal{I}}_n^{(2)} \, f(q_3)=\mathcal{O}(\tau_3)\,,
\end{equation}
and only the $ N_{\text{reg}} = (n + 1)! - 4n! + 4(n - 1)! + 6(n - 3)!$ regular solutions contribute to the supergravity amplitude.\footnote{Details on the counting can be found in \cite{Geyer:2016wjx}, appendix B.} Since the Gram determinant poles can be localised on the degenerate solutions, this precludes unphysical  poles from contributing to the amplitude.

%%%%%%%%%%%%%%%%%%%%%%%%%%%%%%
%%%%%%%%%%%%%%%%%%%%%%%%%%%%%%

\section{Super-Yang-Mills amplitudes} \label{sec:sym}

In the preceding sections, we have succeeded in obtaining an $n$-point formula for supergravity scattering amplitudes at two loops. We started on a genus-two surface but our final result is a formula on the bi-nodal sphere, which provides a dramatic simplification. Though technically much more challenging, our procedure mirrors that followed at one loop \cite{Geyer:2015bja}. We can now proceed to mirror another accomplishment of Ref.~\cite{Geyer:2015bja} by proposing an $n$-point formula for super-Yang-Mills theory, now at two loops. The formula reproduces the four-point results of \cite{Geyer:2016wjx}, and thus gives rise to known expressions for four-particle two-loop amplitudes in super Yang-Mills. 

\subsection{Parke-Taylor factor on the bi-nodal Riemann sphere}

There are three main concerns in achieving our goal of describing super-Yang-Mills amplitudes. The first is that we have no expression on the genus-two surface, as we had in the case of supergravity. So we cannot follow a straightforward derivation from the degeneration limit to a nodal sphere. In fact, this apparent obstruction happens already at one loop. However, one of the lessons of \cite{Geyer:2015bja} is that, while the degeneration is important in order to obtain a formalism on the nodal sphere, the new formalism can then be extended to a variety of theories, in particular gauge theory, without recourse to the higher-genus surface (where it may not even be possible to define those theories). Ref.~\cite{Roehrig:2017gbt} fully exploited this idea at one loop by re-deriving the formulas obtained in \cite{Geyer:2015bja,Geyer:2015jch} directly from a non-local one-loop `gluing operator' representing the node of the sphere. To conclude, we will simply work directly on the (bi-)nodal sphere.

The second concern comes from the description of colour in the ambitwistor string formalism \cite{Mason:2013sva}, which is analogous to that in the heterotic string \cite{Gross:1984dd} (and also Nair's observation \cite{Nair:1988bq}). Colour degrees of freedom are introduced via a current algebra, which leads to Parke-Taylor factors. For instance, at three points,
\begin{equation}
\langle \rho^{a_1}(\sigma_1)\rho^{a_2}(\sigma_2)\rho^{a_3}(\sigma_3) \rangle = 
\frac{\text{tr}([T^{a_{1}},T^{a_{2}}]T^{a_{3}})} {(12)(23)(31)} \,,
\end{equation}
where we denote $(ij)=\sigma_i-\sigma_j$ as usual. There is a difficulty at higher points, where unwanted multi-trace terms appear in the correlation function. Some constructions avoid these terms, but they also have limitations \cite{Casali:2015vta}. In this paper, as in \cite{Mason:2013sva}, we will simply discard the multi-trace terms, since this directly gives a valid formula for gauge-theory amplitudes. It will be useful to have in mind a certain representation of the tree-level result. Suppose we have an $n$-point tree-level amplitude: the colour part of the CHY formula \cite{Cachazo:2013hca} can be written as\footnote{This is the CHY implementation of the Dixon-Del Duca-Maltoni half-ladder basis for the colour dependence \cite{DelDuca:1999rs}. The loop-level case in \eqref{eq:PT2loopfff} is closely related to the procedure detailed in \cite{Ochirov:2016ewn}.}
\begin{equation}
{\mathcal I}^{\text{PT}(0)}_n = \sum_{\gamma \in S_{n-2}} 
\frac{\text{tr}([[\cdots[[[T^{a_{1}},T^{a_{\gamma(2)}}],T^{a_{\gamma(3)}}],T^{a_{\gamma(4)}}],\cdots],T^{a_{\gamma(n-1)}}]T^{a_n})}
{\big(1\,\gamma(2)\,\gamma(3)\,\gamma(4)\cdots \gamma(n-1)\, n\big)}\,, 
\label{eq:PTtree}
\end{equation}
where we denote the Parke-Taylor denominators by $(123\cdots m) = (12)(23)\cdots (m1)$. 

The third concern is that the cross ratio appearing in the supergravity formula \eqref{equ:integrandsugra}, originating in the genus-two fundamental domain, signals a requirement that did not exist at tree level or at one loop. As we mentioned in \cref{sec:unphys-poles}, the practical role of that cross ratio on the nodal sphere is to forbid unphysical factorisation channels. Our proposed formula for gauge theory will satisfy the same factorisation requirement. The solution to this problem was already mentioned in \cite{Geyer:2016wjx}. It suffices to restrict the orderings of the loop insertions in the colour part in the manner to be described momentarily.

We are now in a position to present the $n$-point two-loop formula for super-Yang-Mills theory. It is analogous to the supergravity formula  \eqref{equ:integrandsugra}, and it reads
\begin{equation}
 \mathcal{M}_n=\int \!\frac{\d^{10}\ell_1\,\d^{10}\ell_2}{\ell_1^2\ell_2^2} \,\, \mathfrak{I}_n\,,
\end{equation}
where $\mathfrak{I}_n$ is given by 
\begin{equation}
 \mathfrak{I}_n=
  \int  \hspace{-15pt}{\phantom{\Bigg(}}_{\mathfrak{M}_{0,n+4}}\hspace{-10pt}
\frac{1}{\text{vol SL}(2,\mathbb{C})^2}\,
\,\,\prod_{A}\,\bar\delta\big( \mathcal{E}_A\big)\,\,
\mathcal{I}^{(2)}_n\,{\mathcal I}^{\text{PT}(2)}_n\,.
 \label{equ:integrandsym}
\end{equation}
The final expression for $\mathcal{I}^{(2)}_n$ was obtained in \eqref{equ:int_nodal_final}, and the new object is the colour part,
\begin{equation}
{\mathcal I}^{\text{PT}(2)}_n = \d^{n+4}\sigma_{A}  \sum_{\gamma \in S'_{n+2}} 
\frac{\text{tr}([[\cdots[[[T^{a_{1^+}},T^{a_{\gamma(1)}}],T^{a_{\gamma(2)}}],T^{a_{\gamma(3)}}],\cdots],T^{a_{\gamma(n+2)}}]T^{a_{1^-}}) \, \delta^{a_{1^+},a_{1^-}} \, \delta^{a_{2^+},a_{2^-}}}
{\big(1^+\,\gamma(1)\,\gamma(2)\,\gamma(3)\cdots \gamma(n+2)\, 1^-\big)}\,.
\label{eq:PT2loopfff}
\end{equation}
This formula should be compared with the tree-level analogue \eqref{eq:PTtree}. The two-loop formula is very similar, but has four extra `particles', corresponding to the loop insertions $1^\pm$ and $2^\pm$, whose colour indices we contract for each node. The two insertions $1^\pm$ play now the special role of $1$ and $n$ in \eqref{eq:PTtree}, so that the sum is over permutations of all the remaining $n+2$ insertions. In fact, the sum in \eqref{eq:PT2loopfff} is over a restricted set of permutations (hence the prime in $S'_{n+2}$): we require that the ordering of loop insertions is $(1^+\cdots 2^+ \cdots 2^- \cdots 1^-)$, which leads to $S'_{n+2}$ having $(n+2)!/2$ elements; that is, we drop terms with ordering $(1^+\cdots 2^- \cdots 2^+ \cdots 1^-)$, where the dots represent external particles. The reason for this is that the forbidden terms would lead to unphysical factorisation channels, the same type of unphysical channels that were eliminated by the cross-ratio in the supergravity case. We leave a more detailed exposition of the factorisation argument to future work, where we intend to provide field theory proofs of at least some of our ambitwistor-string-derived formulae.

\subsection{Colour trace decomposition}

While our two-loop Parke-Taylor formula \eqref{eq:PT2loopfff} contains only single traces when seen as an $(n+4)$-particle-like expression, we know it must give rise to single-trace, double-trace and triple-trace contributions. These contributions arise due to the colour index contractions $\delta^{a_{1^+},a_{1^-}}$ and $\delta^{a_{2^+},a_{2^-}}$, together with the use of the completeness relation for the fundamental representation generators of the Lie algebra of SU($N_c$),
\begin{equation}
 (T^a)_{i_1}^{\;\;j_1}\,(T^a)_{i_2}^{\;\;j_2}= \delta_{i_1}^{j_2}\delta_{i_2}^{j_1}-\frac{1}{N_c}\,\delta_{i_1}^{j_1}\delta_{i_2}^{j_2} \,.
\end{equation}
After the use of this identity, the Parke-Taylor formula takes the known form 
\begin{align}
{\mathcal I}^{\text{PT}(2)}_n \;=& \;\; \sum_{\rho}\big(N_c^2 \,C^{\text{P}}_{\rho}+ C^{\text{NP},1}_{\rho}\big)\;
 \big(\tr(\rho)+(-1)^{n}\, \tr(\rho^{-1})\big) \nonumber \\
& 
+\sum_{\rho_1,\rho_2} N_c\,C^{\text{NP},2}_{\rho_1,\rho_2} \;\;
\big(\tr(\rho_1)\, \tr(\rho_2)+(-1)^{n}\, \tr(\rho_1^{-1})\, \tr(\rho_2^{-1})\big)  \nonumber \\
& 
+\sum_{\rho_1,\rho_2,\rho_3}C^{\mathrm{NP},3}_{\rho_1,\rho_2,\rho_3} 
\;\;
\big(\tr(\rho_1)\, \tr(\rho_2)\, \tr(\rho_3)+(-1)^{n}\, \tr(\rho_1^{-1})\, \tr(\rho_2^{-1})\, \tr(\rho_3^{-1})\big) \,.
\label{eq:PT2loop}
\end{align}
where $\text{tr}(12\cdots)\equiv\text{tr}(T^{a_1}T^{a_2}\cdots)$ is a colour trace, and $\{12\cdots m\}^{-1}\equiv\{m\cdots21\}$ denotes the inverse ordering. The sums in \eqref{eq:PT2loop} are over non-cyclic permutations of the $n$ external particles or of partitions of these.

The elements introduced above completely determine the trace decomposition coefficients. For instance, the planar contribution (the leading order in $N_c$) is
\begin{align}
C^{\text{P}}_{\rho} =\left( \sum_{\substack{\rho=(\rho_1,\rho_2,\rho_3)}}
\frac{1+\delta_{|\rho_2|,0}+\delta_{|\rho_2|,n}}{(1^+\,\rho_1\,2^+\,\rho_2\,2^-\,\rho_3\,1^-)}
\,+\, \text{cyc}(\rho) \right)
+ (-1)^n[\,\rho\to \rho^{-1}\,] \,.
\end{align}
The sum runs over all the order-respecting splittings $\rho=(\rho_1,\rho_2,\rho_3)$, where a set $\rho_r$ may be empty. The remaining notations should be clear: $\delta_{|\rho_2|,0}$ is 1 if $\rho_2$ is empty and is 0 otherwise; $\text{cyc}(\rho)$ denotes sum over cyclic permutations of $\rho$; the last term corresponds (up to sign) to the same expression for the inverse ordering. Notice that, while this expression does not resemble the one presented in \cite{Geyer:2016wjx} for $n=4$, they are actually equivalent. We can also identify the terms in the leading non-planar correction:
\begin{align}
C^{\text{NP},2}_{\rho_1,\rho_2}
=& \left( \sum_{\rho_1=(\rho_{1,1},\rho_{1,2})}\frac{3}{(1^+\,\rho_{1,1}\,2^+\,\rho_2\,2^-\,\rho_{1,2}\,1^-)}   
+\sum_{\substack{\tilde{\rho}=\rho_1\shuffle\rho_2\\ \tilde\rho=(\tilde\rho_1,\tilde\rho_2,\tilde\rho_3)\\\rho_1\subset\tilde\rho_1\cup\tilde\rho_3,\,\,\rho_2\not\subset\tilde\rho_2}}\frac{1}{(1^+\,\tilde\rho_1\,2^+\,\tilde\rho_2\,2^-\,\tilde\rho_3\,1^-)}  \right. \nonumber \\
& \qquad \left. +\sum_{\substack{\tilde{\rho}=\rho_1\shuffle\rho_2\\ \tilde\rho=(\tilde\rho_1,\tilde\rho_2,\tilde\rho_3)\\\rho_2\subset\tilde\rho_2,\,\,\rho_2\neq\tilde\rho_2}}\frac{1+\delta_{|\tilde\rho_2|,0}}{(1^+\,\tilde\rho_1\,2^+\,\tilde\rho_2\,2^-\,\tilde\rho_3\,1^-)}
\;\;+\,[\text{cyc}(\rho_1),\text{cyc}(\rho_2),\rho_1 \leftrightarrow \rho_2] \right) \nonumber \\
&+ (-1)^{|\rho_1|}[\,\rho_1\to \rho_1^{-1}\,] + (-1)^{|\rho_2|}[\,\rho_2\to \rho_2^{-1}\,] + (-1)^{n}[\,\rho_1\to \rho_1^{-1},\,\rho_2\to \rho_2^{-1}\,] \,,
\end{align}
where the inversion terms in the last line are only included for $|\rho_1|>2$ and/or $|\rho_2|>2$.

We have obtained these formulae for the trace coefficients $C^{\text{P}}_{\rho}$ and $C^{\text{NP},2}_{\rho_1,\rho_2}$ by inference from the complete result \eqref{eq:PT2loopfff} for four and five particles, and will not attempt a proof here. In fact, there are different representations, due to (KK-type \cite{KK1989}) identities among expressions with Parke-Taylor denominators. For instance, we find that we can also write each trace coefficient as
\begin{equation}
C(1^+,1^-,2^+,2^-) = \frac{1}{4} \big(
c(1^+,1^-,2^+,2^-)+c(1^-,1^+,2^-,2^+)+c(2^+,2^-,1^+,1^-)+c(2^-,2^+,1^-,1^+)
\big)\,, \nonumber
\end{equation}
so that
\begin{align}
c^{\text{P}}_{\rho}(1^+,1^-,2^+,2^-)
=& \;\frac{4}{(1^+\,2^+\,\rho\,2^-\,1^-)}   
+\sum_{\substack{\rho=(\rho_1,\rho_2)}}\frac{1}{(1^+\,\rho_1\,2^+\,2^-\,\rho_2\,1^-)} 
\,+\, \text{cyc}(\rho)\,,
\end{align}
and
\begin{align}
c^{\text{NP},2}_{\rho_1,\rho_2}(1^+,1^-,2^+,2^-)
=& \sum_{\rho_1=(\rho_{1,1},\rho_{1,2})}\frac{2}{(1^+\,\rho_{1,1}\,2^+\,\rho_2\,2^-\,\rho_{1,2}\,1^-)}   
+\sum_{\substack{\tilde{\rho}=\rho_1\shuffle\rho_2\\ \tilde\rho=(\tilde\rho_1,\tilde\rho_2,\tilde\rho_3)\\\rho_2\subset\tilde\rho_2}}\frac{1}{(1^+\,\tilde\rho_1\,2^+\,\tilde\rho_2\,2^-\,\tilde\rho_3\,1^-)} \nonumber \\
&+ \,[\text{cyc}(\rho_1),\text{cyc}(\rho_2),\rho_1 \leftrightarrow \rho_2] \,.
\end{align}

A very helpful consistency test of these formulae was provided by the relations among trace coefficients studied in \cite{Naculich:2011ep,Edison:2011ta,Edison:2012fn}, which we checked up to five points.

%%%%%%%%%%%%%%%%%%%%%%%%%%%%%%
%%%%%%%%%%%%%%%%%%%%%%%%%%%%%%
\section{Discussion} \label{sec:discussion}

In this paper, we have constructed new formulae based on the two-loop scattering equations for the $n$-particle two-loop integrands in supergravity and in super-Yang-Mills theory. We started by constructing a formula derived from the ambitwistor string at genus two in the case of supergravity. We then turned this formula into a much simpler one at genus zero via the residue theorem on the genus-two moduli space. Finally, we proposed an analogous genus-zero formula for the $n$-particle two-loop integrand of super-Yang-Mills theory. A summary of the results was given in section~\ref{subsec:summary}. We stress that we have presented results for the loop integrands, since the ten-dimensional amplitudes are not defined due to the ultraviolet divergence of the loop integration, as expected from these field theories. Loop integrands for theories in fewer spacetime dimensions are obtained via dimensional reduction as usual, including for ${\mathcal N}=4$ super-Yang-Mills theory and ${\mathcal N}=8$ supergravity in four dimensions, if we reduce from 10$d$ on a 6-torus.

Our results provide the two-loop extension of the one-loop formulae constructed in \cite{Geyer:2015bja}. The developments of the one-loop story point therefore towards the next obvious steps. One goal is to present analogous formulae for the two-loop integrands of non-supersymmetric theories, as in \cite{Geyer:2015jch}. The RNS formalism used here, where we consider separately contributions from the various spin structures, is helpful in that regard, since it is clear that the non-supersymmetric theories should arise entirely from the four NS-NS spin structures. This is work in progress. Obviously, our formulae admit further simplifications from the choice of gravitino gauge slice (the marked points $x_1,x_2$) and, in the supersymmetric case, from the sum over all even spin structures. Indeed, it would be interesting to compare our results to the known expressions for the five-point two-loop integrand in both supergravity and super-Yang-Mills theory \cite{Carrasco:2011mn,Gomez:2015uha,Mafra:2015mja}.

Another very interesting direction is to re-derive our formulae from a two-loop `gluing operator' directly on the Riemann sphere, obviating the intricacies of higher-genus surfaces. This was achieved at one loop in \cite{Roehrig:2017gbt}. Indeed, the heavy machinery involved in our calculations suggests that higher-loop results require a different approach. A major motivation for us is that the two-loop case may be sufficient to reveal important parts of the all-loop structure.

The formulae presented here for the loop integrands are of CHY type: they are expressed as moduli integrals on the sphere that localise on the solutions to the two-loop scattering equations. While such formulae have many interesting properties, it is important to obtain standard formulae for the loop integrand, depending only on the kinematic invariants. The goal is the extension of what was achieved in \cite{He:2017spx,Geyer:2017ela} at one loop. As in those works, there is the prospect of clarifying the colour-kinematics duality at loop level \cite{Bern:2008qj,Bern:2010ue}, using the ambitwistor string as a first-principles tool. From the perspective of conventional string theory, the colour-kinematics duality of gauge theory is intimately connected to the monodromy properties of the open string \cite{BjerrumBohr:2009rd,Stieberger:2009hq}, and recent work has analysed these properties at higher genus \cite{Tourkine:2016bak,Hohenegger:2017kqy,Ochirov:2017jby}.

It would also be important to provide proofs for our loop-integrand formulae, e.g., based on factorisation as in \cite{Geyer:2015jch} at one loop. Beyond the loop integrand, the ultimate objective is, of course, to obtain the scattering amplitude, particularly in the four-dimensional case, and hopefully to contribute beyond the state-of-the-art level to the phenomenology-oriented computation of gauge theory amplitudes.

%%%%%%%%%%%%%%%%%%%%%%%%%%%%%%
%%%%%%%%%%%%%%%%%%%%%%%%%%%%%%

\section*{Acknowledgements}
We would like to thank Tim Adamo, Eric D'Hoker, Lionel Mason, Kai Roehrig and Piotr Tourkine for valuable discussions. We would also like to thank the anonymous referee for very helpful comments on the manuscript. This research was supported in part by the National Science Foundation under Grant No. NSF PHY-1125915, as well as the Munich Institute for Astro- and Particle Physics (MIAPP) of the DFG cluster of excellence ``Origin and Structure of the Universe''. We are grateful to the KITP Santa Barbara and the MIAPP Munich, as well as the organisers of the workshops ``Scattering Amplitudes and Beyond'' (KITP) and ``Mathematics and Physics of Scattering Amplitudes'' (MIAPP) for providing hospitality, support and a stimulating atmosphere. This research was also  supported in part by the Perimeter Institute for Theoretical Physics. Research at the Perimeter Institute is supported by the Government of Canada through Industry Canada and by the Province of Ontario through the Ministry of Economic Development \& Innovation. YG gratefully acknowledges support from the National Science Foundation grant PHY-1606531, the W.M. Keck Foundation Fund and the Roger Dashen Membership. RM is supported by a Royal Society University Research Fellowship.

%\newpage
%%%%%%%%%%%%%%%%%%%%%%%%%%%%%%
%%%%%%%%%%%%%%%%%%%%%%%%%%%%%%
\appendix

\section{Useful identities}\label{sec:identities}
\subsection{Identities involving the chiral partition function}
The calculation of the four-point amplitude in \cref{sec:4pt} makes use of various identities for Szeg\H{o} kernels summed over all even spin structures. These were derived in \cite{DHoker:2005vch}, and the interested reader is referred to the original work for details of the proof. We quote them here for completeness and convenience. Consider the following sums over chiral partition functions and Szeg\H{o} kernels:
\begin{subequations}
\begin{align}
 I_1 &= \sum_{\delta} \mathcal{Z}^{\text{chi}}[\delta]\,\, S_\delta(x_1,x_2)\label{equ:I1}\\
 I_2 &= \sum_{\delta} \mathcal{Z}^{\text{chi}}[\delta]\,\, S_\delta(x_1,x_2)S_\delta(z_1,z_2)^2\label{equ:I2}\\
 I_3 &= \sum_{\delta} \mathcal{Z}^{\text{chi}}[\delta]\,\, S_\delta(x_1,x_2)S_\delta(z_1,z_2) S_\delta(z_2,z_3)S_\delta(z_3,z_1)\\
 I_4 &= \sum_{\delta} \mathcal{Z}^{\text{chi}}[\delta]\,\, S_\delta(x_1,z_1)S_\delta(z_1,x_2)\\
 I_5 &= \sum_{\delta} \mathcal{Z}^{\text{chi}}[\delta]\,\, S_\delta(x_1,z_1)S_\delta(z_1,x_2) S_\delta(z_2,z_3)^2\\
 I_6 &= \sum_{\delta} \mathcal{Z}^{\text{chi}}[\delta]\,\, S_\delta(x_1,z_1)S_\delta(z_1,z_2)S_\delta(z_2,x_2)\\
 I_7 &= \sum_{\delta} \mathcal{Z}^{\text{chi}}[\delta]\,\, S_\delta(x_1,z_1)S_\delta(z_1,z_2)S_\delta(z_2,z_3) S_\delta(z_3,x_2)\label{equ:I7}\\
 I_8 &= \sum_{\delta} \mathcal{Z}^{\text{chi}}[\delta]\,\, S_\delta(x_1,z_1)S_\delta(z_1,z_2)S_\delta(z_2,z_3)S_\delta(z_3,z_4) S_\delta(z_4,x_2)\\
 I_9 &= \sum_{\delta} \mathcal{Z}^{\text{chi}}[\delta]\,\, S_\delta(x_1,z_1)S_\delta(z_1,z_2) S_\delta(z_2,x_2)S_\delta(z_3,z_4)^2\\
 I_{10} &= \sum_{\delta} \mathcal{Z}^{\text{chi}}[\delta]\,\, S_\delta(x_1,z_1)S_\delta(z_1,x_2) S_\delta(z_2,z_3)S_\delta(z_3,z_4)S_\delta(z_4,z_2)\\
 I_{11} &= \sum_{\delta} \mathcal{Z}^{\text{chi}}[\delta]\,\, S_\delta(x_1,x_2)S_\delta(z_1,z_2)^2S_\delta(z_3,z_4)^2\\
 I_{12} &= \sum_{\delta} \mathcal{Z}^{\text{chi}}[\delta]\,\, S_\delta(x_1,x_2)S_\delta(z_1,z_2) S_\delta(z_2,z_3)S_\delta(z_3,z_4)S_\delta(z_4,z_1)\,.
\end{align}
\end{subequations}
Only the last two of these sums are non-trivial, the rest vanish,
\begin{subequations}
 \begin{align}
  &I_1=I_2=I_3=I_4=I_5=I_6=I_7=I_8=I_9=I_{10}=0\,,\\
  &I_{11}=I_{12}=-2\mathcal{Z}_0\prod_{i=1}^4\varpi(z_i)\,,\label{equ:Idres}
 \end{align}
\end{subequations}
where we defined $\varpi$ in \eqref{equ:varpi}. Since the right-hand side of \cref{equ:Idres} is independent of the ordering of marked points $z_i$, both $I_{11}$ and $I_{12}$ are invariant under permutations of the marked points.

\subsection{Useful identities for the PCO gauge slice}\label{sec:other_ids}
Throughout \cref{sec:g=2}, we made use of the PCO gauge choice \eqref{equ:varpi} fixing the moduli PCO insertions $x_\alpha$ to coincide with the zeros of a holomorphic $(1,0)$-form $\varpi$ defined by
\begin{equation}
 \varpi(z) \equiv \omega_I(z)\partial_I\vartheta(x_1-\Delta)e^{2i\pi\kappa'\cdot(x_1-\Delta)} = -\,\omega_I(z)\partial_I\vartheta(x_2-\Delta)e^{2i\pi\kappa'\cdot(x_2-\Delta)} \,,
\end{equation}
with $x_1+x_2-2\Delta =2\kappa \in {\mathbb Z^2}\oplus \Omega {\mathbb Z^2}$. Here, we collect useful identities available due to this gauge choice:
\begin{subequations}
\begin{align}
 c_1\omega_I(x_1) &= c_2\omega_I(x_2)\,,\\
 -c_1^2\partial\varpi(x_1) &= c_2^2\partial\varpi(x_2)\,,\label{equ:varpi_id_2}\\
 c_1\omega_{i*}(x_1)-c_2\omega_{i*}(x_2) &= -c_1^2\partial\varpi(x_1)\frac{\Delta_{i*}}{\varpi(z_i)\varpi(z_*)}\,,\label{equ:varpi_id_3}\\
 \mathcal{Z}_0 c_1c_2 \partial\varpi(&x_1)\partial\varpi(x_2)=1 \,,\label{equ:varpi_id_4}
\end{align}
\end{subequations}
where the $c_\alpha$ are defined via $\varpi(z)=c_1 \Delta(x_1,z)=c_2 \Delta(x_2,z)$, and $z_\ast$ is an arbitrary marked point. As above, the interested reader is referred to the original string theory literature \cite{DHoker:2005vch} for details and proofs.

%%%%%%%%%%%%%%%%%%%%%%%%%%%%%%%%%%%%%%%%%%%%%%
%%%%%%%%%%%%%%%%%%%%%%%%%%%%%%%%%%%%%%%%%%%%%%

\section{Modular transformations and \texorpdfstring{$\mathfrak{M}'_{g,n}$}{M'}}\label{app:proof_modspace}

In this section, we prove that the two-loop amplitude can be expressed as an integral over the moduli space $\mathfrak{M}'_{2,n}$ defined in \cref{sec:moduli_space_A}. To see this, we provide the explicit modular transformations that map each of the six terms in the amplitude
\begin{equation}
 \mathcal{M}_n=\sum_{\alpha=1}^6\mathcal{M}_n^{(\alpha)}\,,
\end{equation}
to a different copy of the fundamental domain, all localising $P_\mu$ to $P^{(1)}_\mu= \ell^I_\mu \omega_I +\sum_{i=1}^n k_{i\,\mu}\omega_{i,*} $. The proof relies on the modular invariance of the amplitude proven in  \cref{sec:mod-inv}, which we assume from here on. For convenience, we also remind ourselves of the definition of the terms $\mathcal{M}_n^{(\alpha)}$: each represents the ambitwistor string correlator, with $P$ localised to $P_\mu=P^{(\alpha)}_\mu$ defined by \cref{equ:all_P},
\begin{align}
 P_\mu^{(1)} &= \ell_{1\,\mu} \omega_1 +\ell_{2\,\mu} \omega_2+\sum_{i=1}^n k_{i\,\mu}\omega_{i,*} \,, && P_\mu^{(4)}  = \ell_{2\,\mu} \omega_1 + \big(\ell_{1}  +\ell_{2} \big)_\mu \omega_2 +\sum_{i=1}^n k_{i\,\mu}\omega_{i,*} \,,\nonumber\\
 P_\mu^{(2)} &= \ell_{2\,\mu} \omega_1 +\ell_{1\,\mu} \omega_2+\sum_{i=1}^n k_{i\,\mu}\omega_{i,*} \,,&&P_\mu^{(5)} = \ell_{1\,\mu} \omega_1 + \big(\ell_{1}  +\ell_{2} \big)_\mu\omega_2 +\sum_{i=1}^n k_{i\,\mu}\omega_{i,*} \,,\\
 P_\mu^{(3)} &= \big(\ell_{1}  +\ell_{2} \big)_\mu\omega_1+\ell_{2\,\mu} \omega_2+\sum_{i=1}^n k_{i\,\mu}\omega_{i,*} \,,&& P_\mu^{(6)} = \big(\ell_{1}  +\ell_{2} \big)_\mu\omega_1 +\ell_{1\,\mu} \omega_2 +\sum_{i=1}^n k_{i\,\mu}\omega_{i,*} \,.\nonumber
\end{align}
Clearly, all terms $\mathcal{M}_n^{(\alpha)}$ in the sum are related to $\mathcal{M}_n^{(1)}$ by a redefinition of the loop momenta with trivial Jacobian $J=1$.

\paragraph{$\mathbf{\alpha=2}$:} In \cref{sec:moduli_space_A}, we already encountered the modular transformation relating $\mathcal{M}_n^{(2)}$ to $\mathcal{M}_n^{(1)}$. As a warm-up, let us briefly revisit this here before proceeding. Since $P_\mu^{(2)}$ is related to $P_\mu^{(1)}$ by simply exchanging the holomorphic differentials, consider the modular transformation $M_2$ exchanging the homology cycles,
\begin{equation}
 a=\begin{pmatrix}0 &1 \\ 1 & 0\end{pmatrix}\,,\qquad d=\begin{pmatrix} 0 & 1\\1 & 0\end{pmatrix}\,,\qquad b=c=0\,.
\end{equation}
Expressed in terms of the basis of homology cycles, this transformation corresponds to 
\begin{equation}
 \big(\widetilde{A}_1, \widetilde{B}_1\big) = \big(A_2,B_2\big)\,,\qquad \big(\widetilde{A}_2, \widetilde{B}_2\big) = \big(A_1,B_1\big)\,,
\end{equation}
so it indeed interchanges the cycle $A_1$ and $A_2$, and analogously $B_1$ and $B_2$. From \cref{sec:mod-inv}, we also confirm that this exchanges $\Omega_{11}$ and $\Omega_{22}$ in the period matrix,
\begin{equation}\label{equ:period_alpha=2}
 \widetilde{\Omega} = \begin{pmatrix} \Omega_{22} & \Omega_{12} \\ \Omega_{12} & \Omega_{11} \end{pmatrix}\,,
\end{equation}
as well as the following behaviour of the holomorphic differentials and the loop momenta;
\begin{equation}
 \big(\widetilde{\omega}_1, \widetilde{\omega}_2\big) = \big(\omega_2,\omega_1\big)\,,\qquad \big(\widetilde{\ell}_1, \widetilde{\ell}_2\big) = \big(\ell_2,\ell_1\big)\,.
\end{equation}
Directly substituting $\widetilde{\ell}$ by  $\ell$, the modular transformation $M_2$ thus maps $P_\mu^{(2)}$ to
\begin{equation}
 \widetilde{P}_\mu^{(2)} = \ell_{1\,\mu}\widetilde{\omega}_1+\ell_{2\,\mu}\,\widetilde{\omega}_2+\sum_i k_{i,\mu} \omega_{i,*}\,.
\end{equation}
Moreover, using the original inequalities $0<2\text{Im}(\Omega_{12})\leq\text{Im}(\Omega_{11})\leq\text{Im}(\Omega_{22})$ as well as the modular transformation of the period matrix \eqref{equ:period_alpha=2}, we conclude that $M_2$ maps the integration domain to
\begin{equation}
 0<2\text{Im}(\widetilde{\Omega}_{12})\leq\text{Im}(\widetilde{\Omega}_{22})\leq\text{Im}(\widetilde{\Omega}_{11})\,,
\end{equation}
as claimed in \cref{sec:moduli_space_A}. The contribution $\mathcal{M}_n^{(2)} $ to the amplitude is therefore given by
\begin{equation}
  \mathcal{M}_n^{(2)} =  \mathcal{M}_n^{(1)}\bigg|_{0<2\text{Im}(\Omega_{12})\leq\text{Im}(\Omega_{22})\leq \text{Im}(\Omega_{11})} \,,
\end{equation}
where we reset the notation for the period matrix to $\Omega$ again for convenience.

\paragraph{$\mathbf{\alpha=3}$:} To map $P_\mu^{(3)}$ to $P_\mu^{(1)}$, we are looking for a modular transformation $M_3$ that maps $\ell_1+\ell_2$ to $\ell_1$ while preserving $\ell_2$.  Consider therefore the transformation 
\begin{equation}
 a=\begin{pmatrix}1 &0 \\ 1 & 1\end{pmatrix}\,,\qquad d=\begin{pmatrix} 1 & -1\\0 & 1\end{pmatrix}\,,\qquad b=c=0\,.
\end{equation}
Using again the modular properties reviewed in \cref{sec:mod-inv}, the period matrix transforms into
\begin{equation}\label{equ:period_alpha=3}
 \widetilde{\Omega} = \begin{pmatrix} \Omega_{11} & \Omega_{12}+\Omega_{11} \\ \Omega_{12}+\Omega_{11} & \Omega_{11}+\Omega_{22}+2\Omega_{12} \end{pmatrix}\,,
\end{equation}
while the holomorphic differentials and the loop momenta map to
\begin{equation}
 \big(\widetilde{\omega}_1, \widetilde{\omega}_2\big) = \big(\omega_1+\omega_2,\omega_2\big)\,,\qquad \big(\widetilde{\ell}_1, \widetilde{\ell}_2\big) = \big(\ell_1-\ell_2,\ell_2\big)\,,
\end{equation}
Again substituting $\widetilde{\ell}$ by  $\ell$ while keeping the new differential $\widetilde{\omega}_I$, the modular transformation $M_3$ maps $P_\mu^{(3)}$ to the same form as $P_\mu^{(1)}$,
\begin{equation}
 \widetilde{P}_\mu^{(3)} = \ell_{1\,\mu}\widetilde{\omega}_1+\ell_{2\,\mu}\,\widetilde{\omega}_2+\sum_i k_{i,\mu} \omega_{i,*}\,.
\end{equation}
From the original inequalities $0<2\text{Im}(\Omega_{12})\leq\text{Im}(\Omega_{11})\leq\text{Im}(\Omega_{22})$ as well as the modular transformation of the period matrix \eqref{equ:period_alpha=3}, we again conclude that the integration domain for the modular parameters transforms to
\begin{equation}
  0<\text{Im}(\widetilde{\Omega}_{11})\leq2\text{Im}(\widetilde{\Omega}_{12})\leq\text{Im}(\widetilde{\Omega}_{22})\,,
\end{equation}
in agreement with \cref{sec:moduli_space_A}. 

\paragraph{$\mathbf{\alpha=4}$:}  This is a particularly simple case; since $M_4 = M_3\circ M_2$. We can thus recycle the two modular transformations discussed above by applying first $M_2$ to exchange the coefficients of the holomorphic differentials. This maps $P^{(4)}_\mu$ to $P^{(3)}_\mu$, so a further transformation $M_3$ leads back to 
\begin{equation}
 \widetilde{P}_\mu^{(4)} = \ell_{1\,\mu}\widetilde{\omega}_1+\ell_{2\,\mu}\,\widetilde{\omega}_2+\sum_i k_{i,\mu} \omega_{i,*}\,.
\end{equation}
In particular, $M_4$ maps the period matrix to
\begin{equation}
 \widetilde{\Omega} = \begin{pmatrix} \Omega_{22} & \Omega_{12}+\Omega_{22} \\ \Omega_{12}+\Omega_{22} & \Omega_{11}+\Omega_{22}+2\Omega_{12} \end{pmatrix}\,,
\end{equation}
and thus we confirm that after a modular transformation, the amplitude is integrated over
\begin{equation}
  0<\text{Im}(\widetilde{\Omega}_{11})\leq2\text{Im}(\widetilde{\Omega}_{22})\leq\text{Im}(\widetilde{\Omega}_{12})\,.
\end{equation}

\paragraph{$\mathbf{\alpha=5}$:} Note  that $\alpha=5$ closely resembles $\alpha=3$ discussed above: we are interested in a modular transformation $M_5$ that maps $P_\mu^{(5)}$ to $P_\mu^{(1)}$, and hence $\ell_1+\ell_2$ to $\ell_2$ while preserving $\ell_1$.  $M_3$ provided a similar map, but reversed the roles of $\ell_1$ and $\ell_2$.  This suggests that we can simply take $M_5=M_3^t$, where ${}^t$ denotes the transpose,
\begin{equation}
 a=\begin{pmatrix}1 &1 \\ 0 & 1\end{pmatrix}\,,\qquad d=\begin{pmatrix} 1 & 0\\-1 & 1\end{pmatrix}\,,\qquad b=c=0\,.
\end{equation}
Under $M_5$, the period matrix transforms as
\begin{equation}\label{equ:period_alpha=5}
 \widetilde{\Omega} = \begin{pmatrix}  \Omega_{11}+\Omega_{22}+2\Omega_{12}& \Omega_{12}+\Omega_{22} \\ \Omega_{12}+\Omega_{22} &\Omega_{22} \end{pmatrix}\,,
\end{equation}
and we confirm that both the holomorphic differentials and the loop momenta behave as expected,
\begin{equation}
 \big(\widetilde{\omega}_1, \widetilde{\omega}_2\big) = \big(\omega_1,\omega_2+\omega_1\big)\,,\qquad \big(\widetilde{\ell}_1, \widetilde{\ell}_2\big) = \big(\ell_1,\ell_2-\ell_1\big)\,,
\end{equation}
The modular transformation $M_3$ therefore maps $P_\mu^{(5)}$ to the same form as $P_\mu^{(1)}$. Moreover, using \eqref{equ:period_alpha=5} and $0<2\text{Im}(\Omega_{12})\leq\text{Im}(\Omega_{11})\leq\text{Im}(\Omega_{22})$,  the integration domain for the modular parameters transforms to
\begin{equation}
  0<\text{Im}(\widetilde{\Omega}_{22})\leq2\text{Im}(\widetilde{\Omega}_{11})\leq\text{Im}(\widetilde{\Omega}_{12})\,,
\end{equation}
in agreement with \cref{sec:moduli_space_A}. 

\paragraph{$\mathbf{\alpha=6}$:}  The modular transformation mapping $\mathcal{M}_n^{(6)}$ to  $\mathcal{M}_n^{(1)}$ can again be composed of the modular transformations $M_5$ and $M_2$;  $M_6 = M_5\circ M_2$. As above, this is best understood at the level of the field $P$, where $M_2$ interchanges the coefficients of the holomorphic differentials, thereby mapping $P^{(6)}_\mu$ to $P^{(5)}_\mu$. Since we just discussed this case, we only state the important transformation properties under $M_6$. In particular, the period matrix behaves as
\begin{equation}
 \widetilde{\Omega} = \begin{pmatrix}  \Omega_{11}+\Omega_{22}+2\Omega_{12}& \Omega_{12}+\Omega_{11} \\ \Omega_{12}+\Omega_{11} &\Omega_{11} \end{pmatrix}\,.
\end{equation}
This confirms that applying $M_6$ maps the integration domain for the modular parameters to the following copy of the fundamental domain:
\begin{equation}
  0<\text{Im}(\widetilde{\Omega}_{22})\leq2\text{Im}(\widetilde{\Omega}_{12})\leq\text{Im}(\widetilde{\Omega}_{11})\,,
\end{equation}

\paragraph{} This concludes the proof. In summary, after applying a modular transformation $M_\alpha$, all terms $\mathcal{M}_n^{(\alpha)}$ localise $P_\mu$ on $ P_\mu^{(1)} = \ell^I_\mu \omega_I +\sum_{i=1}^n k_{i\,\mu}\omega_{i,*} $, but are formulated over six different copies of the fundamental domain $\mathfrak{M}_g$,
\begin{align}
 \mathcal{M}_n^{(1)} &\equiv  \mathcal{M}_n^{(1)}\bigg|_{0<2\text{Im}(\Omega_{12})\leq\text{Im}(\Omega_{11})\leq \text{Im}(\Omega_{22})} && \mathcal{M}_n^{(4)} =  \mathcal{M}_n^{(1)}\bigg|_{0<\text{Im}(\Omega_{11})\leq\text{Im}(\Omega_{22})\leq 2\text{Im}(\Omega_{12})}\nonumber\\
  \mathcal{M}_n^{(2)} &=  \mathcal{M}_n^{(1)}\bigg|_{0<2\text{Im}(\Omega_{12})\leq\text{Im}(\Omega_{22})\leq \text{Im}(\Omega_{11})} && \mathcal{M}_n^{(5)} =  \mathcal{M}_n^{(1)}\bigg|_{0<\text{Im}(\Omega_{22})\leq\text{Im}(\Omega_{11})\leq 2\text{Im}(\Omega_{12})}\\
   \mathcal{M}_n^{(3)} &=  \mathcal{M}_n^{(1)}\bigg|_{0<\text{Im}(\Omega_{11})\leq2\text{Im}(\Omega_{12})\leq \text{Im}(\Omega_{22})} && \mathcal{M}_n^{(6)} =  \mathcal{M}_n^{(1)}\bigg|_{0<\text{Im}(\Omega_{22})\leq2\text{Im}(\Omega_{12})\leq \text{Im}(\Omega_{11})}\,.\nonumber
\end{align}

%%%%%%%%%%%%%%%%%%%%%%%%%%%%%%%%%%%%%%%%%%%%%%
%%%%%%%%%%%%%%%%%%%%%%%%%%%%%%%%%%%%%%%%%%%%%%

\section{The separating degeneration \texorpdfstring{$q_{12}=0$}{q12=0} and uniqueness of \texorpdfstring{$f(q_{12})$}{f(q)}}\label{sec:f}
This appendix provides the proof for the uniqueness of $f(q_{12})$ when requiring that the maximal non-separating boundary divisor $\mathfrak{D}^{\text{max}}_{2,n}$ remains the only simple pole of the integrand. As a first step, we prove that introducing a factor of
\begin{equation}
 f(q_{12})=\frac{1}{1-q_{12}}
\end{equation}
does not introduce a pole at the separating  degeneration $\mathfrak{D}^{\text{sep}}_{2,n}$. This degeneration, corresponding to $\Omega_{12}\rightarrow 0$, has been extensively studied in string theory \cite{fay1973theta, DHoker:1988ta, Verlinde:1986kw, Polchinski:1988jq}, and the ambitwistor string discussion here proceeds in close analogy. 

Just as for the string, it will be convenient to use a so-called `plumbing fixture' to explicitly parametrise the moduli space near the separating boundary divisor. As discussed around \cref{equ:sep-degen}, the separating degeneration $\mathfrak{D}^{\text{sep}}$ splits the Riemann surface into two components, in this case two tori $\Sigma_I$ with an additional puncture encoding the node on each component,
\begin{equation}
 \mathfrak{D}^{\text{sep}}_{2,n}\cong \widehat{\mathfrak{M}}_{1,n_1+1}\times\widehat{\mathfrak{M}}_{1,n_2+1}\,,
\end{equation}
and $n=n_1+n_2$.  To parametrise the moduli space near the boundary, let us introduce coordinates $z_I$ on each torus $\Sigma_I$, such that $z_I=0$ will be the nodal point $y_I$ in the degeneration limit, and remove an open neighbourhood $U_I=\{|z_I|<\tau^{1/2}\}$, where $|\tau|<1$ is a coordinate on the unit disk. The two tori, with $U_I$ removed, can now be glued together using an annulus $A_\tau=\{w\in\mathbb{C}\big|\,|\tau|^{1/2}<|w|<|\tau|^{-1/2}\}$ with 
\begin{equation}
 w=\begin{cases}\tau^{1/2}z_1^{-1} & \text{if }|\tau|^{1/2}<|w|<1\,,\\ \tau^{-1/2}z_2 & \text{if }1<|w|<|\tau|^{-1/2}\,.\end{cases}
\end{equation}
In the family of surfaces $\big(\Sigma_1\backslash U_1\big) \cup A_\tau \cup\big(\Sigma_2\backslash U_2\big)$ constructed in this way, the separating degeneration $\mathfrak{D}^{\text{sep}}_{2,n}$ is  given by the singular surface $\tau=0$, where $\Omega_{12} \propto \tau$. Following \cite{fay1973theta} and \cite{DHoker:1988ta}, the asymptotics of the period matrix in this singular limit are given by
\begin{equation}
 \Omega = \begin{pmatrix}\Omega_{11} & 0 \\ 0 & \Omega_{22}\end{pmatrix}+\mathcal{O}(\tau)\,,
\end{equation}
where $\Omega_{II}$ are the modular parameters of the tori $\Sigma_I$. Moreover, the genus-two holomorphic differentials 
$\omega_I$ approach the ones on the two tori  \cite{fay1973theta, DHoker:1988ta},
\begin{equation}\label{equ:hol-diff_sep}
 \omega_I(z)=\begin{cases}\omega_I^{(1)}(z) + \mathcal{O}(\tau) &\text{if } z \in\Sigma_I\,, \\ \mathcal{O}(\tau) &\text{otherwise,}\end{cases}
\end{equation}
where we denoted the holomorphic differentials on the tori by $\omega_I^{(1)}(z) $ for $I=1,2$ respectively. Moreover, the prime form around the separating boundary divisor becomes  \cite{DHoker:1988ta}, 
\begin{equation}\label{equ:prime_sep}
 E(z,w|\Omega)=\begin{cases}E_I^{(1)}(z,w|\Omega_{II}) & \text{if }z,w\in\Sigma_I \,,\\
                E_1^{(1)}(z,y_1|\Omega_{11}) \,w \tau^{-3/4} & \text{if }z\in\Sigma_1,\, w\in A_\tau \,,\\
                E_2^{(1)}(z,y_2|\Omega_{22}) \, \tau^{-1/4} & \text{if }z\in\Sigma_2,\, w\in A_\tau \,,\\
                E_1^{(1)}(z,y_1|\Omega_{11}) E_2^{(1)}(y_2,w|\Omega_{22}) \, \tau^{-1/2} & \text{if }z\in\Sigma_1,\, w\in \Sigma_2 \,,
               \end{cases}
\end{equation}
where $E_I^{(1)}(z,w|\Omega_{II})$ are the prime forms on the respective tori $\Sigma_I$, and $y_I$ denote the extra puncture encoding the node on each torus. In particular, this implies the following asymptotics for the meromorphic differentials
\begin{equation}\label{equ:merom_sep}
 \omega_{w_1,w_2}(z)=\begin{cases} \omega^{(1)}_{w_1,w_2}(z)&\text{if } z,w_1,w_2 \in\Sigma_I\,,  \\ \omega^{(1)}_{w_1,y_I}(z)+\mathcal{O}(\tau)&\text{if }z,w_1 \in \Sigma_I,\,w_2\in \Sigma_J\,,\\\d z/z + \mathcal{O}(\tau)& \text{if } z\in A_\tau,\,w_1\in \Sigma_1,w_2\in \Sigma_2\,,\\
 \mathcal{O}(\tau) &\text{otherwise}\,,\\ \end{cases}
\end{equation}
where $\omega_{w_1,w_2}^{(1)}(z)$ are the meromorphic differentials on the tori, and $\d w/w$ denotes the differential on the annulus.  Upon distributing the marked points on the two tori, \cref{equ:hol-diff_sep,equ:prime_sep} provide all the asymptotics needed to study $P_\mu$ in the separating degeneration,\footnote{Notice that $\omega_{i,\ast}^{(1)}-\omega_{y_1,\ast}^{(1)}=\omega_{i,y_1}^{(1)}$, since both sides of the equation have the same residues. There can be no holomorphic contribution, since we defined the Abelian differentials of the third kind as having vanishing $A$-periods.}
\begin{subequations}
 \begin{align}
  P_\mu(z) & = \ell_{1\,\mu}\omega^{(1)}_1(z) + \sum_{i\in\Sigma_1}k_{i,\mu}\omega_{i,y_1}^{(1)}(z)+\mathcal{O}(\tau)  &&\text{for } z\in \Sigma_1\,,\\
  P_\mu(z) & = \ell_{2\,\mu}\omega^{(1)}_2(z) + \sum_{j\in\Sigma_2}k_{j,\mu}\omega_{j,y_2}^{(1)}(z)+\mathcal{O}(\tau) && \text{for } z\in \Sigma_2\,,\\
  P_\mu(w) & =K_\mu\frac{\d w}{w} +\mathcal{O}(\tau)&& \text{for } w\in A_\tau\,,
 \end{align}
\end{subequations}
%\begin{subequations}
% \begin{align}
%  P_\mu(z) & = \ell_{1\,\mu}\omega^{(1)}_1(z) + \sum_{i\in\Sigma_1}k_{i,\mu}\omega_{i,*}^{(1)}(z)-K_\mu \omega_{y_1,*}^{(1)}(z)+\mathcal{O}(\tau)  &&\text{for } z\in \Sigma_1\,,\\
%  P_\mu(z) & = \ell_{2\,\mu}\omega^{(1)}_2(z) + \sum_{j\in\Sigma_2}k_{j,\mu}\omega_{j,*}^{(1)}(z)+K_\mu \omega_{y_2,*}^{(1)}(z)+\mathcal{O}(\tau) && \text{for } z\in \Sigma_2\,,\\
%  P_\mu(w) & =K_\mu\frac{\d w}{w} +\mathcal{O}(\tau)&& \text{for } w\in A_\tau\,,
% \end{align}
%\end{subequations}
where $K_\mu = \sum_{i\in\Sigma_1}k_{i,\mu}$ is the momentum flowing through the cylinder. The scattering equations thus descend to the separating degeneration as expected, with $n_I$ particle scattering equations on the torus $\Sigma_I$, as well as a modular parameter scattering equation $u_{II}=0$ enforcing $P^2=0$ on $\Sigma_I$. The remaining scattering equation is naturally associated to the annulus (see also  \cite{Adamo:2015hoa}), and can be expressed as
\begin{equation}
 u_{3} = K^2+\mathcal{O}(\tau)\,,
\end{equation}
on the support of the other constraints. Therefore, neither the integrand nor the scattering equations contribute to a pole in $\Omega_{12}\propto \tau$, and the full amplitude \eqref{equ:ampl_schematic} can have at most a simple pole in $1-q_{12}$. To establish that the integral actually vanishes as $\tau\rightarrow 0$, consider again the  integrand as defined in \cref{equ:chiral-int_final}. Since the amplitude is independent of the PCO gauge slice, let us choose both $x_\alpha$ to be located on the connecting cylinder. This is the best we can do after  chosing $x_\alpha$ to be the zeros of the differential $\varpi(z)$ due to the consistency condition
\begin{equation}\label{equ:c12_consistency}
 \frac{c_2}{c_1}=\frac{\omega_1(x_1)}{\omega_1(x_2)}=\frac{\omega_2(x_1)}{\omega_2(x_2)}\,.
\end{equation}
Clearly, this is only satisfied if both $x_\alpha$ lie on the cylinder or on the same torus.\footnote{Of course, we could in principle choose a different PCO gauge that does not require $\varpi(x_\alpha)=0$. Note, however, that the representation of the integrand in \cref{sec:nodalRS_integrand} relies on \cref{equ:c12_consistency}, so amplitudes in a PCO gauge with $\varpi(x_\alpha)\neq0$ do not localise on the nodal Riemann sphere.}
Choosing the former with both $x_\alpha$ on the cylinder, the Szeg\H{o} kernels $S_\delta(x_\alpha,z_i)$ vanish to  order $\mathcal{O}(\tau^{1/4})$ due to the asymptotics of the prime form \eqref{equ:prime_sep},\footnote{See also \cite{Tuite:2010mq} for further details on the degeneration of the Szeg\H{o} kernels, and how to obtain them from a sewing mechanism of lower-genus Riemann surfaces.} while the component $A_{x_1\,x_2}=\px(x_1,x_2)S_\delta(x_1,x_2)$ of the Pfaffian behaves as $\mathcal{O}(\tau)$ on the support of the scattering equation $u_3$, 
\begin{equation}
 \px(x_1,x_2) = K^2\frac{\d x_1\,\d x_2}{x_1\,x_2}+\mathcal{O}(\tau)=\mathcal{O}(\tau)\,.
\end{equation}
The leading order contribution to the pfaffian $\pf\big(M_\delta\big)$  is therefore of order $\mathcal{O}(\tau^{1/2})$,  with the rows and columns associated to $x_\alpha$ contributing $\tau^{1/4}$ each. Moreover, the partition functions \eqref{equ:Zchi_kappa} are of order $\mathcal{O}(1)$ in this gauge, and the chiral integrand thus behaves as $\mathcal{I}_n^{\text{chi}}=\mathcal{O}(\tau^{1/2})$ for each spin structure. Consequently, $\mathcal{I}_n^{\text{chi}} \tilde{\mathcal{I}}_n^{\text{chi}}$ vanishes on the separating degeneration as $\mathcal{O}(\tau)$, confirming that  $f\big(q_{12}\big)=\mathcal{O}(\tau^{-1})$ does not introduce a new pole in the integrand.\footnote{If we had chosen instead the PCO gauge where both $x_\alpha$ lie on the same torus, the Pfaffian would have been of order one, while the partition function contributes $\mathcal{O}(\tau^{1/2})$ for each spin structure. Evidently, this again leads to $\mathcal{I}_n^{\text{chi}}=\mathcal{O}(\tau^{1/2})$.}

Evidently, the discussion above is specific to $q_{12}\rightarrow 1$, and the integrand will {\it not} vanish for other values of $q_{12}$. Proving the uniqueness of $f$ is then straightforward: \eqref{equ:choice_f} is the only function satisfying \eqref{equ:condition_f} with at most a simple pole in $1-q_{12}$ that vanishes as $q_{12}\rightarrow\infty$, as required to retain $\mathfrak{D}^{\text{non-sep}}_{2,n}$ as the only pole.\hfill$\square$

\paragraph{}Since this will play an important role in establishing the absence of degenerate solutions on the bi-nodal Riemann sphere, note that it is sufficient to consider the contribution from a single spin structure to establish the absence of a pole in $\tau$. The full integrand -- including the sum over spin structures -- actually vanishes to higher order in $\tau$ if no additional punctures are present on one of the tori, because all $n$-point amplitudes for $n<4$ vanish in type II supergravity. On the bi-nodal sphere, this argument ensures the absence of a certain type of unphysical pole, discussed in detail in \cref{sec:unphys-poles}, and thereby provides an important check on the amplitude.

%%%%%%%%%%%%%%%%%%%%%%%%%%%%%%%%%%%%%%%%%%%%%%
%%%%%%%%%%%%%%%%%%%%%%%%%%%%%%%%%%%%%%%%%%%%%%

\section{Degeneration of the \texorpdfstring{Szeg\H{o}}{Szego} kernels and the Partition functions}\label{sec:degen}
In this section, we give  explicit expressions for the Szeg\H{o} kernels and the partition functions of even spin structures near the non-separating boundary divisor, up to the relevant orders in the degeneration parameters. These degeneration formulae underly our results in \cref{sec:contour_argument}, and lead in particular to the representation \eqref{equ:int_nodal_final} of the two-loop chiral integand from the bi-nodal Riemann sphere.

Beyond the scope of this article, the non-separating degeneration also plays an important role in superstring theory, for example in the field theory limit (see e.g. \cite{Tourkine:2013rda}) or for modular graph functions \cite{Green:2008uj, DHoker:2013fcx, DHoker:2017pvk}. Due to the strong similarity between the ambitwistor string and string theory, the expressions given here may prove useful in these contexts as well.

\subsection{Degeneration of the \texorpdfstring{Szeg\H{o}}{Szego} Kernels}\label{sec:Szego}
We will focus first on the degeneration of the Szeg\H{o} kernels. Throughout this section, we will work to order $o(q_1, q_2)$ since this is the highest pole present in the partition functions $\mathcal{Z}^{\text{chi}}[\delta]$; see below.
As discussed in \cref{sec:contour_argument}, all subleading terms in the asymptotics of the prime form cancel \eqref{equ:degen_prime-form},
\begin{equation}
 E(z,w) = \frac{z-w}{\sqrt{dz}\sqrt{dw}} +\,\,o(q_1,q_2)\,,
\end{equation}
and thus the degeneration of the Szeg\H{o} kernels depends only on the behaviour of the theta function near the non-separating boundary divisor. Using the expansion \cref{equ:theta_expansion} for the theta functions, the Szeg\H{o} kernels for the NS-NS spin structures only differ by signs and can be summarised conveniently as follows,
\begin{subequations}
\begin{align}
  S_{\delta_1}(z,w) &= S^{(0,0)}_{\text{NS}}(z,w)+q_1\,S^{(1,0)}_{\text{NS}}(z,w)+q_2\,S^{(0,1)}_{\text{NS}}(z,w)+q_1q_2\,S^{(1,1)}_{\text{NS}}(z,w)+o(q_1,q_2)\,,\\
  S_{\delta_2}(z,w) &= S^{(0,0)}_{\text{NS}}(z,w)+q_1\,S^{(1,0)}_{\text{NS}}(z,w)-q_2\,S^{(0,1)}_{\text{NS}}(z,w)-q_1q_2\,S^{(1,1)}_{\text{NS}}(z,w)+o(q_1,q_2)\,,\\
  S_{\delta_3}(z,w) &= S^{(0,0)}_{\text{NS}}(z,w)-q_1\,S^{(1,0)}_{\text{NS}}(z,w)+q_2\,S^{(0,1)}_{\text{NS}}(z,w)-q_1q_2\,S^{(1,1)}_{\text{NS}}(z,w)+o(q_1,q_2)\,,\\
  S_{\delta_4}(z,w) &= S^{(0,0)}_{\text{NS}}(z,w)-q_1\,S^{(1,0)}_{\text{NS}}(z,w)-q_2\,S^{(0,1)}_{\text{NS}}(z,w)+q_1q_2\,S^{(1,1)}_{\text{NS}}(z,w)+o(q_1,q_2)\,.
\end{align}
\end{subequations}
To highlight the relations among the different spin structures, we have used  the following definitions for the `NS-NS Szeg\H{o} kernels' at the relevant orders in the expansion:
\begin{subequations}
\begin{align}
  S^{(0,0)}_{\text{NS}}(z,w)&=\frac{\sqrt{\d z\,\d w}}{(zw)},,\\
  S^{(1,0)}_{\text{NS}}(z,w)&=q_3\frac{(zw)(1^+1^-)^2 \,\sqrt{\d z\,\d w}}{(z1^+)(z1^-)(w1^+)(w1^-)}=q_3\,\omega_{1^+\!,1^-}(z)\, \omega_{1^+\!,1^-}(w)\,\big( S^{(0,0)}_{\text{NS}}(z,w)\big)^{-1}\,,\\
  S^{(0,1)}_{\text{NS}}(z,w)&=q_3\frac{(zw)(2^+2^-)^2 \,\sqrt{\d z\,\d w}}{(z2^+)(z2^-)(w2^+)(w2^-)}=q_3\,\omega_{2^+\!,2^-}(z)\, \omega_{2^+\!,2^-}(w)\,\big( S^{(0,0)}_{\text{NS}}(z,w)\big)^{-1}\,,\\
  S^{(1,1)}_{\text{NS}}(z,w)&=q_3^2\,S^{(1,0)}_{\text{NS}}(z,w)S^{(0,1)}_{\text{NS}}(z,w)\frac{\big((z1^+)(w2^+)(1^-2^-)+(z2^-)(w1^-)(1^+2^+)\big)^2}{(zw)(1^+2^+)(1^-2^-)(1^+2^-)(1^-2^+)\,\sqrt{\d z\,\d w}}\,.
\end{align}
\end{subequations}
Similarly, for the R-NS and NS-R cases, the Szeg\H{o} kernels only differ by a sign in the subleading order,
\begin{subequations}
\begin{align}
  S_{\delta_5}(z,w) &= \frac1{2}\left(S^{(0,0)}_{\text{R2}}(z,w)+q_1\,S^{(1,0)}_{\text{R2}}(z,w)\right)+o(q_1,q_2)\,,\\
  S_{\delta_6}(z,w) &= \frac1{2}\left(S^{(0,0)}_{\text{R2}}(z,w)-q_1\,S^{(1,0)}_{\text{R2}}(z,w)\right)+o(q_1,q_2)\,,\\
  S_{\delta_7}(z,w) &= \frac1{2}\left(S^{(0,0)}_{\text{R1}}(z,w)+q_2\,S^{(0,1)}_{\text{R1}}(z,w)\right)+o(q_1,q_2)\,,\\
  S_{\delta_8}(z,w) &= \frac1{2}\left(S^{(0,0)}_{\text{R1}}(z,w)-q_2\,S^{(0,1)}_{\text{R1}}(z,w)\right)+o(q_1,q_2)\,.
\end{align}
\end{subequations}
To improve the readability of the formulas, we have again defined `R-NS Szeg\H{o} kernels' for the respective loops  (R1 and R2),
\begin{subequations}
\begin{align}
  S^{(0,0)}_{\text{R2}}(z,w)&=\frac{\sqrt{\d z\,\d w}}{(zw)}\left(\sqrt{\frac{(z2^+)(w2^-)}{(z2^-)(w2^+)}}+\sqrt{\frac{(z2^-)(w2^+)}{(z2^+)(w2^-)}}\right)\,,\\
  S^{(1,0)}_{\text{R2}}(z,w)&=q_3\,S^{(1,0)}_{\text{NS}}(z,w)\left(\sqrt{\frac{(1^+2^+)(1^-2^+)(z2^-)(w2^-)}{(1^+2^-)(1^-2^-)(z2^+)(w2^+)}}+\sqrt{\frac{(1^+2^-)(1^-2^-)(z2^+)(w2^+)}{(1^+2^+)(1^-2^+)(z2^-)(w2^-)}}\right)\,,\\
  S^{(0,0)}_{\text{R1}}(z,w)&=\frac{\sqrt{\d z\,\d w}}{(zw)}\left(\sqrt{\frac{(z1^+)(w1^-)}{(z1^-)(w1^+)}}+\sqrt{\frac{(z1^-)(w1^+)}{(z1^+)(w1^-)}}\right)\,,\\
  S^{(0,1)}_{\text{R1}}(z,w)&=q_3\,S^{(0,1)}_{\text{NS}}(z,w)\left(\sqrt{\frac{(2^+1^+)(2^-1^+)(z1^-)(w1^-)}{(2^+1^-)(2^-1^-)(z1^+)(w1^+)}}+\sqrt{\frac{(2^+1^-)(2^-1^-)(z1^+)(w1^+)}{(2^+1^+)(2^-1^+)(z1^-)(w1^-)}}\right)\,.
\end{align}
\end{subequations}
Finally, for the Ramond-Ramond Szeg\H{o} kernels at spin structures $\delta_9$ and $\delta_{10}$, it will be useful to define the following shorthand notation for (square-roots of) cross-ratios involving the marked points as well as the nodes;
\begin{equation}
  v_1=\sqrt{\frac{(z1^+)(w1^-)}{(z1^-)(w1^+)}}\,,\qquad v_2=\sqrt{\frac{(z2^+)(w2^-)}{(z2^-)(w2^+)}}\,,\qquad \text{and recall that } q_3 = \frac{(1^+2^+)(1^-2^-)}{(1^+2^-)(1^-2^+)}\,.
\end{equation}
Using this, the R-R Szeg\H{o} kernels become
\begin{subequations}
\begin{align}
  S_{\delta_9}(z,w) &= \frac{\sqrt{\d z\,\d w}}{2\,(zw)}\Bigg(+\frac{1-q_3^{1/2}}{1-q_3}\left(\frac{v_1}{v_2}+\frac{v_2}{v_1}\right)-\frac{q_3-q_3^{1/2}}{1-q_3}\left(v_1v_2+\frac{1}{v_1v_2}\right) \Bigg)+o(q_1,q_2)\,,\\
  S_{\delta_0}(z,w) &= \frac{\sqrt{\d z\,\d w}}{2\,(zw)}\Bigg(-\frac{1+q_3^{1/2}}{1-q_3}\left(\frac{v_1}{v_2}+\frac{v_2}{v_1}\right)+\frac{q_3+q_3^{1/2}}{1-q_3}\left(v_1v_2+\frac{1}{v_1v_2}\right) \Bigg)+o(q_1,q_2)\,.
\end{align}
\end{subequations}

\subsection{Degeneration of the partition function}\label{sec:partition}
The behaviour of the integrand on the non-separating boundary divisor is governed by two factors: the Pfaffians and the partition functions $\mathcal{Z}^{\text{chi}}[\delta]$. The Szeg\H{o} kernels discussed in the preceding section, together with  $P_\mu$ near the boundary divisor, fully determine the form of the Pfaffian. Here, we focus on the degeneration of the partition function \eqref{equ:Zchi_unsimpl},
\begin{equation}
 \mathcal{Z}^{\text{chi}}[\delta]=\frac{\vartheta[\delta](0)^5\,\vartheta(D_b)\,\prod_{r<s}E(y_r,y_s)\prod_r \sigma(y_r)^3}{Z^{15/2}\,\vartheta[\delta](D_\beta)\,E(x_1,x_2)\,\prod_\alpha\sigma(x_\alpha)\,\det \omega_I\omega_J(y_r)}\,.
\end{equation}
All relevant formulae have already been established, and we can use  \cref{equ:Verlinde} in conjunction with  the expansions of the theta function \eqref{equ:theta_expansion} and the prime form \eqref{equ:degen_prime-form}  to arrive at our results below. Just as for the Szeg\H{o} kernels, we find that the NS-NS partition functions contain the same terms -- up to signs --  in the expansion around the boundary divisor:
\begin{subequations}
\begin{align}
 \mathcal{Z}^{\mathrm{chi}}[\delta_1]&=\mathcal{Z}^{(0,0)}_{\mathrm{NS}}+q_1^{-1}\mathcal{Z}^{(-1,0)}_{\mathrm{NS}}+q_2^{-1}\mathcal{Z}^{(0,-1)}_{\mathrm{NS}}+(q_1q_2)^{-1}\mathcal{Z}^{(-1,-1)}_{\mathrm{NS}}+o(q_1,q_2)\,,\\
 \mathcal{Z}^{\mathrm{chi}}[\delta_2]&=\mathcal{Z}^{(0,0)}_{\mathrm{NS}}+q_1^{-1}\mathcal{Z}^{(-1,0)}_{\mathrm{NS}}-q_2^{-1}\mathcal{Z}^{(0,-1)}_{\mathrm{NS}}-(q_1q_2)^{-1}\mathcal{Z}^{(-1,-1)}_{\mathrm{NS}}+o(q_1,q_2)\,,\\
 \mathcal{Z}^{\mathrm{chi}}[\delta_3]&=\mathcal{Z}^{(0,0)}_{\mathrm{NS}}-q_1^{-1}\mathcal{Z}^{(-1,0)}_{\mathrm{NS}}+q_2^{-1}\mathcal{Z}^{(0,-1)}_{\mathrm{NS}}-(q_1q_2)^{-1}\mathcal{Z}^{(-1,-1)}_{\mathrm{NS}}+o(q_1,q_2)\,,\\
 \mathcal{Z}^{\mathrm{chi}}[\delta_4]&=\mathcal{Z}^{(0,0)}_{\mathrm{NS}}-q_1^{-1}\mathcal{Z}^{(-1,0)}_{\mathrm{NS}}-q_2^{-1}\mathcal{Z}^{(0,-1)}_{\mathrm{NS}}+(q_1q_2)^{-1}\mathcal{Z}^{(-1,-1)}_{\mathrm{NS}}+o(q_1,q_2)\,.
\end{align}
\end{subequations}
This can  be summarised more compactly as
\begin{equation}
  \mathcal{Z}^{\mathrm{chi}}[\delta]=\sum_{n_1,n_2\in\{0,1\}}(-1)^{\delta''_1 n_1+\delta''_2 n_2}q_1^{-n_1}q_2^{-n_2}\mathcal{Z}^{(-n_1,-n_2)}_{\mathrm{NS}}\,,\qquad\text{for }\delta\in\{\delta_1,\delta_2,\delta_3,\delta_4\}\,.
\end{equation}
Again, we have introduced `NS-NS partition functions' at the respective orders to keep the notation compact and highlight the similarities among the spin structures,
\begin{subequations}
 \begin{align}
 & \mathcal{Z}^{(-1,-1)}_{\mathrm{NS}}=\frac{\sqrt{\d x_1 \d x_2}}{(2i\pi)^4(x_1x_2)}\frac{q_3^{-2}}{\omega_{1^+\!,1^-}(x_1)\, \omega_{1^+\!,1^-}(x_2)\,\omega_{2^+\!,2^-}(x_1)\,\omega_{2^+\!,2^-}(x_2)}\,,\\
 & \mathcal{Z}^{(-1,0)}_{\mathrm{NS}}\hspace{6pt}=\frac{\sqrt{\d x_1 \d x_2}}{(2i\pi)^4(x_1x_2)}\frac{q_3^{-1}}{\omega_{1^+\!,1^-}(x_1)\, \omega_{1^+\!,1^-}(x_2)}Z_8^{(-1,0)}\,,\\
 & \mathcal{Z}^{(0,-1)}_{\mathrm{NS}}\hspace{6pt}=\frac{\sqrt{\d x_1 \d x_2}}{(2i\pi)^4(x_1x_2)}\frac{q_3^{-1}}{\omega_{2^+\!,2^-}(x_1)\,\omega_{2^+\!,2^-}(x_2)}Z_8^{(0,-1)}\,,\\
  &\mathcal{Z}^{(0,0)}_{\mathrm{NS}}\hspace{12pt}=10\,q_3\big(1+3q_3+q_3^2\big)\,\mathcal{Z}^{(-1,-1)}_{\mathrm{NS}} + \frac{\sqrt{\d x_1 \d x_2}}{(2i\pi)^4(x_1x_2)}\Big(2Z_3^{(-1,0)}Z_3^{(0,-1)}-Z^{(0,0)}\Big)\,,
 \end{align}
\end{subequations}
where the factors of $Z_a^{(-1,0)}$, $Z_a^{(0,-1)}$ and $Z^{(0,0)}$ are given by
 \begin{align*}
  &Z_a^{(-1,0)}=\frac{a}{\omega_{2^+\!,2^-}(x_1)\omega_{2^+\!,2^-}(x_2)}-\frac{\big((x_12^+)(x_22^+)(2^-1^+)(2^-1^-)-(x_12^-)(x_22^-)(2^+1^+)(2^+1^-)\big)^2}{(2^+2^-)^2(2^+1^+)(2^+1^-)(2^-1^+)(2^-1^-)\,\,\d x_1 \d x_2}\,,\\
  &Z_a^{(0,-1)}=\frac{a}{\omega_{1^+\!,1^-}(x_1)\omega_{1^+\!,1^-}(x_2)}-\frac{\big((x_11^+)(x_21^+)(1^-2^+)(1^-2^-)-(x_11^-)(x_21^-)(1^+2^+)(1^+2^-)\big)^2}{(1^+1^-)^2(1^+2^+)(1^+2^-)(1^-2^+)(1^-2^-)\,\,\d x_1 \d x_2}\,,\\
  &Z^{(0,0)}=\frac{\Bigg(\big((x_11^+)(x_21^+)(x_12^+)(x_22^+)(1^-2^-)^2\big)^2+\big(\sigma_{1^+}\leftrightarrow \sigma_{1^-}\big)+\big(\sigma_{2^+}\leftrightarrow \sigma_{2^-}\big)+\begin{pmatrix}\sigma_{1^+}\leftrightarrow \sigma_{1^-}\\ \sigma_{2^+}\leftrightarrow\sigma_{2^-}\end{pmatrix}\Bigg)}{(1^+1^-)^2(2^+2^-)^2\,(1^+2^+)(1^+2^-)(1^-2^+)(1^-2^-)\,\,\d x_1^2 \d x_2^2}\,.
 \end{align*}
 As expected, the partition functions carry form degree $-3/2$ in both PCO insertion points $x_\alpha$, exactly balanced by the Pfaffians. To see this, note that each term in a Pfaffian is proportional to either $\px(x_1,x_2)S_\delta(x_1,x_2)$ or the product $\prod_{\alpha=1,2}P(x_\alpha)\cdot v_{i_\alpha}\,S_\delta(x_\alpha,z_{i_\alpha})$, with $v_{i_\alpha}\in\{k_{i_\alpha},\epsilon_{i_\alpha}\}$, and thus carries form degree $+3/2$ in each $x_\alpha$.

Similarly to the NS-NS case, the expansions of the  NS-R and R-NS partition functions differ only by relative signs in the subleading order,
 \begin{subequations}
\begin{align}
  \mathcal{Z}^{\mathrm{chi}}[\delta_5]&=\mathcal{Z}_{\mathrm{R2}}^{(0,0)}+q_1^{-1}\mathcal{Z}_{\mathrm{R2}}^{(-1,0)}+\,o(q_1,q_2)\,,\\
  \mathcal{Z}^{\mathrm{chi}}[\delta_6]&=\mathcal{Z}_{\mathrm{R2}}^{(0,0)}-q_1^{-1}\mathcal{Z}_{\mathrm{R2}}^{(-1,0)}+\,o(q_1,q_2)\,,\\
  \mathcal{Z}^{\mathrm{chi}}[\delta_7]&=\mathcal{Z}_{\mathrm{R1}}^{(0,0)}+q_2^{-1}\mathcal{Z}_{\mathrm{R1}}^{(0,-1)}+\,o(q_1,q_2)\,,\\
  \mathcal{Z}^{\mathrm{chi}}[\delta_8]&=\mathcal{Z}_{\mathrm{R1}}^{(0,0)}-q_2^{-1}\mathcal{Z}_{\mathrm{R1}}^{(0,-1)}+\,o(q_1,q_2)\,.
\end{align}
 \end{subequations}
 To define the `R-NS partition functions' $\mathcal{Z}^{-n_1,-n_2}_{\mathrm{R1}}$ and $\mathcal{Z}^{-n_1,-n_2}_{\mathrm{R2}}$, it will be useful to introduce two further square-roots of cross-ratios, in this case involving the PCO gauge insertion points $x_\alpha$ and the nodes,
  \begin{equation}
  v_1^\pm = \sqrt{\frac{(x_11^+)(x_21^+)(2^\pm1^-)^2}{(x_11^-)(x_21^-)(2^\pm1^+)^2}}\,,\qquad\text{ and } v_2^\pm = \sqrt{\frac{(x_12^+)(x_22^+)(1^\pm2^-)^2}{(x_12^-)(x_22^-)(1^\pm2^+)^2}}\,.
 \end{equation}
 Using this, the terms in the R-NS partition functions are given by
\begin{subequations}
 \begin{align}
  & \mathcal{Z}_{\mathrm{R2}}^{(0,0)}\hspace{7pt} = -4\, q_3^2\,\mathcal{Z}_{\mathrm{NS}}^{(-1,-1)} \left(\sqrt{v_2^+v_2^-}+\frac{1}{\sqrt{v_2^+v_2^-}}\right)^{-1}\,,\\
  &\mathcal{Z}_{\mathrm{R2}}^{(-1,0)} = -4 \frac{\sqrt{\d x_1 \d x_2}}{(2i\pi)^4(x_1x_2)}\frac{q_3^{-1}}{\omega_{2^+\!,2^-}(x_1)\omega_{2^+\!,2^-}(x_2)}\hat Z_{\mathrm{R2}}^{(-1,0)}\,,\\
  & \mathcal{Z}_{\mathrm{R1}}^{(0,0)} \hspace{6pt}= -4 \,  q_3^2\,\mathcal{Z}_{\mathrm{NS}}^{(-1,-1)} \left(\sqrt{v_1^+v_1^-}+\frac{1}{\sqrt{v_1^+v_1^-}}\right)^{-1}\,,\\
  &\mathcal{Z}_{\mathrm{R1}}^{(-1,0)} = -4 \frac{\sqrt{\d x_1 \d x_2}}{(2i\pi)^4(x_1x_2)}\frac{q_3^{-1}}{\omega_{1^+\!,1^-}(x_1)\omega_{1^+\!,1^-}(x_2)}\hat Z_{\mathrm{R1}}^{(0,-1)}\,,
 \end{align}
\end{subequations}
For the sake of readability, we introduced a short-hand notation for the Ramond part of the subleading term of the partition function,
\begin{subequations}
 \begin{align}
 \hat Z_{\mathrm{R2}}^{(-1,0)}&=\frac{5}{ \omega_{1^+\!,1^-}(x_1)\omega_{1^+\!,1^-}(x_2)}\;\frac{v_2^+ +v_2^-}{v_2^+v_2^-+1}
  -\frac{v_2^+}{v_2^+v_2^-+1} \;Z_{\text{R2}}^{\text{cr}}\,,\\
  \hat Z_{\mathrm{R1}}^{(0,-1)}&=\frac{5}{ \omega_{2^+\!,2^-}(x_1)\omega_{2^+\!,2^-}(x_2)}\;\frac{v_1^+ +v_1^-}{v_1^+v_1^-+1}
  -\frac{v_1^+}{v_1^+v_1^-+1} \;Z_{\text{R1}}^{\text{cr}}\,.
 \end{align}
\end{subequations}
and defined the following product of cross-ratios:
\begin{align*}
 Z_{\text{R2}}^{\text{cr}}&=\frac{(x_11^+)^2(x_21^+)^2(1^-2^+)(1^-2^-)\,\,\big((x_12^-)(x_22^-)(1^-2^+)^2+(x_12^+)(x_22^+)(1^-2^-)^2\big)+\big(1^+\leftrightarrow 1^-\big)}{(1^+1^-)^2(1^+2^-)(1^-2^+)\,\,\big((x_12^+)(x_22^+)(1^+2^-)(1^-2^-)+(x_12^-)(x_22^-)(1^+2^+)(1^-2^+)\big)\,\d x_1\d x_2}\,,\\
  Z_{\text{R1}}^{\text{cr}}&=\frac{(x_12^+)^2(x_22^+)^2(2^-1^+)(2^-1^-)\,\,\big((x_11^-)(x_21^-)(2^-1^+)^2+(x_11^+)(x_21^+)(2^-1^-)^2\big)+\big(2^+\leftrightarrow 2^-\big)}{(2^+2^-)^2(2^+1^-)(2^-1^+)\,\,\big((x_11^+)(x_21^+)(2^+1^-)(2^-1^-)+(x_11^-)(x_21^-)(2^+1^+)(2^-1^+)\big)\,\d x_1\d x_2}\,.
\end{align*}
 To conclude, we give the (single) term in the expansion of the Ramond-Ramond partition functions to the relevant order,

\begin{subequations}
 \begin{align}
 \mathcal{Z}^{\mathrm{chi}}[\delta_9]&= -4 \,q_3^2\,  \mathcal{Z}_{\mathrm{NS}}^{(-1,-1)} \frac{\left(1+ q_3^{1/2}\right)^{5}}{q_3\left(v^{+-}+\frac{1}{v^{+-}}\right)+ q_3^{1/2}\left(v^{++}+v^{--}\right)}+\,o(q_1,q_2)\,,\\
 \mathcal{Z}^{\mathrm{chi}}[\delta_0]&= -4 \,q_3^2\,  \mathcal{Z}_{\mathrm{NS}}^{(-1,-1)} \frac{\left(1- q_3^{1/2}\right)^{5}}{q_3\left(v^{+-}+\frac{1}{v^{+-}}\right)- q_3^{1/2}\left(v^{++}+v^{--}\right)}+\,o(q_1,q_2)\,,
 \end{align}
\end{subequations}
where we used the following definition for the cross-ratios,
\begin{subequations}
 \begin{align}
 v^{++}&=\frac{(1^-2^-)^2}{(1^+2^-)(1^-2^+)}\sqrt{\frac{(x_11^+)(x_21^+)(x_12^+)(x_22^+)}{(x_11^-)(x_21^-)(x_12^-)(x_22^-)}}\,,\\
 v^{--}&=\frac{(1^+2^+)^2}{(1^+2^-)(1^-2^+)}\sqrt{\frac{(x_11^-)(x_21^-)(x_12^-)(x_22^-)}{(x_11^+)(x_21^+)(x_12^+)(x_22^+)}}\,,\\
 v^{+-}&=\frac{(1^-2^+)}{(1^+2^-)}\sqrt{\frac{(x_11^+)(x_21^+)(x_12^-)(x_22^-)}{(x_11^-)(x_21^-)(x_12^+)(x_22^+)}}\,.
\end{align}
 \end{subequations}

\bibliography{twistor-bib}
\bibliographystyle{JHEP}

\end{document}